\documentclass[aps,pra,amssymb,twocolumn,showpacs,a4paper]{revtex4-1}

\usepackage{amsmath,amssymb,epsfig,graphicx,color,framed}
\usepackage[english]{babel}
\usepackage{hyperref}
\usepackage{amsfonts} 
\usepackage{amsmath}
\usepackage{amssymb}
\usepackage{graphicx}
\usepackage{hyperref}
\usepackage{color}
\usepackage{natbib}  
\usepackage{bm} 
\usepackage{isomath} 
\usepackage{braket}

\renewcommand{\H}{{\cal H}}
\newcommand{\vect}[1]{\vectorsym{#1}} 

\begin{document}

\title{Quantum dissipation of planar harmonic systems: Maxwell-Chern-Simons theory}

\author{Antonio A. Valido}
\email{a.valido@iff.csic.es}
\affiliation{Instituto de F\'isica Fundamental IFF-CSIC, Calle Serrano 113b, 28006 Madrid, Spain \\
QOLS, Blackett Laboratory, Imperial College London, London SW7 2AZ, United Kingdom}

\date{\today}


\begin{abstract}
Conventional Brownian motion in harmonic systems has provided a deep understanding of a great diversity of dissipative phenomena. We address a rather fundamental microscopic description for the (linear) dissipative dynamics of two-dimensional harmonic oscillators that contains the conventional Brownian motion as a particular instance. This description is derived from first principles in the framework of the so-called Maxwell-Chern-Simons electrodynamics, or also known, Abelian topological massive gauge theory. Disregarding backreaction effects and endowing the system Hamiltonian with a suitable renormalized potential interaction, the conceived description is equivalent to a minimal-coupling theory with a gauge field giving rise to a fluctuating force that mimics the Lorentz force induced by a particle-attached magnetic flux. We show that the underlying symmetry structure of the theory (i.e. time-reverse asymmetry and parity violation) yields an interacting vortex-like Brownian dynamics for the system particles. An explicit comparison to the conventional Brownian motion in the quantum Markovian limit reveals that the proposed description represents a second-order correction to the well-known damped harmonic oscillator, which manifests that there may be dissipative phenomena intrinsic to the dimensionality of the interesting system.
\end{abstract}

\maketitle

\section{Introduction}
The study of the physical process whereby an interesting system reaches asymptotically a stationary state following a dissipative dynamics is ubiquitous in several areas of physics such as quantum thermodynamics \cite{weiss20121}, condensed matter physics \cite{garciaripoll20091,sieberer20161}, or cosmology \cite{unruh19891,gautier20121,calzetta20081,anisimov20091,boyanovsky20151}. Unfortunately, this constitutes an intricate open-system theory problem  \cite{weiss20121,devega20171,breuer20021,grabert19881,rivas20121,caldeira20141} for which there is no a "universal" recipe that could successfully provide a rigorous solution. Indeed only a few specific instances can be exactly solved, among which, the (linear) quantum Brownian motion in harmonic systems (generically known as the damped harmonic oscillator \cite{grabert19841,riseborough19851,haake19851,boyanovsky20171,hu19921,alamoudi19981,alamoudi19991,ford19881}) represents a prominent example \cite{hanggi20051,philbin20121}. One of the most fruitful approaches to the latter rest on assuming that the microscopic Hamiltonians describing both the environment and system-environment interaction basically consist of a large set of non-interacting harmonic oscillators linearly coupled to the system. This is commonly refereed to as the Feynman-Vernon \cite{feynman19631}, Caldeira-Leggett \cite{caldeira19831} or independent-oscillator model \cite{grabert19881,ford19651,ford19881}, and recently, it has been employed to investigate quantum thermometry \cite{correa20151,hovhannisyan20181,correa20171}, and non-equilibrium quantum thermodynamics or information properties  \cite{valido20151,valido20132,boyanovsky20172,hsiang20181,charalambous20181,venkataraman20141}.

Although this standard model may look somehow artificial, it resemblances to the Pauli-Fierz Hamiltonian \cite{ruggenthaler20181,rokaj20181} in the strict dipole-approximation when the latter describres simple charged particles interacting with Maxwell electromagnetic fields \cite{kohler20131,valido20131,ford19881,rzazewski19761,efimov19941,reuther20101}. That is, the independent-oscillator model is essentially a particular instance of the non-relativistic Maxwell electrodynamics \cite{cohen19971,landau19711}. Remarkably, along with the usual Maxwell contribution, the action of the two-dimensional Abelian electrodynamics admits a Chern-Simons kinetic term \cite{jackiw19902} which preserves the essential ingredients demanded for a sensible (Abelian) gauge theory \cite{dunne19991}: space-time locality, as well as local $U(1)$-gauge and Lorentz invariance. This is the so-called Maxwell-Chern-Simons electrodynamics \cite{deser19821} or Abelain topological massive gauge theory \cite{dunne19991,matsuyama19901}, and has been successfully applied to study new forms of gauge field mass generation \cite{deser19821,dunne19991}, the dynamics of vortices \cite{dunne19901,horvathy20091}, or the statistics transmutation \cite{matsuyama19901} which have recently found appealing applications in quantum computation theory \cite{pachos20121}. Then two immediate question arises as to which kind of microscopic description is brought by this more fundamental gauge theory, and further, whether it could shed new light on two-dimensional dissipative dynamics. 

Motivated by these natural questions, the present work is devoted to extensively examine the Abelian topological massive gauge theory from the perspective of the quantum open-system theory. More concretely, we address the dissipative dynamics in the low-energy regime of a two-dimensional system composed of charged harmonic oscillators minimally interacting with a Maxwell-Chern-Simons electromagnetic field acting as a heat bath. Starting from first principles, we derive a low-lying Hamiltonian that provides a reliable and (numerically) solvable dissipative microscopic description within the Langevin-equation framework \cite{gautier20121,alamoudi19981,hanggi20051}. Interestingly, the Chern-Simons effects give rise to a Lorentz-like fluctuating force which represents an alternative to the geometric magnetism \cite{campisi20121} in the context of recently extended environments \cite{guo20161,yao20171}. Unlike previous treatments, we show that the components of such Chern-Simons (electric) force are non-commutative owing to the "topological" nature of the underlying theory, and cause an (ordinary) Hall response of the system particles that recalls the dissipative Hofstadter model \cite{callan19921,novais20051}. We also show that this response is enable to generate stationary correlations between the transversal degrees of freedom in the quantum regimen, which may eventually induce new kinds of environmental-mediated entanglement between the system particles different from the standard dissipative models \cite{valido20131}. Moreover, the Chern-Simons kinetic term endows the Brownian motion with unusual statistical features that enrich the dissipative dynamics, for instance it prompts an anti-symmetric 1/f noise in the classical Markovian Langevin equation that closely resemblances to the low-frequency magnetic flux noise in superconducting circuits \cite{anton20131,lee20081}. Our main concern is to analyze the main characteristics of the novel dissipative dynamics provided by the Chern-Simons action as compared to the conventional Brownian motion. Let us stress that our motivation as well as approach is significantly distinct to most previous treatments within quantum open-system theory to the best of our knowledge \cite{weiss20121,devega20171,breuer20021,grabert19881,rivas20121,caldeira20141}, in particular, those related to the Brownian motion of charged particles moving in the presence of external magnetic fields \cite{gupta20111,czopnik20011,li19901,chun20181}. The present work is in the line to explore the intriguing interplay between dissipation and the latent symmetry structure of the interesting problem (e.g. the influence of time-reverse symmetry or parity conservation on the spectral density), in much the same fashion as Refs.\cite{cobanera20161,diehl20111,viyuela20121}. 

The present paper is organized as follows. In Sec.\ref{SGID} the quantum canonical Hamiltonian governing the whole dynamics is obtained via a Coulomb gauge quantization procedure by starting from the action characteristic of the Maxwell-Chern-Simons electrodynamics of a harmonic $n$-particle system. From this, in Sec.\ref{SQD} we derive the dissipative microscopic Hamiltonian which is the basis of the present work, and extensively discussed its properties. The reduced dynamics of the system particles is addressed in terms of the Langevin equation formalism in Secs. \ref{SGLE}, \ref{PSNDR}, and \ref{SSP}, as well as we study the asymptotic properties of the fluctuation-dissipation relation and the conditions under which the system relax towards a thermal equilibrium state. The Secs. \ref{MLDS} and \ref{SHOS} provide an explicit example of the proposed dissipative description applied to study the Markovian dynamics. Finally, we summarize and draw the main conclusions in Sec.\ref{OCR}.

\section{Gauge invariant description}\label{SGID}
As stated in the introduction, we consider the most general action in Euclidean planar geometry of a $U(1)$ gauge-invariant system composed of $n$ harmonic oscillators coupled to a homogeneous and isotropic gauge field $A_{\mu}=(A_{0},\vect A)$. This is given by
\begin{eqnarray}
S&=&S_{HO}+\int d^{2}\vect x dt \ \mathcal{L}_{MCS}(\vect x,t),
\label{LDAMCS}
\end{eqnarray}
where $S_{HO}$ is the usual action for the reduced harmonic system and $\mathcal{L}_{MCS}$ represents the Lagrangian density of the Maxwell-Chern-Simons electrodynamics \cite{dunne19991,deser19821,jackiw19902,pachos20121}, i.e.
\begin{eqnarray}
\mathcal{L}_{MCS}&=& \frac{1}{2}\big(\vect E^{2}-B^{2}\big)+\frac{\kappa}{2}\epsilon^{\mu\nu\lambda}A_{\mu}\partial_{\nu}A_{\lambda}+\vect{A} \cdot\vect{J}+A_{0}\rho , 
\nonumber 
\end{eqnarray}
with $\epsilon^{\mu\nu\lambda}$ being the completely antisymmetric tensor (i.e. $\epsilon_{012}=1$ and $\epsilon_{ij}=\epsilon_{0ij}$). Here $B$ and $\vect E$ are the magnetic and electric fields (i.e. $B=\epsilon_{\alpha\beta}\partial_{\alpha}A_{\beta}$ and $E_{\alpha}=-\dot{A}_{\alpha}-\partial_{\alpha}A_{0}$), whereas $\rho$ and $\vect{J}$ are respectively the charge and current densities of the harmonic $n$-particle system. The second term in the Lagrangian density (\ref{LDAMCS}) describes the Chern-Simons action whose strength is given by the coupling constant $\kappa$, while the first one is the usual Maxwell kinetic term. Importantly, we shall show that well-known results for the damped harmonic oscillator  \cite{ford19881,caldeira19831,hanggi20051,riseborough19851,haake19851,valido20131} are recovered in any step of the treatment by taking the limit $\kappa\rightarrow 0$. Throughout this work, Latin indices (running from 1 to $n$) are reserved to the system harmonic oscillators, and unless stated otherwise, we use Greek letters as well as Einstein convention of repeated indices for the two spatial dimensions. We use the natural units $c=\hbar=k_{B}=1$.

We shall consider that the total density matrix for the harmonic $n$-particle system and gauge field decouples at the initial time $t_{0}$, and further, the field is in a canonical equilibrium state $\hat\rho_{\beta}\propto e^{-\beta \hat \H_{MSC}}$ with $\hat \H_{MSC}$ being the free Hamiltonian of the Maxwell-Chern-Simons gauge field (defined below), whilst the system may be in an arbitrary state. The restriction to free-correlation initial conditions is not crucial for the subsequent treatment, rather it is an extensively used assumption \cite{grabert19881,breuer20021,hanggi20051,haake19851} that provides a better exposition. Intuitively, this in agreement with preparing the system separately and brought into contact with the gauge field sufficiently fast such that the subsequent dynamics is governed by the Maxwell-Chern-Simons action (\ref{LDAMCS}). As a result, $\hat\rho_{\beta}$ will completely characterize the statistical properties of the gauge field operators, and, as we shall see, the system particle operators in the asymptotic time limit as well.
 
Without loss of generality we assume that all harmonic oscillators possess identical mass $m$, while distinct (bare) frequencies $\omega_{i}$ for $i\in \left\lbrace 1,n\right\rbrace $. Moreover, the charge and current densities can be expressed in terms of a function $\varphi$ that models the charge distribution of each harmonic oscillator \cite{jackiw19901},
\begin{eqnarray}
\rho(\vect x)&=&e\sum_{i=1}^{n}\varphi(\vect x- \vect q_{i}(t)),  \nonumber \\
\vect{J}(\vect x)&=&e\sum_{i=1}^{n}\varphi(\vect x- \vect q_{i}(t))\dot{\vect q}_{i}(t),
\label{CDE}
\end{eqnarray}
where $0< e$ determining the coupling strength to the gauge field $\vect A$, and $\vect q_{i}(t)$ denotes the spatial coordinate of the $i$-th harmonic oscillator. For seek of simplicity, we shall assume an identical $\varphi$ for all particles.

From the Lagrangian density described by the general action (\ref{LDAMCS}), one obtains the following expressions for the canonical momentum of the $i$-th harmonic oscillator $\vect p_{i}$ and gauge field $\vect \Pi$, \cite{landau19711} 
\begin{eqnarray}
\Pi_{0}&=&0, \label{FCCI} \\
\Pi_{\alpha}&=&\dot{ A}_{\alpha}+\partial_{\alpha}A_{0}+\frac{\kappa}{2}\epsilon_{\alpha \beta} A_{\beta}, \nonumber \\
\vect p_{i}&=&m \dot{\vect q}_{i}+e\int d^2\vect x \ \varphi(\vect x- \vect q_{i})\vect A(\vect x),
\end{eqnarray}
as well as the Gauss law \cite{deser19821,jackiw19901,jackiw19902,matsuyama19901}
\begin{equation}
\nabla\vect \Pi+\frac{\kappa{\color{red}}}{2}\nabla\times\vect A-\rho=0,
\label{FCCII}
\end{equation}
which upon surface integration unveils that the harmonic $n$-particle system possesses a magnetic-like flux of strength proportional to $n e/\kappa$ \cite{deser19821,dunne19991}. The latter may be seen by realizing that the contribution from $\nabla\vect \Pi$ vanishes since this represents the longitudinal electric field which here exponentially decays owing to the photon mass \cite{pachos20121}. One may show that the classical Hamiltonian obtained via canonical procedure from the action (\ref{LDAMCS}) reads \cite{devecchi19951}
\begin{eqnarray}
\H&=&\sum_{i=1}^{n}\bigg(\frac{1}{2m}\left(\vect p_{i}-e\int d^2\vect x \varphi(\vect x- \vect q_{i})\vect A(\vect x)\right)^{2}+ V(\vect q_{i}) \bigg)  \nonumber \\
&+&\frac{1}{2}\int d^{2}\vect x \bigg(\vect{\Pi}\cdot\vect{\Pi}-\kappa \vect{\Pi}\times\vect{A}+(\nabla\times \vect{A})^{2}+\frac{\kappa^{2}}{4}\vect{A}\cdot\vect{A}   \nonumber \\
&+&A_{0}\Big(\nabla\cdot\vect{\Pi}+\frac{\kappa}{2}\nabla\times\vect{A}-\rho\Big)\bigg).
\label{CHE}
\end{eqnarray}
where $V$ stands for the isotropic confining harmonic potential of the oscillators, which we shall take as $V(\vect q_{i})=\frac{1}{2}m\omega_{i}^{2}(\vect q_{i}-\bar{\vect q}_{i})^{2}$ for simplicity, with $\bar{\vect q}_{i}$ being the equilibrium position of the $i$th harmonic oscillator.

We now quantize the Hamiltonian (\ref{CHE}) preserving the gauge invariance by following the conventional Coulomb gauge quantization procedure \cite{deser19821,devecchi19951}. By writing the gauge field variables (\ref{CHE}) in terms of the longitudinal $(\vect A^{||},\vect \Pi^{||})$ and transversal components $(\vect A^{\perp},\vect \Pi^{\perp})$, this means to set $\vect A^{||}$ equal to zero, whilst $\vect \Pi^{||}$ is evaluated by demanding the Gauss law (\ref{FCCII}) as a non-dynamical constraint \cite{deser19821,iengo19921,matsuyama19901,bazeia19971}, i.e.
\begin{equation}
\vect{\Pi}^{\parallel}(\vect x)=\nabla_{\vect x}\int d \vect y \ G_{c}(\vect x-\vect y)\left(\rho(\vect y)-\frac{\kappa}{2} \nabla \times\vect{A}_{\perp}(\vect y) \right),
\label{CII}
\end{equation}
where $G_{c}(\vect x)$ is the two-dimensional Coulomb Green's function that satisfy $\nabla^{2}G_{c}(\vect x)=\delta^{2}(\vect x)$, i.e. $G_{c}(\vect x-\vect y)=(2\pi)^{-1}\log |\vect x-\vect y|$ \cite{dunne19991,jackiw19901,bazeia19971}. The quantization of the Hamiltonian is achieved by firstly imposing the Coulomb gauge equal-time commutation relations \citep{deser19821,iengo19921},
\begin{eqnarray}
\left[ \hat A_{\beta}^{\perp}(\vect y), \hat \Pi_{\alpha}^{\perp}(\vect x)\right]&=&i\delta_{\alpha \beta}^{\perp}(\vect x-\vect y),  \nonumber \\
\left[\hat q_{i},\hat p_{i}\right]&=&i\delta_{ij}.
\label{ECMR}
\end{eqnarray}
where $\delta_{\alpha \beta}^{\perp}(\vect x-\vect y)$ denotes the transverse delta function \cite{matsuyama19901},
\begin{eqnarray}
\delta_{\alpha \beta}^{\perp}(\vect x-\vect y)&=&\left(\delta_{\alpha \beta}-\frac{\partial_{\alpha}^{(\vect x)}\partial_{\beta}^{(\vect x)}}{\nabla_{\vect x}^{2}} \right)\delta^{(2)}(\vect x-\vect y) \nonumber \\
&=&\mathcal{P}_{\alpha \beta}(\vect x)\delta^{(2)}(\vect x-\vect y),
\end{eqnarray}
and $\mathcal{P}_{\alpha \beta}(\vect x)$ is called the transverse projective operator. Notice that all other commutators vanish identically. The quantum canonical Hamiltonian governing all the system-field dynamics is then obtained from (\ref{CHE}) after the separation of the transverse and longitudinal components of the quantum gauge field and the replacement of the gauge-fixing constraints (see Appendix\ref{app1} for further deatils). By substituting the expressions for the charge densities (\ref{CDE}) and the Coulomb Green's function in the obtained Hamiltonian (given by \ref{HMCS1}), it is simply to verify that the latter can be cast as follows
\begin{eqnarray}
\hat \H&=&\sum_{i=1}^{n}\bigg(\frac{1}{2m}\left(\hat{\vect p}_{i}-e \int d^2 \vect x \ \varphi(\vect x- \hat{\vect q_{i}})\hat{\vect A}^{\perp}(\vect x)\right)^{2}  \label{HMCS2} \\
&+&\hat  V(\hat{\vect q}_{i}) +\sum_{j=1}^{n}\hat V_{c}(\hat{\vect q}_{i}-\hat{\vect q}_{j})\bigg) +\hat \H_{CS}+ \hat \H_{MCS} ,
\nonumber
\end{eqnarray}
with 
\begin{equation}
\hat \H_{MCS}=\frac{1}{2}\int d^2 \vect x\Big(\hat{\vect{\Pi}}^{\perp}\cdot\hat{\vect{\Pi}}^{\perp}+\hat{A}^{\perp}_{\alpha}\left( -\nabla^2 +\kappa^{2}\right)\hat{ A}^{\perp}_{\alpha}\Big),
\label{HGF}
\end{equation}
and where we have defined the system-field interaction term characteristic of the Chern-Simons action,
\begin{equation}
\hat\H_{CS}=\frac{e\kappa}{2\pi}\sum_{i=1}^{n}\int d^2\vect x d^2\vect y  \ \varphi(\vect x-\hat{\vect q}_{i}) \log|\vect x-\vect y|\nabla \times \hat{\vect{A}}^{\perp}(\vect y),
\label{HMCS3}
\end{equation}
and the commonly known (two-dimensional) Coulomb potential \cite{moura20011},
\begin{equation}
\hat V_{c}(\hat{\vect q}_{i}-\hat{\vect q}_{j})
=\frac{-e^{2}}{4\pi}\sum_{i,j=1}^{n}\int d^2\vect x d^2\vect y \log|\vect x-\vect y|\varphi(\vect x-\hat{\vect q}_{i})\varphi(\vect y-\hat{\vect q}_{j}).
\label{COUP}
\end{equation}
Note that (\ref{HMCS3}) and (\ref{COUP}) emerge from the interaction between the system particles and the longitudinal part of the MSC gauge field via the Gauss law, so they are fundamental for avoiding a gauge-invariance breaking. Here, $\hat \H_{MCS}$ models the Maxwell-Chern-Simons (MSC) environmental Hamiltonian, whilst all the system-environment interaction is mediated by (\ref{HMCS3}) and the minimal coupling to the gauge field appearing in (\ref{HMCS2}). In this way, the proposed description has two main characteristic that distinguish it from previous treatments \cite{campisi20121,gupta20111,yao20171}: (i) the MSC environmental spectrum is gaped by $\kappa$ due to the Chern-Simons action endows the environmental quasiparticle excitations with a "toplogical" mass $\kappa$, and additionally, (ii) the Chern-Simons action attaches a magnetic-like flux to each system particle \cite{dunne19991,deser19821} (see Eq.(\ref{FCCII})) that give rise to an effective charge-flux coupling between the harmonic oscillator mediated by $\hat\H_{CS}$. In the next section, we build the dissipative microscopic description upon the canonical Hamiltonian (\ref{HMCS2}).  

For latter purposes it is convenient to express the environmental Hamiltonian (\ref{HGF}) in terms of the quasiparticle excitations of the MCS gauge field \cite{deser19821},
\begin{equation}
\hat \H_{MCS}=\sum_{\vect k}\omega(\vect k)a^{\dagger}(\vect k)a(\vect k)+E_{0},
\end{equation}
where $a^{\dagger}(\vect k)$ ($a(\vect k)$) stands for the creation (annihilation) operator for the gauge field mode $\vect k\in(2\pi/L)\mathbb{Z}^2$ and excitation frequency
\begin{equation}
\omega(\vect k)=\sqrt{|\vect k|^{2}+\kappa^{2}},
\label{DRB}
\end{equation}
and $E_{0}=\sum_{\vect k}\frac{\omega(\vect k)}{2}$ is the usual vacuum expectation value of the gauge field. Since $E_{0}$ does not play a crucial role in the dissipative dynamics, this can be disregarded in the future treatment by redefining the quasiparticle operators. Furthermore, it is advantageous to express the canonical variables of the MCS gauge field in terms of the complete set of polarized plane waves,
\begin{eqnarray}
\hat{A}^{\perp}_{\alpha}(\vect x)&=&\sum_{\vect k}\frac{L^{-1}e^{i\vect k\cdot\vect x }}{2\pi\sqrt{2\omega(\vect k)}}(\varepsilon_{\alpha}(\vect k) \hat a(\vect k)+\varepsilon_{\alpha}^{\dagger}(-\vect k)\hat a^{\dagger}(-\vect k)), \nonumber \\
\hat \Pi^{\perp}_{\alpha}(\vect k)&=&\sum_{\vect k}\frac{(iL)^{-1}e^{i\vect k\cdot\vect x }}{2\pi\sqrt{2\omega^{-1}(\vect k)}}(\varepsilon_{\alpha}(\vect k)\hat a(\vect k)-\varepsilon^{\dagger}_{\alpha}(-\vect k)\hat a^{\dagger}(-\vect k)), \nonumber \\
\label{ADCH}
\end{eqnarray}
where we have introduced the spatial Fourier transform of the generalized polarization vector $\vect{ \varepsilon}(\vect k)$, whose components satisfy \cite{deser19821},
\begin{eqnarray}
\varepsilon_{\alpha}(\vect k)&=&\frac{i\epsilon_{\alpha\beta}k_{\beta}}{|\vect k|}e^{i\frac{\kappa}{|\kappa|}\theta(\vect k)},  \nonumber \\
\varepsilon_{\alpha}(\vect k)\varepsilon_{\beta}^{\dagger}(\vect k)&=&\mathcal{P}_{\alpha\beta}(\vect k), \nonumber \\
\theta(\vect k) &=& \tan^{-1}\left( \frac{k_{2}}{k_{1}}\right) .
\label{PVP}
\end{eqnarray}
The phase term $e^{\pm i\frac{\kappa}{|\kappa|}\theta}$ reflects the spin-$1$ property of the quasiparticle excitations of the free MCS electrodynamics which guarantees the Poincare algebra is satisfied \cite{deser19821,devecchi19951}. Although such phase must be taken account in order to provide an appropriate creation-destruction algebra (endowed with the usual equal-time commutation relations), we shall see that this has no apparent effect in the asymptotic dissipative dynamics of the harmonic $n$-particle system. This evidences that the gauge field can be treated as a scalar electromagnetic field for practical purposes \cite{moura20011}. 

The physical results of the dissipative model (\ref{HMCS2}) should not depend upon details of the particle charged distribution, which is ideally model by the Dirac delta function for point-like particles. Since we are considering a harmonic confinement of the particles, it is advisable to assume a Gaussian distribution for the charged distribution of the $i$-th particle, i.e.
\begin{equation}
\varphi(\vect x-\hat{\vect q}_{i})=\frac{1}{4\pi\sigma}e^{-\frac{|\vect x-\hat{\vect q}_{i}|^{2}}{4\sigma}},
\end{equation}
where $0 <\sigma$ determines the width of the distribution. For seek of simplicity we shall assume the same width for all the harmonic oscillators. With this choice we may recover the point-particle situation by taking the limit $\sigma\rightarrow 0^{+}$, i.e. $\varphi(\vect x-\hat{\vect q}_{i})\rightarrow \delta^{(2)}(\vect x-\hat{\vect q}_{i})$.

For completeness we would like now to briefly discuss the case when we consider the Chern-Simons electrodynamics alone. By rescaling $\hat{\vect A}^{\perp}\rightarrow \hat{\vect A}^{\perp}/\lambda$, and keeping fixed $\kappa/\lambda^2$ and $e/\lambda$ after taking the limit $\lambda^2\rightarrow \infty$, the Maxwell term disappears from the action (\ref{LDAMCS}), leaving us with the pure Chern-Simons electrodynamics coupled to the harmonic $n$-particle system. Going further to the canonical Hamiltonian governing the whole system-field dynamics, one may verify that $\hat \H_{MCS}$ and $\hat \H_{CS}$ (describing the MSC environment and Chern-Simons system-field interaction, respectively) also disappear from (\ref{HMCS2}). This is not surprising, it is a consequence of the fact that the Chern-Simons action does not modify the energy because it is first-order in time derivatives \cite{jackiw19902,deser19821,dunne19991}. So there would be no environmental dynamics supporting an irreversible transference of energy coming from the reduced system. Indeed $\hat{\vect A}^{\perp}$ takes the form of a statistical gauge field that can be properly absorbed in the matter field in order to produce the desired statistics transmutation \cite{matsuyama19901}, for instance the anyonic statistics \cite{dunne19991}. Consequently, this issue prevents us to consider the Chern-Simons electrodynamics alone as a legitimate microscopic model to describe dissipative dynamics. 

\section{Quantum dissipation}\label{SQD}
As similarly occurs in the case of a charged harmonic oscillator coupled to the classical Maxwell electromagnetic field, obtaining an analytical, exact treatment of the open-system dynamics of the reduced system (\ref{HMCS2}) is likely out of reach \cite{ford19851,valido20131}. Yet, a reliable and rich dissipative description may be provided by doing two approximations well understood and motivated in the theory of quantum open-systems and classical electrodynamics, that eventually turns the Hamiltonian (\ref{HMCS2}) (governing all the quantum dynamics) into a quadratic operator in the canonical variables $\hat{\vect q}_{i}$ and $\hat a(\vect k)$, for $i\in \left\lbrace 1,n \right\rbrace $ and $\vect k\in \mathbb{R}^2$. 

Concretely, since we are dealing with confined particles, it proves convenient to consider both approximations: the small displacement of harmonic oscillators in combination with the usual dipole approximation of the gauge field. Let us emphasize that the long-wavelength limit is ubiquitous in most investigations in quantum optics, atomic physics, and quantum chemistry \cite{ruggenthaler20181}. In this way, the $i$-th harmonic oscillator is assumed to move around the equilibrium position $\bar{\vect q}_{i}$ of the confining potential $V(\hat{\vect q_{i}})$ previously defined, such that we may take the small displacement approximation up to first order $\hat{\vect q}_{i}\rightarrow \bar{\vect q}_{i}+\hat{\vect q}_{i}$ in the Chern-Simons interaction Hamiltonian (\ref{HMCS3}), i.e.
\begin{eqnarray}
&&\hat \H_{CS}\simeq \frac{e\kappa}{8\pi^2\sigma}\sum_{i=1}^{n}\int d^{2}\vect x' e^{-\frac{|\vect x'|^{2}}{4\sigma}}\int d^{2} \vect y \ \nabla\times \hat{\vect A}^{\perp}(\vect y)  \nonumber  \\
&\times & \left(\log |\vect y-\vect x'-\bar{\vect q}_{i}|+\frac{(\vect y-\vect x'-\bar{\vect q}_{i})\cdot \hat{\vect q}_{i} }{|\vect y-\vect x'-\bar{\vect q}_{i}|^{2}}\right)    \label{HCSA}\\
&=&\sum_{{\vect k}}\sum_{i=1}^{n}g_{\beta}(\vect k,\bar{\vect q}_{i}) \left(  \varepsilon_{\beta}(\vect k) \hat a(\vect k)+\varepsilon^{\dagger}_{\beta}(-\vect k)\hat a^{\dagger}(-\vect k)\right) \nonumber \\
&+& \sum_{{\vect k}}\sum_{i=1}^{n}f_{\alpha\beta}(\vect k,\bar{\vect q}_{i})  \hat q^{\alpha}_{i}\left(  \varepsilon_{\beta}(\vect k) \hat a(\vect k)+\varepsilon^{\dagger}_{\beta}(-\vect k)\hat a^{\dagger}(-\vect k)\right), \nonumber 
\end{eqnarray}
where we have replaced the Fourier decomposition of the gauge field (\ref{ADCH}) and defined the complex coefficients,
\begin{eqnarray}
g_{\beta}(\vect k,\bar{\vect q}_{i})&=&\frac{iL^{-1} e\kappa\epsilon_{\gamma\beta}k_{\gamma}}{4\pi^{3}\sqrt{2\omega(\vect k)}}\int \frac{d^{2}\vect x }{4\sigma} \nonumber \\
&\times& \int  \  d^{2} \vect y \ e^{-\frac{|\vect x|^{2}}{4\sigma}} e^{i\vect k\cdot \vect y}\log|\vect y-\vect x-\bar{\vect q}_{i}| \nonumber \\
&=&\frac{ie\kappa\epsilon_{\gamma\beta}k_{\gamma}}{2\pi L\sqrt{2\omega(\vect k)}}\frac{e^{i\vect k\cdot \bar{\vect q}_{i}}e^{-\sigma |\vect k|^{2}}}{|\vect k|^2},    \label{GEIT} \\
 f_{\alpha\beta}(\vect k,\bar{\vect q}_{i})&=&\partial_{\bar{q}_{i}^{\alpha}} g_{\beta}(\vect k,\bar{\vect q}_{i})  \nonumber \\
&=&-\frac{e\kappa\epsilon_{\gamma\beta}k_{\gamma}k_{\alpha}}{2\pi L\sqrt{2\omega(\vect k)}}\frac{e^{i\vect k\cdot \bar{\vect q}_{i}}e^{-\sigma |\vect k|^{2}}}{|\vect k|^2}.    \label{FEIT} 
\end{eqnarray}
We must understand the spatial Fourier transform of the logarithmic function as the solution of the homogeneous two-dimensional Poisson equation \cite{bazeia19971}. Paying attention to the first term on the right-hand side in (\ref{HCSA}), this can be recognized as a displacement effect upon the environmental quasiparticle operators, denoted by $\hat{D}_{BR}$, as a consequence of the backreaction of the harmonic $n$-particle system on the MSC environment. Importantly, the second term is closely similar in structure to the vector potential associated to certain magnetic flux $\Phi(\bar{\vect q}_{i})$ attached to the particle charge distribution $\varphi(\vect x-\hat{\vect q}_{i})$ \cite{dunne19991,moura20011}, which shall be called Chern-Simons flux. This result is in complete agreement with the previous discussions in Sec.\ref{SGID}. Specifically, the Chern-Simons interaction (\ref{HMCS3}) can be rewritten as a combination of these two contributions,
\begin{equation}
\hat \H_{CS}\simeq\hat{D}_{BR}-e\sum_{i=1}^{n}\hat{\vect q}_{i}\cdot\hat{\vect E}_{CS}(\bar{\vect q}_{i}),
\label{HCSAI}
\end{equation}
where $\hat{\vect E}_{CS}(\bar{\vect q}_{i})$ is interpreted as the  electric field self-consistently induced by an electric charge $\kappa \Phi(\bar{\vect q}_{i})$ according to the induction Faraday's law \cite{deser19821,li20181,jackiw19901}. This shall be refereed to as the Chern-Simons electric field and provides the desired interaction Hamiltonian in dipole approximation. It is interesting to note that the Levi-Civita symbol appearing in the definition of the coefficients (\ref{GEIT}) and (\ref{FEIT}) signals that the time-reversal symmetry breaks down, as well as the backreaction term and Chern-Simons electric field inherit the axial symmetry from the Chern-Simons kinetic term \cite{deser19821}. It is important to note as well that the bilinear structure of (\ref{HCSA}) and (\ref{HCSAI}) is independent of the particular choice of $\varphi$, wherein the specific form of the coefficients $g_{\beta}$ and $f_{\alpha\beta}$ only depends  on this. Following the same procedure we may obtain an small-displacement expression of the Coulomb potential (\ref{COUP}), i.e. $\hat V_{c}\simeq\sum_{i,j=1}^{n} \left( \hat V_{c}^{0}(\bar{\vect q}_{i}-\bar{\vect q}_{j})+\hat {\vect V}_{c}'(\bar{\vect q}_{i}-\bar{\vect q}_{j})\cdot(\hat{\vect q}_{i}-\hat{\vect q}_{j})\right)$, where $\hat V_{c}^{0}(\bar{\vect q}_{i}-\bar{\vect q}_{j})$ is a constant operator for given values of the particle central positions that may be removed from the Hamiltonian without perturbing the dissipative dynamics.

On the other side, the dipole approximation in the Fourier decomposition of the gauge field (\ref{ADCH}) yields,
\begin{eqnarray}
\int & d^2 \vect x &  \ \varphi(\vect x- \hat{\vect q_{i}})\hat{\vect A}^{\perp}(\vect x)\simeq  \label{GFDA} \\
&\simeq & \int d^2 \vect y \ \varphi(\vect y)\hat{\vect A}^{\perp}(\vect y+\bar{\vect q_{i}}) \nonumber \\
&=&\sum_{\vect k}\frac{e^{i\vect k\cdot\bar{\vect q}_{i}}e^{-\sigma |\vect k|^2}}{2\pi L\sqrt{2\omega(\vect k)}}\left( \vect \varepsilon(\vect k) \hat a(\vect k)+\vect \varepsilon^{\dagger}(-\vect k)\hat a^{\dagger}(-\vect k)\right)  \nonumber \\ 
&=&\hat{\vect A}_{dip}(\bar{\vect q}_{i}), \nonumber
\end{eqnarray}
assuming that $e^{i\vect k\cdot \hat{\vect q}_{i}}\approx 1$ for $i\in\{1,n\}$ \cite{ford19881,cohen19971}(i.e. the system particles mainly interact with the field low-energy modes in comparison with the oscillator bare frequencies).

Substituting these results (\ref{HCSA}) and (\ref{GFDA}) in (\ref{HMCS2}), we obtain a quadratic Hamiltonian which constitutes a first-order approximation to the low-lying dissipative dynamics (\ref{HMCS2}). At this point, the latter Hamiltonian can be brought into a suitable form by performing the Goeppert-Mayer transformation \cite{cohen19971},
\begin{equation}
\hat U_{GM}=e^{-ie\sum_{i=1}^{n}\hat{\vect q}_{i}\cdot\hat{\vect A}_{dip}(\bar{\vect q}_{i})}, 
\label{GMTE}
\end{equation}
which basically consists of replacing (according to the Baker-Hausdorff-Campbell formula)
\begin{eqnarray}
\hat{\vect p}_{i}&&\rightarrow \hat{\vect p}_{i}+e\hat{\vect A}_{dip}(\bar{\vect q}_{i}), \nonumber \\
\hat a(\vect k)&&\rightarrow \hat a(\vect k)+i\frac{e \ \varepsilon_{\alpha}^{\dagger}(\vect k)e^{-\sigma|\vect k|^2}}{2\pi L\sqrt{2\omega(\vect k)}}\sum_{i=1}^{n}e^{-i\vect k \cdot \bar{q_{i}}} \hat{q}_{i}^{\alpha},
\nonumber
\end{eqnarray}
whereas the other terms remain invariant as they commute with $\hat U_{GM}$. Now, by grouping together the interaction terms we redefine the system-environment coupling coefficient as follows,
\begin{eqnarray}
&&h_{\alpha}(\vect k,\bar{\vect q}_{i})=-\frac{\varepsilon_{\beta}(\vect k)}{\omega(\vect k)}f_{\alpha\beta}(\vect k,\bar{\vect q}_{i})+i\frac{e \ \varepsilon_{\alpha}(\vect k)e^{-\sigma|\vect k|^{2}}e^{i\vect k\cdot\bar{\vect q}_{i}}}{2\pi L\sqrt{2\omega(\vect k)}} \nonumber  \\
&=&\frac{e \ e^{i\vect k\cdot\bar{\vect q}_{i}}e^{-\sigma|\vect k|^{2}}}{2\pi L\sqrt{2\omega(\vect k)}}\left( \frac{\kappa\epsilon_{\gamma\beta}k_{\gamma}k_{\alpha}\varepsilon_{\beta}(\vect k)}{|\vect k|^2\omega(\vect k)}+i\varepsilon_{\alpha}(\vect k)\right), \label{EIT}
\end{eqnarray}
and introduce the renormalized potential interaction between system particles, 
\begin{eqnarray}
V_{ij}^{\alpha\beta}&=&\delta_{ij}\delta_{\alpha\beta}m\omega_{i}^2 \label{QPE}  \\
&+&  \frac{e^2}{4\pi^2 L^2}\sum_{\vect k}\mathcal{P}_{\alpha\beta}(\vect k)e^{-2\sigma|\vect k|^2}\cos(\vect k\cdot(\bar{\vect q}_{i}-\bar{\vect q}_{j})),  \nonumber \\
V_{i}^{\alpha}&=&\sum_{j= 1}^{n}\bigg(2(V_{c}'(\bar{\vect q}_{i}-\bar{\vect q}_{j}))^{\alpha} \nonumber \\
&+& \sum_{\vect k}\Big( h_{\alpha}(\vect k,\bar{\vect q}_{i})g_{\beta}^{\dagger}(\vect k,\bar{\vect q}_{j}) \varepsilon_{\beta}^{\dagger}(\vect k) \nonumber \\ 
&+&h_{\alpha}^{\dagger}(\vect k,\bar{\vect q}_{i})g_{\beta}(\vect k,\bar{\vect q}_{j}) \varepsilon_{\beta}(\vect k)\Big)\bigg) ,
\label{VSPE}
\end{eqnarray}
where the second line of (\ref{QPE}) corresponds to the so-called dipole self-energy associated to $\hat{\vect A}_{dip}$ \cite{rokaj20181}, and (\ref{VSPE}) contains the Coulomb potential along with a backreaction contribution. Taking a closer look to (\ref{EIT}), one may see that the particle-charge-distribution width $(2\sigma)^{-\frac{1}{2}}$ plays the role of a frequency cut-off on the system-environment interaction. Consequently, the Chern-Simons influence on the dissipative dynamics becomes weak (or strong) when $\sqrt{2\sigma}\kappa\ll 1$ (or $1\ll \sqrt{2\sigma}\kappa$), this will be seen more clearly in the definition of the spectral density in Sec.\ref{SGLE}. Furthermore, the breaking of time-reversal and parity symmetries are now hidden in the coupling coefficients (\ref{EIT}) and the linear potential $V_{i}^{\alpha}$. 

Finally, after some manipulation once Eqs. (\ref{EIT}), (\ref{QPE}), and (\ref{VSPE}) are substituted, we arrive to the desired microscopic Hamiltonian which is the basis of the present work,
\begin{eqnarray}
&\hat \H'&=\sum_{i=1}^{n} \frac{\hat{\vect p}_{i}^{2}}{2m}+\sum_{i=1}^{n}V_{i}^{\alpha}\hat{q}_{i}^{\alpha}+\frac{1}{2}\sum_{j,i=1}^{n}\Big( V_{ij}^{\alpha\beta} \nonumber \\
&-&\sum_{\vect k}\omega(\vect k)(h_{\alpha}(\vect k,\bar{\vect q}_{i})h_{\beta}^{\dagger}(\vect k,\bar{\vect q}_{j})+c.c.)\Big)\hat{q}_{i}^{\alpha}\hat{q}_{j}^{\beta} \nonumber \\
&+&\sum_{\vect k}\omega(\vect k)\bigg| \hat a^{\dagger}(\vect k)-\sum_{i=1}^{n}\left( h_{\alpha}(\vect k,\bar{\vect q}_{i})\hat q_{i}^{\alpha}-g_{\beta}(\vect k,\bar{\vect q}_{i})\frac{\varepsilon_{\beta}(\vect k)}{\omega(\vect k)}\right)  \bigg|^{2} \nonumber \\
&-&\sum_{i,j=1}^{n}\sum_{\vect k}\frac{\mathcal{P}_{\alpha\beta}(\vect k)}{\omega(\vect k)}g_{\alpha}^{\dagger}(\vect k,\bar{\vect q}_{i})g_{\beta}(\vect k,\bar{\vect q}_{j}).
\label{HMCSF2}
\end{eqnarray}
where the subscript "c.c." stands for Hermitian conjugation. According to the thermodynamic limit, the gauge field is considered to be composed of an infinite number of modes, then we may take the limit of a dense spectrum of field frequencies in (\ref{HMCSF2}) whenever convenient, and replace the discrete momentum sum by an integral following the prescription $\sum_{\vect k}\rightarrow L^2\int_{-\infty}^{\infty}d^2\vect k$. Doing this, we provide explicit expressions for (\ref{QPE}) and (\ref{VSPE}) in the Appendix\ref{app4}.

Let us now discuss some important properties of the dissipative Hamiltonian (\ref{HMCSF2}). First, it is immediate to see that disregarding the Chern-Simons effects (that is, $\hat{D}_{BR}\rightarrow 0$ and $\hat{E}^{\alpha}_{CS}(\bar{\vect q}_{i})\rightarrow 0$ for arbitrary $i\in \{1,n\}$) in (\ref{HMCSF2}) returns the independent-oscillator model  \cite{hanggi20051,valido20131,ford19881}. Conversely, (\ref{HMCSF2}) exhibits the symmetry structure characteristic of the underlying Chern-Simons action: parity breaking and time-reversal asymmetry, as similarly occurs in systems subjected to external magnetic fields or recent extended environments \cite{yao20171}. We shall see that such symmetry affords the appearance of a dissipative vortex-like dynamics driven by a Lorentz force arising from the aforementioned particle-attached Chern-Simons flux. Concretely, by direct comparison to the so-called blackbody radiation bath \cite{ford19881}, we identify a pseudo-electric field $\hat{\mathcal{E}}^{\alpha}_{i} (t)$ responsible for the environmental force acting upon the harmonic $i$th oscillator,
\begin{equation}
\hat{\mathcal{E}}^{\alpha}_{i} (t)=e \left( -\frac{\partial}{\partial t} \hat{A}_{dip}^{\alpha}(\bar{\vect q}_{i},t)+\hat{E}_{CS}^{\alpha}(\bar{\vect q}_{i},t)\right) + \hat{F}_{BR}^{\alpha}(\bar{\vect q}_{i},t),\label{SFFE} 
\end{equation}
where $\hat{E}_{CS}^{\alpha}(\bar{\vect q}_{i},t)$ and $\hat{A}_{dip}^{\alpha}(\bar{\vect q}_{i},t)$ are respectively obtained by replacing $a(\vect k)\rightarrow a(\vect k)e^{-i\omega(\vect k)(t-t_{0})}$ in the definitions (\ref{HCSAI}) and (\ref{GFDA}), whilst $ \hat{\vect F}_{BR}(\bar{\vect q}_{i},t)$ identifies with a backreaction force obtained from substituting $a(\vect k)\rightarrow h_{\alpha}(\vect k,\bar{\vect q}_{i})e^{-i\omega(\vect k)(t-t_{0})}$ in the expression of $\hat{D}_{BR}$.

The first term of the right-hand side of (\ref{SFFE}) bears the dissipation mechanism in the conventional Brownian motion, and it coincides identically with the vector potential contribution to the electric field corresponding to the Maxwell electrodynamics action alone (i.e. $\kappa=0$) in dipole approximation. Hence, the second term may be interpreted as the Lorentz force due to the Chern-Simons electric field $\hat{\vect E}_{CS}(\bar{\vect q}_{i})$, whereas the backreaction force $\hat{\vect F}_{BR}(\bar{\vect q}_{i},t)$ contains the previously mentioned effective shift in the environmental quasiparticle operators. Contrary to the so-called "initial slip" term in the standard microscopic model that just depends on the initial harmonic oscillator positions \cite{hanggi20051}, this backreaction displacement is independent of the initial state of the harmonic system. In Sec.\ref{PSNDR}, we shall see that the latter gives rise to non-stochastic fluctuations which eventually cancel out in the time asymptotic limit. 
 
Unlike the dipole approximation taken in (\ref{GFDA}), the validity of the small displacement approximation described in (\ref{HCSA}) is intricate to elucidate just by looking at (\ref{HCSAI}). This can be better assessed by requiring the Hamiltonian (\ref{HMCSF2}) to be a positive definite operator \cite{haake19851,valido20131,ford19881} in order it has a lower-bounded spectrum preventing "runaway" solutions \cite{coleman19621}, and thus, it gives rise to simple dissipative dynamics (which preserves $\omega_{i}$ as the bare frequencies for the system particles). As shown in detail in Appendix\ref{app3}, such condition is found to be equivalent to the following inequality,
\begin{widetext}
\begin{eqnarray}
\frac{1}{2}\sum_{j,i=1}^{n}\Bigg(V_{ij}^{\alpha\beta}&-&\delta_{2\alpha}\delta_{2\beta}\frac{e^{-\frac{|\bar{\vect q}_{i}-\bar{\vect q}_{j}|^2}{8\sigma}}}{8\pi\sigma}+\frac{e^2}{8}\bigg((\delta_{\alpha\beta}-2\delta_{1\beta}\delta_{1\alpha})\frac{\kappa H_{1}\left(i\kappa |\Delta\bar{q}_{ij}| \right) }{|\Delta\bar{q}_{ij}|}+\delta_{1\alpha}\delta_{1\beta} \kappa^2 H_{0}\left( i\kappa |\Delta\bar{q}_{ij}|\right) \bigg)\Bigg)\hat{q}_{i}^{\alpha}\hat{q}_{j}^{\beta}\nonumber \\
&+&\sum_{i,j=1}^{n}\Bigg(2 \left( V_{c}'(\Delta\bar{q}_{ij})\right)_{\alpha} - \delta_{1\alpha}\frac{e^2\kappa^{2}}{2\pi}\mathsf{J}_{1}(|\Delta\bar{q}_{ij}|)\Bigg)\hat{q}_{i}^{\alpha} \geq \frac{e^2\kappa^2}{4\pi}\sum_{i,j=1}^{n}\mathsf{J}_{0}(|\Delta\bar{q}_{ij}|) 
\label{DHPCE}
\end{eqnarray}
\end{widetext}
where $\Delta\bar{q}_{ij}=\bar{\vect q}_{i}-\bar{\vect q}_{j}$ and we have introduced the auxiliary functions
\begin{equation}
\mathsf{J}_{l}(|\Delta\bar{q}_{ij}|)=\int_{0}^{\infty} dk \ \frac{k^{l+1}e^{-2\sigma k^2} }{k^2\left(k^{2}+\kappa^2 \right)}J_{l}(k|\bar{\vect q}_{i}-\bar{\vect q}_{j}|) ,
\nonumber
\end{equation}
with $J_{i}(x)$ denoting the $i$-order Bessel function of the first kind in the variable $x$, and $H_{j}(x)=i^{j+1}e^{2\sigma\kappa^2 }\left( H_{j}^{(2)}(-x)+(-1)^{j+1}H_{j}^{(1)}(x)\right) $, with $H_{i}^{(j)}(x)$ being the $i$-order Hankel function of the $j$-th kind \cite{gradshteyn20141}. It is worthwhile to note that the integral involved in the definition of $\mathsf{J}_{l}$ may present an infrared divergence (i.e. $k\rightarrow 0$) owing to the two-dimensional Coulomb Green function blows up at the origin (see Eq.(\ref{GEIT})). This is a feature characteristic of the Maxwell-Chern-Simons electric and magnetic fields that requires adequate regularization schemes \cite{moura20011,deser19821}. 

Although the positive condition may be looked rather complicated for supporting an intuitive interpretation at first sight, the right-hand side of (\ref{DHPCE}), which emerges exclusively as a consequence of the backreaction on the MSC environment, reflects a repulsive effect between the system particles that challenges with the confining harmonic potential. To see this more clearly, let us focus in the single harmonic oscillator system (i.e. $n=1$). Hence, it can be shown that the formidable inequality (\ref{DHPCE}) boils down to
\begin{eqnarray}
\sum_{\alpha=1,2}\bigg(m\omega_{1}^2&-& \frac{e^2\kappa^2}{8\pi}\Gamma\left( 0, 2\sigma\kappa^2\right)e^{2\sigma\kappa^2} \bigg)\hat{q}_{1}^{\alpha}\hat{q}_{1}^{\alpha} \geq \frac{e^2\kappa^2}{2\pi}R_{0}^2 \nonumber,
\end{eqnarray}
where $R_{0}^2=\sum_{i,j=1}^{n}\mathsf{J}_{0}(0) $, and $\Gamma(0,x)$ denotes the incomplete Euler Gamma function \cite{gradshteyn20141}. Clearly, the above condition may be interpreted as the single harmonic oscillator is enforced to follow a fluctuating motion around a circular area of radius larger than certain $R$ given by
\begin{equation}
\frac{R^{2}}{2\sigma}=\frac{R_{0}^2}{\frac{4\pi m\sigma\omega_{1}^{2}}{e^2\kappa^2}\left(1-\frac{1}{8\pi}\frac{e^2 \kappa^2\Gamma\left( 0, 2\sigma\kappa^2\right)e^{2\sigma\kappa^2}}{m\omega_{1}^2}\right) }.
\label{PCREI}
\end{equation}
To be this result consistent with the small displacement approximation considered previously, we demand the length of $R$ to be sufficiently small in comparison to the width of the particle charged distribution $2\sigma$, which implies
\begin{equation}
\frac{e^2}{2m\sigma^2\omega^2_{1}}\ll \frac{1}{\sigma\kappa^2} .
\label{PCREII}
\end{equation}
Expression (\ref{PCREII}) entails that there must exist a trade-off between the system-environment interaction strength and system particle bare frequencies. The physical intuition behind the latter is that the environment could drive the particle to reach highly excited states for an arbitrary large coupling, which would eventually lead to break down the small displacement approximation assumed in (\ref{HCSA}). For instance, for a strong Chern-Simons action $1 \ll \sigma\kappa^2$, we may approximate $\Gamma\left( 0, 2\sigma\kappa^2\right) e^{2\sigma\kappa^2}\sim (2\sigma\kappa^2)^{-1}$, and then, the positive condition (\ref{PCREI}) holds for $1\ll m\sigma\omega_{1}^{2}/e^2$, which is equivalent to (\ref{PCREII}). In this way, the small displacement approximation again requires that the system-environment coupling must pay off the repulsive counteract of a strong backreaction effect. From this point onward we work within the parameter domain where expression (\ref{DHPCE}) holds, and therefore, the subsidiary condition (\ref{PCREII}) is always satisfied for the bare frequencies of the $n$ harmonic oscillators. In particular, this result is a manifestation of the issue that the Maxwell-Chern-Simons theory works better for developing models of confined particle systems \cite{caruso20131,dunne19991}.

Our final remark is that the Hamiltonian (\ref{HMCSF2}) can be regarded as a \textit{gauge-invariant} microscopic description by construction, since it was derived from a gauge-invariant Hamiltonian (\ref{HMCS2}). This is a major difference with previous treatments \cite{yao20171}, where it is not guaranteed the gauge-invariance for a given choice of the system-environment coupling coefficients. Remarkably, (\ref{HMCSF2}) looks very similar to an environmental minimal-coupling Hamiltonian (in dipole approximation) \cite{kohler20131} with gauge field 
\begin{equation}
\hat{A}_{MSC}^{\alpha}(\bar{\vect q}_{i})=\frac{i}{e}\sum_{\vect k}(h_{\alpha}(\vect k,\bar{\vect q}_{i})\hat a(\vect k)+c.c.), \nonumber
\end{equation}
and associated electric field,
\begin{equation}
\hat{E}_{MCS}^{\alpha}(\bar{\vect q}_{i},t)=-\frac{\partial}{\partial t} \hat{A}_{dip}^{\alpha}(\bar{\vect q}_{i},t)+\hat{E}_{CS}^{\alpha}(\bar{\vect q}_{i},t). \label{MCSEF}
\end{equation}
Concretely, the Hamiltonian (\ref{HMCSF2}) is equivalent to a minimal-coupling theory of $n$ harmonic oscillators with the gauge field $\hat{\vect A}_{MSC}(\bar{\vect q}_{i})$ provided we disregard the backreaction effects and endow the system Hamiltonian with a renormalized potential interaction which cancels the environmental influence on the conservative dynamics, i.e.
\begin{eqnarray}
\hat \H_{RN}&=&-2\sum_{j= 1}^{n}(V_{c}'(\bar{\vect q}_{i}-\bar{\vect q}_{j}))^{\alpha}\hat{q}_{i}^{\alpha}+\frac{1}{2}\sum_{i,j= 1}^{n}\Big(\delta_{ij}\delta_{\alpha\beta}m\omega_{i}^2  \nonumber\\
&-&V_{ij}^{\alpha\beta}+\sum_{\vect k}\omega(\vect k)(h_{\alpha}(\vect k,\bar{\vect q}_{i})h_{\beta}^{\dagger}(\vect k,\bar{\vect q}_{j})+c.c.)\Big)\hat{q}_{i}^{\alpha}\hat{q}_{j}^{\beta},\nonumber
\end{eqnarray}
where the first term is the familiar Coulomb contribution. This statement can be explicitly verified by absorbing the gauge field in the canonical conjugate momentum by means of a gauge transformation $\hat U=e^{ie\sum_{i=1}^{n}\hat{\vect q}_{i}\cdot\hat{\vect A}_{MCS}(\bar{\vect q}_{i})}$ upon the Hamiltonian (\ref{HMCSF2}), once we have dropped the backreaction terms (i.e. $g_{\beta}(\vect k,\bar{\vect q}_{i})\rightarrow 0 $) and introduced the renormalization $\hat \H_{RN}$. The associated Maxwell-Chern-Simons electric field (\ref{MCSEF}) features non-commutative components (see Eq.(\ref{MCSEFCCR}) in Appendix\ref{app4} for further details), i.e. 
\begin{equation}
\left[  \hat E^{\alpha}_{MCS}(\bar{\vect q}_{i}), \hat E^{\beta}_{MCS}(\bar{\vect q}_{j})\right]\propto -i \kappa \epsilon_{\alpha\beta}. \label{MCSCCRI}
\end{equation} 
Interestingly, this property is shared with the electric field of the free Maxwell-Chern-Simons electrodynamics (i.e. without matter-field interaction) \cite{dunne19991}, and has several consequences in the dissipative dynamics illustrated in Sec.\ref{PSNDR}.

We spend the following sections to justify that the Hamiltonian (\ref{HMCSF2}) regards a legitimate microscopic description to simulate the relaxation process towards a thermal equilibrium state (see \ref{SGLE}, \ref{PSNDR} and \ref{SSP}) despite the approximations taken to derive it, as well as we provide an explicit comparison with the popular damped harmonic oscillator \cite{hanggi20051,ford19881,caldeira19831,unruh19891,hu19921,grabert19881}
in the Markovian Langevin dynamics limit (see  \ref{MLDS} and \ref{SHOS}). Before proceeding with our treatment, it is convenient to recall the $\omega$-variable Fourier transform $\tilde{r}(\omega)$ of a time-dependent function $r(t)$,
\begin{eqnarray}
\tilde{r}(\omega)=\frac{1}{2\pi}\int_{-\infty}^{\infty}dt \  e^{i\omega t}r( t),
\nonumber
\end{eqnarray}
and its corresponding real and imaginary parts,
\begin{eqnarray}
\text{Re}\ \tilde{r}(\omega)=\frac{\tilde{r}(\omega)+\tilde{r}^{\dagger}(\omega)}{2}, \ \ \text{Im}\ \tilde{r}(\omega)=\frac{\tilde{r}(\omega)-\tilde{r}^{\dagger}(\omega)}{2i},
\nonumber
\end{eqnarray}
where $\tilde{r}^{\dagger}(\omega)$ represents the complex conjugate of $\tilde{r}(\omega)$.
\subsection{Generalized Lanvegin equation} \label{SGLE}

Having determined the dissipative Hamiltonian (\ref{HMCSF2}) along with the quasiparticle excitations of the gauge field, we may turn the attention to the non-equilibrium dynamics of the harmonic $n$-particle system. Starting from the Hamiltonian (\ref{HMCSF2}) we derive the following Heisenberg equations for the $i$th-oscillator position and momentum operators,
\begin{eqnarray}
\dot{\hat{q}}_{i}^{\alpha}&=&\frac{\hat{p}_{i}^{\alpha}}{m}, \label{HEM1} \\
\dot{\hat{p}}_{i}^{\alpha}&=&-2\sum_{j= 1}^{n}(V_{c}'(\bar{\vect q}_{i}-\bar{\vect q}_{j}))^{\alpha}-\sum_{j=1}^{n}V_{ij}^{\alpha\beta}\hat{q}_{j}^{\beta}  \label{HEM2} \\
&+&\sum_{\vect k}\omega(\vect k)\left(h_{\alpha}(\vect k,\bar{\vect q}_{i})\hat a(\vect k)+h_{\alpha}^{\dagger}(\vect k,\bar{\vect q}_{i})\hat a^{\dagger}(\vect k) \right)  ,
\nonumber
\end{eqnarray}
as well as for the quasiparticle creation operator of the gauge field,
\begin{eqnarray}
\dot{\hat a}^{\dagger}(\vect k)&=&i\omega(\vect k)\hat a^{\dagger}(\vect k) \nonumber \\
&+&i\sum_{i=1}^{n}\left(g_{\beta}(\vect k,\bar{\vect q}_{i})\varepsilon_{\beta}(\vect k)- \omega(\vect k)h_{\beta}(\vect k,\bar{\vect q}_{i})\hat{q}_{i}^{\beta}\right) . \nonumber
\end{eqnarray}
It is straightforward to obtain the formal solution of the latter equation by using the standard Green's function method, since it constitutes an inhomogeneous linear system of differential equations. First, we obtain for the quasiparticle operators of the gauge field 
\begin{eqnarray}
\hat a^{\dagger}(\vect k,t)&=&\left( \hat a^{\dagger}(\vect k,t_{0})+\sum_{i=1}^{n}\frac{\varepsilon_{\beta}(\vect k) }{\omega(\vect k)}g_{\beta}(\vect k,\bar{\vect q}_{i})\right) e^{i\omega(\vect k)(t-t_{0})}  \nonumber \\
&-&i\omega(\vect k)\sum_{i=1}^{n}h_{\beta}(\vect k,\bar{\vect q}_{i})\int_{t_{0}}^{t}e^{i\omega(\vect k)(t-\tau)}\hat{q}_{i}^{\beta}(\tau)d\tau \nonumber \\
&-&\sum_{i=1}^{n}\frac{\varepsilon_{\beta}(\vect k)}{\omega(\vect k)}g_{\beta}(\vect k,\bar{\vect q}_{i}),\label{AnnihiS}
\end{eqnarray}
where $t_{0}^{+}\leq t$ in order to be physically consistent with the considered initial preparation. Inserting the solution (\ref{AnnihiS}) into equation (\ref{HEM2}) and manipulating the subsequent result, one gets to the desired generalized Langevin equation,
\begin{eqnarray}
m\frac{d^{2}\hat q_{i}^{\alpha}}{dt^{2}}&+&\sum_{j=1}^{n}V_{ij}^{\alpha\beta}\hat{q}_{j}^{\beta} +V_{i}^{\alpha} \label{GLE} \\
&-&\sum_{j=1}^{n}\int_{t_{0}}^{t}\Sigma_{ij}^{\alpha\beta}(t-\tau)\hat q_{j}^{\beta}(\tau)d\tau =\hat{\mathcal{E}}^{\alpha}_{i} (t), \nonumber
\end{eqnarray}
where we have identified the pseudo-electric field $\hat{\mathcal{E}}^{\alpha}_{i}$ defined in (\ref{SFFE}) as the fluctuating force, and the generalized susceptibility or \textit{self-energy} as the retarded Green's function,
\begin{eqnarray}
\Sigma_{ij}^{\alpha\beta}(t-t')&=&i\Theta\left( t-t'-|\Delta \bar{\vect q}_{ij}|\right) \left\langle \left[ \hat{\mathcal{E}}^{\alpha}_{i} (t), \hat{\mathcal{E}}_{j}^{\beta\dagger} (t')\right]   \right\rangle_{\hat{\rho}_{\beta}},  \nonumber \\ 
\label{MKE}
\end{eqnarray}
where $\Theta(t)$ denotes the Heaviside step function. Clearly, the linear potential $V_{i}$ represents an non-stochastic force affecting mainly the mean average position of the system particles, so it could be neglected from the future discussion by doing a suitable renormalization of the harmonic oscillators. 

Although equation (\ref{GLE}) may look similar at first sight to the quantum Langevin equation in presence of magnetic fields \cite{gupta20111,czopnik20011,li19901}, both equations significantly differ in the statistical and analytical properties of the corresponding fluctuating force and retarded self-energy. On one side, the backreaction effects in the pseudo-electric force (\ref{SFFE}) prevents the dissipative dynamics to fulfill the fluctuation-dissipation theorem at all times, in contrast to the conventional Brownian motion. On the other side, the breaking of time-reversal and parity symmetry in the present context induces an imaginary contribution to the (field) spectral density that has no counterpart in the independent-oscillator model \cite{valido20131,ford19881}. This deeply modifies the analytical structure of the Fourier transform of the retarded self-energy, which can be compactly written as follows
\begin{eqnarray}
\tilde{\Sigma}_{ij}^{\alpha\beta}(\omega)=\mathcal{R}\tilde{\Sigma}_{ij}^{\alpha\beta}(\omega)+i\mathcal{I}\tilde{\Sigma}_{ij}^{\alpha\beta}(\omega), 
\label{FTSE}
\end{eqnarray}
where $\mathcal{R}\tilde{\Sigma}_{ij}^{\alpha\beta}(\omega)$ ($\mathcal{I}\tilde{\Sigma}_{ij}^{\alpha\beta}(\omega)$) must not be confused with the real (imaginary) part previously defined. Despite of this, we would like to remark that the self-energy $\tilde{\Sigma}_{ij}^{\alpha\beta}(\omega)$ exhibits the general properties required to produce a reliable dissipative dynamics: (causality condition) it is analytic in the upper-half $\omega$-complex plane, and further, (reality condition \cite{ford19881,li19901}) it holds 
\begin{equation}
\left( \tilde{\Sigma}_{ij}^{\beta\alpha}\right)^{\dagger} (\omega)= \tilde{\Sigma}_{ji}^{\alpha\beta}(\omega).
\label{RSERC}
\end{equation}
Basically, these properties are encoded by the (field) spectral density arising from the Maxwell-Chern-Simons electrodynamics, denoted by $\mathcal{J}_{\alpha\beta}$, and which reduces to the well-known spectral density of the independent-oscillator model for zero Chern-Simons constant. 

Let us draw more attention to the properties of the retarded self-energy. The Heaviside step function in the expression of the retarded self-energy (\ref{MKE}) guaranties the dissipative dynamics to be consistent with the initial decoupling of the harmonic particle system and gauge field, and further, it gives the usual pole prescription in the frequency domain mentioned above: $\Sigma_{ij}^{\alpha\beta}(t)$ is an analytic function in the upper-half complex plane. Moreover, it is in agreement with the fact that the pseudo-electric fields $\hat{\mathcal E}^{\alpha}_{i} (t)$ and $\hat{\mathcal E}^{\beta}_{j} (t)$ must commute for space-like separations, i.e. $[\hat{\mathcal E}^{\alpha}_{i} (t_{0}), \hat{\mathcal E}^{\beta}_{j}(t)] = 0$ if $|\bar{\vect q}_{i}-\bar{\vect q}_{j}| > |t-t_{0}|$. This is usually known as microscopic causality, and for instance, it is fulfilled for the free Maxwell electromagnetic field \cite{rzazewski19761}. As a consequence, we show in Appendix \ref{app4} that the self-energy satisfies a generalized Kramers-Kronig identity \cite{valido20131,philbin20121}, i.e.
\begin{eqnarray}
\mathcal{R}\tilde{\Sigma}_{ij}^{\alpha\beta}(\omega)&=&\mathcal{H}\big[\mathcal{I}\tilde{\Sigma}_{ij}^{\alpha\beta}(\omega')\big](\omega) \nonumber \\
&=&\frac{1}{\pi}P\int_{-\infty}^{+\infty} \frac{\mathcal{I}\tilde{\Sigma}_{ij}^{\alpha\beta}(\omega')}{\omega'-\omega}d\omega' \label{KKR},
\end{eqnarray}
where $\mathcal{H}[f(x)](\omega)$ denotes the Hilbert transform of the function $f(x)$ in the variable $\omega$, and $P$ is the Cauchy principal value.

By replacing the pseudo-electric field (\ref{SFFE}) in (\ref{MKE}) and taking the dense spectrum limit, the retarded self-energy can be cast in terms of the environmental spectral density as follows (see Apendix\ref{app4} for further details),
\begin{widetext}
\begin{equation}
\Sigma_{ij}^{\alpha\beta}(t-t')=\frac{2}{\pi}\Theta\left( t-t'-|\Delta \bar{\vect q}_{ij}|\right)\int_{0}^{\infty}d\omega \ \Big(\text{Re} \{ \mathcal{J}_{\alpha\beta}(\omega,\Delta\bar{q}_{ij})\} \sin(\omega(t-t'))+\text{Im} \{ \mathcal{J}_{\alpha\beta}(\omega,\Delta\bar{q}_{ij})\} \cos(\omega(t-t')) \Big),
\label{MKEII}
\end{equation}
whereas the spectral density takes the form,
\begin{eqnarray}
\mathcal{J}_{\alpha\beta}(\omega,\Delta\bar{q}_{ij} )&=&\left( \frac{e}{2}\right) ^{2}e^{-2\sigma \left( \omega^{2}-\kappa^{2}\right) }Y_{\alpha\beta}\left( |\Delta\bar{q}_{ij}|,\sqrt{\omega^{2}-\kappa^{2}} \right), \ \text{with} \  \kappa\leq  \omega\label{SPD}
\end{eqnarray}
and 
\begin{eqnarray}
 \vect{Y}(a,b)&=&\left( \begin{array}{cc}
\kappa^2 J_{0}(ab)  +\frac{b}{a}J_{1}(ab)& i\kappa\omega J_{0}(ab)  \\
-i\kappa \omega J_{0}(ab) & \omega^{2} J_{0}(ab)  -\frac{b}{a}J_{1}(ab)
\end{array}\right) ,
\nonumber
\end{eqnarray}
where we may clearly observe that the off-diagonal elements arise exclusively from the Chern-Simons action. This specific form of the spectral density deserves some attention. Expression (\ref{SPD}) shares some features with the usual (bath) spectral density of the damped harmonic oscillator model \cite{alamoudi19981,alamoudi19991}: (i) it features a broad gaped spectrum ($\kappa< \omega < \infty$) that may span the harmonic oscillator frequencies, and further, (ii) the strength of the system-field coupling decays (exponentially) for sufficiently large frequencies compared to the aforementioned frequency cut-off given by $(2\sigma)^{-\frac{1}{2}}$. Although the latter eventually prevents from ultraviolet divergence issues in the non-equilibrium particle dynamics for most interesting cases, it is worthwhile to notice that the specific case of point particles (i.e. $\sigma\rightarrow 0^{+}$) is not free from this divergence. This may be seen as a consequence of the well-known self-energy problems that suffers the point-particle electrodynamics \cite{jackiw19902}. Furthermore, this shows that the cut-off factor of the spectral density is mainly determined by the choice of the particle charged distribution $\varphi(\hat{\vect q})$. Accordingly, the spectral density constitutes a $2n\times 2n$ Hermitian matrix (i.e. $\mathcal{J}_{\alpha\beta}(\omega,\Delta\bar{q}_{ij} )=\left( \mathcal{J}_{\beta\alpha}\right)^\dagger (\omega,\Delta\bar{q}_{ji} )$), which immediately implies (\ref{RSERC}), and therein, the retarded self-energy is a $2n\times 2n$ real matrix in the time domain.

The specific form of the spectral density (\ref{SPD}) also reveals interesting properties related to the dissipative dynamics. For instance, the fact that the spectral density is highly oscillatory in the frequency domain unveils that the harmonic system may undergo a strong non-Markovian dissipative dynamics \cite{devega20171}, rendering a richer quantum dissipative evolution than the conventional Brownian motion. Furthermore, the diagonal elements of the spectral density manifest an anisotropic influence to the transversal spatial degrees of freedom of distant harmonic oscillators. Nevertheless, this effect cancels out when the particles are very closed or localized in identical positions, which can be seen by taking the asymptotic limit $|\Delta\bar{q}_{ij}|\rightarrow 0$ in (\ref{SPD}). By virtue of the off-diagonal shape of the spectral density, we may also realize that the new dissipative Chern-Simons effects are mainly encoded in the Fourier cosine transform appearing in the retarded self-energy definition (\ref{MKEII}). This result is consistent with previous treatments \cite{yao20171,gupta20111} about Brownian motion in the presence of magnetic fields, where it was shown that either a Berry's geometric magnetic or uniform magnetic field produces a "transversal" contribution to the memory kernel. Interestingly, we shall show in Sec.\ref{MLDS} that in the Markovian Langevin limit the off-diagonal elements of the retarded self-energy turn into an effective interaction which is akin to applying a non-conservative rotating force upon the harmonic oscillators, which is in agreement with the fact that such contribution is promoted by the time-reversal asymmetry and parity violation. 

A further simplified expression between the retarded self-energy and spectral density is obtained by performing the Fourier transform in (\ref{MKEII}) (the details of the derivation can be found in Appendix\ref{app4}). Doing this we arrive at the following identity 
\begin{eqnarray}
\mathcal{I}\tilde{\Sigma}_{ij}^{\alpha\beta}(\omega)=\left\langle \left[  \tilde{\mathcal{E}}_{i}^{\alpha} (\omega), \tilde{\mathcal{E}}^{\beta\dagger}_{j} (\omega)\right]   \right\rangle=\frac{1}{2\pi}\left( \Theta(\omega) \mathcal{J}_{\alpha\beta}^{\dagger}(\omega,\Delta\bar{q}_{ij}) -\Theta(-\omega)\mathcal{J}_{\alpha\beta}(-\omega,\Delta\bar{q}_{ij})\right)  , \label{IFSE} 
\end{eqnarray}
which completely characterizes the dissipative effects. It is well-known that a system, whose open-system dynamics is governed by a given quantum Langevin equation, will reach an asymptotic thermal equilibrium state if the dissipative effects are related to the fluctuations of the environmental noise via the so-called fluctuation-dissipation theorem \cite{valido20131,valido20151,pagel20131}, e.g.
\begin{eqnarray}
\frac{\left\langle \left\lbrace \tilde{\xi}^{\alpha}_{i} (\omega),\tilde{\xi}_{j}^{\beta\dagger} (\omega')\right\rbrace  \right\rangle_{\hat{\rho}_{\beta}}}{2\mathcal{I}\tilde{\Sigma}_{ij}^{\alpha\beta}(\omega)} =\delta(\omega-\omega')\left(1+2n(\omega, \beta^{-1}) \right), 
\label{FDTI}
\end{eqnarray}
where $\hat{\xi}^{\alpha}_{i}(t)$ represents a mean zero stationary Gaussian noise (i.e. a quantum Brownian noise) and $1+2n(\omega, \beta^{-1})=\coth\left(\beta\omega/2\right) $. This relation manifests that the fluctuating force is only due to thermal fluctuations, which is the case for the independent-oscillator model \cite{caldeira19831,ford19881,hanggi20051} (e.g. see the case of the electromagnetic field \cite{efimov19941}). Going back to the expression (\ref{SFFE}), the aforementioned backreaction contribution $\hat{\vect F}_{BR}(\bar{\vect q}_{i},t)$ to the pseudo-electric field constitutes a non-stochastic force that breaks down the fluctuation-dissipation theorem (\ref{FDTI}) for the MSC environment initially in a canonical equilibrium state $\hat{\rho}_{\beta}$. As shown in Appendix\ref{app4}, the statistics of the pseudo-electric force is related to the dissipative effects via the following formidable equation,
\begin{eqnarray}
\frac{1}{2}\left\langle \left\lbrace \tilde{\mathcal{E}}_{i}^{\alpha} (\omega), \tilde{\mathcal{E}}^{\beta\dagger}_{j} (\omega')\right\rbrace  \right\rangle_{\hat \rho_{\beta}} &=& \delta(\omega-\omega')\left( 1+2 n\left( \omega, \beta^{-1}\right) \right)\mathcal{I}\tilde{\Sigma}_{ij}^{\alpha\beta}(\omega)  \label{CFFE}\\
&+&\Theta(\omega)\Theta(\omega')\left( \tilde{\mathcal{G}}_{ij}^{\alpha\beta}\right)^{\dagger}(\omega,\omega',t_{0}) +\Theta(-\omega)\Theta(-\omega')\tilde{\mathcal{G}}_{ij}^{\alpha\beta} (-\omega,-\omega',t_{0})\nonumber \\
&+& \Theta(\omega)\Theta(-\omega')\left(\tilde{\mathcal{F}}_{ij}^{\alpha\beta}\right)^{\dagger}(\omega,-\omega',t_{0}) +\Theta(-\omega)\Theta(\omega') \tilde{\mathcal{F}}_{ij}^{\alpha\beta} (-\omega,\omega',t_{0}), \nonumber
\end{eqnarray}
where we have defined the following non-stochastic spectral functions for $\kappa\leq\omega,\omega'$
\begin{eqnarray}
\tilde{\mathcal{G}}_{ij}^{\alpha\beta}(\omega,\omega',t_{0})=\frac{e^4\kappa^2}{16\omega\omega'}e^{i(\omega-\omega')t_{0}}e^{-2\sigma\left( \omega^{2}+\omega'^{2}-2\kappa^{2}\right) }\sum_{l,m=1}^{n}R_{\alpha\beta}\left( \sqrt{\omega^{2}-\kappa^{2}},\sqrt{\omega'^{2}-\kappa^{2}},|\bar{\vect q}_{i}+\bar{\vect q}_{l}|,|\bar{\vect q}_{j}+\bar{\vect q}_{m}|\right) ,
\label{GGE}
\end{eqnarray}
and 
\begin{eqnarray}
\tilde{\mathcal{F}}_{ij}^{\alpha\beta}(\omega,\omega',t_{0})= -\frac{e^4\kappa^2}{16\omega\omega'}e^{i(\omega+\omega')t_{0}}e^{-2\sigma\left( \omega^{2}+\omega'^{2}-2\kappa^{2}\right) }\sum_{l,m=1}^{n}T_{\alpha\beta}\left( \sqrt{\omega^{2}-\kappa^{2}},\sqrt{\omega'^{2}-\kappa^{2}},|\bar{\vect q}_{i}+\bar{\vect q}_{l}|,|\bar{\vect q}_{j}+\bar{\vect q}_{m}|\right), \label{FFE} 
\end{eqnarray}
as well as we have introduced the matrices $\vect R$ and $\vect T$, e.g.
\begin{eqnarray}
R_{11}(a,b,c,d)&=&\frac{1}{acbd}u\left(\kappa a c ,2\frac{\kappa\omega}{|\kappa|}-2\kappa ,ac\right)  u\left(\kappa b d ,2\frac{\kappa\omega}{|\kappa|}-2\kappa ,bd\right),  \nonumber \\
R_{12}(a,b,c,d)&=&\frac{i}{acbd} u\left(-\kappa a c ,-2\frac{\kappa\omega}{|\kappa|}-2\kappa ,ac\right)  u\left(- b d \omega',-\frac{\kappa^{2}}{|\kappa|}+2\omega',bd\right) ,\nonumber \\
R_{21}(a,b,c,d)&=&\frac{i}{acbd}u\left( a c \omega',-\frac{\kappa^{2}}{|\kappa|}-2\omega',ac\right) u\left(\kappa b^2 d ,-2\frac{\kappa\omega}{|\kappa|}-2\kappa ,bd\right), \nonumber \\
R_{22}(a,b,c,d)&=&\frac{1}{acbd}u\left(a c \omega,\frac{\kappa^{2}}{|\kappa|}-2\omega,ac\right)  u\left(b d \omega',\frac{\kappa^{2}}{|\kappa|}-2\omega',bd\right) ,
\label{SDGM}
\end{eqnarray}
\end{widetext}
and the auxiliary function $u(x,y,z)=x J_{1}(z)+yJ_{2}(z) $. Due to the detailed representation of $\vect T$ is lengthy and not crucial for the future discussion, we move it to the Appendix \ref{app5} (see equation (\ref{AEFTM}), as well as the derivation of the fluctuation-dissipation relation (\ref{CFFE}). At this point, it is important to realize that both non-stochastic spectral functions, (\ref{GGE}) and (\ref{FFE}), are integrable functions in the frequency domain, and further, they decay as fast as an exponential function for large arguments of $\omega$ and $\omega'$. For instance, it is readily to see that the matrix elements of $\vect R$ reduce to a finite and continuous algebraic function for small arguments $|\bar{\vect q}_{i/j}+\bar{\vect q}_{l}|\sqrt{\omega^{2}-\kappa^{2}}\ll 1$, by using the asymptotic expressions of the Bessel functions, i.e. $J_{\alpha}(z)\sim \Gamma(\alpha+1)^{-1}(z/2)^{\alpha}$ for $z\ll 1$. The matrix $\vect T$ is found to feature this property as well.

From the derivation of (\ref{CFFE}) follows that the first line is just due to the Maxwell and Chern-Simons electric contributions to the fluctuating force. That is, we can identify the Lorentz force rendered by these electric fields with a stochastic thermal noise, i.e.
\begin{eqnarray}
\hat \xi_{i}^{\alpha}(t)&=& e \hat{E}_{MCS}^{\alpha}(\bar{\vect q}_{i},t),
\label{ASEF}
\end{eqnarray}
where $\hat{E}_{MCS}^{\alpha}(\bar{\vect q}_{i},t)$ was defined in (\ref{MCSEF}). Recall that $\hat \xi_{i}^{\alpha}(t)$ is an unbiased random operator (i.e. $\left\langle \hat \xi_{i}^{\alpha}(t)\right\rangle =0 $) that satisfies the fluctuation-dissipation relation illustrated in (\ref{FDTI}) even though the Chern-Simons electric field exhibits time-reversal asymmetry \cite{sieberer20161}. On the other hand, the second and third line represents the non-stochastic fluctuations owing to the backreaction force in (\ref{SFFE}). Concretely, they come from non-stationary terms involving $\hat a(\vect k)\hat a(\vect k)$ and $\hat a^{\dagger}(\vect k)\hat a^{\dagger}(\vect k)$, which represent non-conservative energy processes taking place in the gauge field at the initial time $t_{0}$. Interestingly, paying attention to (\ref{GGE}) and (\ref{FFE}), we may observe that such fluctuations become highly oscillatory in the long time limit $t-t_{0}\rightarrow\infty$, which could make their contribution to the stationary dynamics neglectable. Indeed, we illustrate in the next section how the non-stochastic fluctuations asymptotically cancel out in the strict limit by appealing to the fact that the aforementioned spectral functions (\ref{GGE}) and (\ref{FFE}) have both a broad bandwidth and rapid decayment at large frequencies compared to the particle frequencies $\omega_{i}$, recovering in turn the fluctuation-dissipation theorem (\ref{FDTI}). Before to continue, it is worthwhile to mention that these fluctuations would effectively disappear from the noise statistics (and thus, the theorem (\ref{FDTI}) would be valid for the fluctuating force $\hat{\mathcal{E}}_{i}^{\alpha} (t)$ during the whole time evolution) if we would have access to the initial preparation of the Maxwell-Chern-Simons electromagnetic field and its initial state could be tuned to
\begin{eqnarray}
\hat{\rho}_{\beta} \propto \text{exp}\left(-\beta\sum_{\vect k}\omega(\vect k)\hat a_{d}^{\dagger}(\vect k)\hat a_{d}(\vect k) \right),
\nonumber
\end{eqnarray}
instead of the canonical equilibrium state $\hat\rho_{_{\beta}}\propto e^{-\beta\hat H_{MCS}}$. That is, the environmental annihilation (creation) operator would be initially replaced by a shifted operator $\hat a_{d}$ ($\hat a_{d}^{\dagger}$) which effectively counteracts the backreaction effects. As this is not the case for most interesting physical situations, in the next sections we find useful to briefly present the concrete arguments that justify the microscopic model (\ref{HMCSF2}) may reproduce a relaxation process towards a thermal equilibrium state despite of this issue.

\subsection{Properties of the fluctuating force and retarded self-energy}\label{PSNDR}
Let us draw attention to the retarded self-energy and force fluctuations governing the quantum Langevin dynamics in the asymptotic time limit. From equation (\ref{CFFE}), we may envisage the two-point autocorrelation function of the fluctuating force in the time domain for $t'\leq t$, i.e.
\begin{eqnarray}
\left\langle \left\lbrace  \hat{\mathcal{E}}_{i}^{\alpha} (t), \hat{\mathcal{E}}_{j}^{\beta\dagger} (t')\right\rbrace  \right\rangle_{\hat{\rho}_{\beta}}&=& \left\langle \left\lbrace \hat \xi_{i}^{\alpha} (t), \hat \xi_{j}^{\beta\dagger} (t')\right\rbrace  \right\rangle_{\hat{\rho}_{\beta}} \label{ENE} \\
&+&\Upsilon_{ij}^{\alpha\beta}(t,t',t_{0}) +\Xi_{ij}^{\alpha\beta}(t,t',t_{0}),
\nonumber 
\end{eqnarray}
where $\vect{\hat\xi}_{i}$ describes the previously mentioned thermal noise upon the $i$-th harmonic oscillator given by (\ref{ASEF}), whereas the non-stochastic fluctuations $\Upsilon_{ij}^{\alpha\beta}(t,t',t_{0}) $ and $\Xi_{ij}^{\alpha\beta}(t,t',t_{0})$ represent the inverse Fourier transform of the second and third line of the right-hand side of equation (\ref{CFFE}), respectively. 

Owning to the functions (\ref{GGE}) and (\ref{FFE}) are continuously differentiable for $\kappa\leq x< \infty$ as well as they exponentially decay for values $x$ larger than $(2\sigma)^{-\frac{1}{2}}$, $\Upsilon_{ij}^{\alpha\beta}(t,t',t_{0}) $ and $\Xi_{ij}^{\alpha\beta}(t,t',t_{0})$ can be computed for any finite value $t_{0}$, though we may need to resort to numerical computation methods in most interesting cases. In particular, these fluctuations can be evaluated in the asymptotic time limit $t-t_{0}\rightarrow \infty$ by appealing to the so-called Riemann-Lebesgue lemma \cite{chandrasekharan20121}, which is illustrated in Appendix \ref{app5}, and has been employed in the study of the stationary properties of the damped harmonic oscillator \cite{valido20151,pagel20131}. Essentially, this lemma states that the factor $e^{\pm i(\omega\pm\omega')t_{0}}$ appearing in (\ref{GGE}) and (\ref{FFE}) becomes so highly oscillatory that the integral of the corresponding inverse Fourier transform averages out to zero over the bandwidth of the MSC environment. As a result, it follows from the Riemann-Lebesgue lemma that both $\Upsilon_{ij}^{\alpha\beta}(t,t',t_{0}) $ and $\Xi_{ij}^{\alpha\beta}(t,t',t_{0})$ asymptotically vanishes in the long time limit $t-t_{0}\rightarrow \infty$. We elaborate on this discussion in Appendix\ref{app5}.
In this way, the asymptotic dynamics of the harmonic $n$-particle system will be dominated mainly by the thermal fluctuations, i.e. $\hat{\mathcal{E}}_{i}^{\alpha} (t)\rightarrow \hat\xi_{i}^{\alpha}(t)$ for $t-t_{0}\rightarrow \infty$, and thus, the time asymptotic dissipative dynamics will follow a fluctuation-dissipation relation (\ref{FDTI}), as we wanted to show. We would like to emphasize that this result is general and just rests on the basic properties of the spectral function: it exhibits a broad bandwidth, and a finite and continuous coupling strength between the MSC environment and system particles. 

\begin{figure*}[ht]
\begin{center}
\includegraphics[scale=0.35]{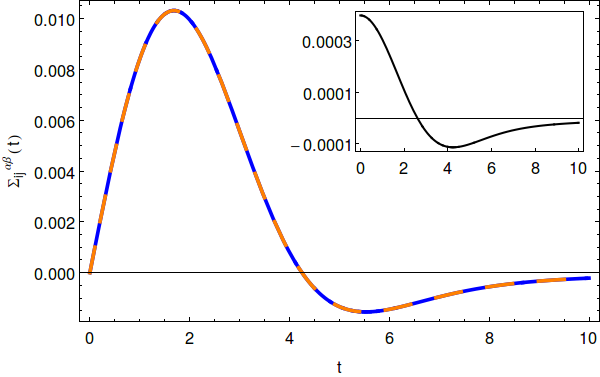}
\includegraphics[scale=0.35]{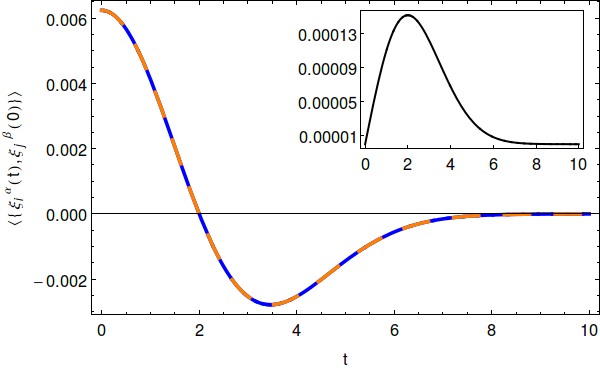}
\caption{(color online). Left: The elements of the retarded self-energy as a function of time. The solid blue and dashed orange lines correspond respectively to $\Sigma_{ij}^{11}(t)$ and $\Sigma_{ij}^{22}(t)$, whereas $\Sigma_{ij}^{12}(t)$ is represented by the solid black line in the inset. Rigth: The plot illustrates the elements of the thermal fluctuations in the zero-temperature regime as a function of time. The diagonal correlations for $\alpha=1,2$ are given by the solid blue and dashed orange lines, respectively. In the inset, the solid black line depicts the off-diagonal correlation $\left\langle \left\lbrace  \hat{\mathcal{E}}_{i}^{1} (t), \hat{\mathcal{E}}_{j}^{2\dagger} (t')\right\rbrace  \right\rangle$. In both pictures, we have fixed $e=1$, $\sqrt{2\sigma}\kappa\simeq0.01$, and $|\Delta\bar{q}_{ij}|/\sqrt{2\sigma}\simeq0.01$. \label{Fig1}} 
\end{center}
\end{figure*}
We pay attention to the properties of the thermal noise $\hat\xi_{i}^{\alpha}(t)$ in what follows. By performing the inverse Fourier transform in equation (\ref{FDTI}) after substituting the expression of the retarded self-energy in terms of the spectral density (\ref{IFSE}), we obtain 
\begin{widetext}
\begin{equation}
\left\langle \left\lbrace \hat \xi_{i}^{\alpha} (t'+\tau), \hat \xi^{\beta\dagger}_{j} ( t')\right\rbrace  \right\rangle=\frac{1}{\pi}\int_{0}^{\infty} d \omega \  \left( 1+2 n\left(  \omega, \beta^{-1}\right) \right)\Big(\text{Re}\left\lbrace  \mathcal{J}_{\alpha\beta}(\omega,\Delta\bar{q}_{ij}) \right\rbrace\cos(\tau\omega)  + \text{Im}\left\lbrace  \mathcal{J}_{\alpha\beta}(\omega,\Delta\bar{q}_{ij}) \right\rbrace\sin(\tau\omega)  \Big). \label{TNE}
\end{equation} 
Observe that the Chern-Simons effects give rise to the off-diagonal contribution contained by the Fourier sine transform term appearing in (\ref{TNE}), as similarly occurs for extended Caldeira-Legget environments \cite{yao20171}. This term encodes all the thermal fluctuations emerging from the Chern-Simons electric field $\hat{\vect E}_{CS} $ acting upon the transversal spatial degrees of freedom. In Secs. \ref{MLDS} and \ref{SHOS}, it is shown that such contribution may be interpreted as an ordinary Hall response associated to $\hat{\vect E}_{CS} $, and interestingly, it may generate long-time correlations between the transversal degrees of freedom of the system particles in the asymptotic equilibrium state.

A profound analysis of the self-energy (\ref{MKEII}) and thermal fluctuations (\ref{TNE}) inferred from the MSC environment is beyond the scope of the present work. Rather we will focus the attention to the realistic physical situation when the Chern-Simons action strength is weak, i.e. $\sqrt{2\sigma}\kappa\ll 1$, and all harmonic oscillators are very close to each other compared with the particle charge distribution, i.e. $\frac{|\Delta\bar{q}_{ij}|}{\sqrt{2\sigma}}\ll 1$ for arbitrary $i$ and $j$. Under these assumptions the definition for the spectral density (\ref{SPD}) may be brought into the simplified expression,
\begin{eqnarray}
\vect{\mathcal{J}}(\omega,\Delta\bar{q}_{ij} )&\simeq& \left( \frac{e}{4}\right)^{2} e^{-2\sigma \omega^{2}} \Bigg( \left( \begin{array}{cc}
\frac{1}{4}\left( 8 \kappa^2 + 2 \left( 4 - \kappa^2 \left( |\Delta\bar{q}_{ij}|^2 - 8 \sigma\right) \right)  \omega^2\right)  & i 4  \kappa\omega \\
-i 4 \kappa \omega  & \frac{1}{4}(8 \kappa^2 + 2(4 + \kappa^2 (|\Delta\bar{q}_{ij}|^2 + 8 \sigma)) \omega^2)\\
\end{array}\right)  \nonumber \\ 
&-&\left(\begin{array}{cc}
(1 + 2 \kappa^2 \sigma) & \frac{4i\kappa}{\omega} \kappa\\
 -\frac{4i\kappa}{\omega} \kappa & 3(1 + 2 \kappa^2 \sigma)\\
\end{array}\right) \left( \frac{|\Delta\bar{q}_{ij}|\omega^2}{2}\right)^2\Bigg), \ \text{with} \ 0\leq \omega.
\label{SPDA}
\end{eqnarray}
Eq.(\ref{SPDA}) is obtained by using the asymptotic form of the Bessel functions of the first kind for small arguments, and then, performed a Taylor series expansion around $|\Delta\bar{q}_{ij}|=0$. Now, replacing (\ref{SPDA}) in (\ref{MKEII}) and using the standard tables of integration \cite{gradshteyn20141}, one may obtain closed-form formulas for the retarded self energy with $0\leq t$. For instance, the off-diagonal elements takes the following form
\begin{eqnarray}
\Sigma_{ij}^{12}(t)= -\Sigma_{ij}^{21}(t) \simeq \frac{e^2\kappa}{64 \pi \sigma^2} \Big(|\Delta\bar{q}_{ij}|^2(-1 + x^2) + 8 \sigma +\frac{x\sqrt{\pi}}{2}  (|\Delta\bar{q}_{ij}|^2 (3-2 x^2) -16\sigma) e^{-x^2}\text{erfi}(x)\Big), \label{SEAIII}
\end{eqnarray}
\end{widetext}
where $x=\frac{t}{\sqrt{8\sigma}}$ and $\text{erfi}(x)=i^{-1}\text{erf}(ix)$, with $\text{erf}(x)$ being  the error function in the variable $x$ \cite{gradshteyn20141}. The computed form for the other terms can be found in the appendix\ref{app6}, see equations (\ref{SEAI}) and (\ref{SEAII}). The components of the retarded self-energy as functions of time are depicted in figure\ref{Fig1}. Paying attention to (\ref{SEAIII}), we may observe that the quantum Langevin dynamics of the harmonic $n$-particle system presents an intricate non-Markovian memory kernel, which exhibits an algebraic behavior at small times, whilst it is dominated by an exponential decayment with vanishing time $(2\sigma)^{-\frac{1}{2}}$ in the long time. Such non-Markovianity is a clear signature of a rich dissipative dynamics \cite{devega20171}. Interestingly, the expression for the off-diagonal component (\ref{SEAIII}) reveals that the fluctuating forces acting upon transversal spatial components of the harmonic oscillators do not commute at $t=0$, rather it takes a finite value proportional to the so-called topological mass $\kappa $ \cite{dunne19991}. Taking into account Eq.(\ref{SFFE}), this feature can be traced back to the fact that the Maxwell-Chern-Simons electric field $\hat{\vect E}_{MCS}(\bar{\vect q}_{i})$ responsible for the dissipative dynamics has non-commutative components (see Eq.(\ref{MCSCCRI})), as pointed out in Sec.\ref{SQD}. It is well-known in the free Maxwell-Chern-Simons electrodynamics \cite{dunne19991,deser19821} that such non-commutative property for the electric fields arises from the latent topological features of the microscopic theory, thus this feature of the memory kernel can be thought of as a topological trademark in the present dissipative dynamics \cite{cobanera20161}.

Figure \ref{Fig1} also illustrates the thermal fluctuations (\ref{TNE}) obtained in the zero-temperature limit after replacing the spectral density by (\ref{SPDA}). The exact representation of the diagonal and off-diagonal elements can be found in the Appendix\ref{app6} (see (\ref{TFAI}),(\ref{TFAII}) and (\ref{TFAIII})). Observe that the time-dependent thermal fluctuations share a similar behavior with the retarded self-energy. Moreover, both diagonal components takes almost on the same values due to the apparent anisotropy of the spectral density vanishes for closed particles, as was just discuss in the previous section. Interestingly, we shall see in Sec.\ref{MLDS} that the transversal contribution (see the insest) at high temperatures may be identified with the fluctuations of an antisymmetric $1/f$ noise in the Markovian Langevin dynamics limit.

In summary, on one hand we have shown that the stationary dissipative effects characterized by (\ref{IFSE}) and the stochastic electric force (\ref{ASEF}) follow a generalized fluctuation-dissipation relation (\ref{FDTI}). Remarkably, the retarded self-energy exhibits a \textit{non-commutative} feature  characteristic of the topologically massive gauge theory. On the other hand, we have seen that the Maxwell-Chern-Simons field induces non-stochastic fluctuations which may dominate the initial dissipative dynamics, but nevertheless their influence on the long-time dynamics can be disregarded, and thus, the system could eventually reach a thermal equilibrium state determined by the initial gauge-field temperature $\beta^{-1}$. We shall discuss in the following section under which conditions any initially prepared configuration of the harmonic $n$-particle system decays into a thermal state at temperature $\beta^{-1}$ due to the interaction with the MCS environment.

\subsection{Equilibrium structure of propagators}\label{SSP}

We now turn the attention to the equilibrium properties of the harmonic $n$-particle system at late times. It proves convenient to analyze this by means of the (one-particle) Green's function $G_{ij}^{\alpha\beta}(t,t')=\left\langle T_{\mathcal{C}}\hat Q_{i}^{\alpha}(t)\hat Q_{j}^{\beta}(t')  \right\rangle $ defined for a pair of particle position operators $\hat{\vect Q}_{i}$ and $\hat{\vect Q}_{j}$, where $T_\mathcal{C}$ denotes the closed time path or Schwinger-Keldysh contour \cite{sieberer20161}. This is also known as the contour-ordered propagator and may be conveniently expressed as follows,
\begin{equation}
G_{ij}^{\alpha\beta}(t,t')=\Delta^{\alpha\beta}_{ij}(t,t')-\frac{i}{2}\text{sgn}_{T_\mathcal{C}}(t-t')\Lambda_{ij}^{\alpha\beta}(t,t'),
\nonumber
\end{equation}
in terms of the spectral and statistical correlators, respectively
\begin{eqnarray}
\Lambda^{\alpha\beta}_{ij}(t,t')&=&i\left\langle \left[\hat Q_{i}^{\alpha}(t),\hat Q_{j}^{\beta}(t') \right]  \right\rangle, \label{SPE} \\ 
\Delta^{\alpha\beta}_{ij}(t,t')&=&\frac{1}{2}\left\langle \left\lbrace \hat Q_{i}^{\alpha} (t),\hat Q_{j}^{\beta}(t') \right\rbrace   \right\rangle. \label{SCE}
\end{eqnarray}
Interestingly, in a thermal equilibrium state the contour-ordered propagator becomes time-translational invariant, i.e. $G_{ij}^{\alpha\beta}(t,t')=G_{ij}^{\alpha\beta}(t-t')$, and more importantly, the above correlators are related such that it is satisfied the Kubo-Martin-Schwinger (KMS) boundary condition \cite{sieberer20161,anisimov20091}, 
\begin{equation}
G_{ij}^{\alpha\beta}(t-t'+i\beta)|_{t<t'}=G^{\alpha\beta}_{ij}(t-t')|_{t'<t}.
\label{KMSC}
\end{equation}
By means of arguments analogous to derive the long-time behavior of the environmental fluctuations, we shall show that the harmonic $n$-particle system asymptotically approaches to a stationary state retrieving the KMS relation (\ref{KMSC}) with $\beta^{-1}$ being the initial gauge-field temperature, under certain conditions consistent with the spectral density (\ref{SPD}).

In order to evaluate the asymptotic time solution of the contour-ordered propagator we firstly solve the Cauchy problem for the quantum Langevin equation (\ref{GLE}). Since the latter constitutes a linear integral-differential equation, its solution can be straightforwardly obtained via either the Laplace or Fourier transform methods \cite{valido20151,alamoudi19981,alamoudi19991}. In this context, the solution can be conveniently expressed in terms of the (Kadanoff-Baym) retarded Green's function matrix $\vect G_{R}(t)$ of the harmonic $n$-particle system \citep{gautier20121,alamoudi19991}, whose entries are completely determined by the time Fourier transform
\begin{equation}
\left( \tilde{\vect G}_{R}^{-1}\right)_{ij}^{\alpha\beta}(\omega) =-\delta_{ij}\delta_{\alpha\beta}m (\omega+i0^{+})^{2}+V_{ij}^{\alpha\beta} -\tilde{\Sigma}_{ij}^{\alpha\beta}(\omega+i0^{+}),
\label{GRF}
\end{equation}
where $\tilde{\vect G}_{R}^{-1}$ denotes the $2n\times2n$ matrix inverse. Additionally, the homogeneous solution $\hat q_{i,h}^{\alpha}(t)$ of equation (\ref{GLE}) is obtained from,
\begin{equation}
\sum_{j=1}^{n}\int_{t_{0}}^{t}d\tau \ \left(\vect G^{-1}_{R}\right)_{ij}^{\alpha\beta}(t-\tau)\hat q_{j,h}^{\beta}(\tau)=0,
\label{HEF}
\end{equation}
by setting the initial conditions $\hat q_{i,h}^{\alpha}(t_{0})=\hat q_{i}^{\alpha}(t_{0})$ and $\dot{\hat q}_{ih}^{\alpha}(t_{0})=\dot{\hat q}_{i}^{\alpha}(t_{0})$ (which imply $\vect G_{R}(t_{0})=\vect I_{2n}$ and $\dot{\vect G}_{R}(t_{0})=m^{-1}\vect I_{2n}$, with $\vect I_{2n}$ being the $2n\times2n$ identity matrix). Notice that $\tilde{\vect G}_{R}(-\omega)=\tilde{\vect G}_{R}^{\dagger}(\omega) $ thanks to the properties of the retarded self-energy (see Eqs. (\ref{KKR}) and (\ref{IFSE})). Then the solution of equation (\ref{GLE}) formally reads,
\begin{eqnarray}
\hat q_{i}^{\alpha}(t)&=&\hat q_{i,h}^{\alpha}(t)  \label{GLES} \\
&+&\sum_{j=1}^{n}\int_{t_{0}}^{t}\left( \vect G_{R}\right)_{ij}^{\alpha\beta}(t-\tau)\left(\hat \xi_{j}^{\beta}(\tau) -V_{j}^{\beta}\right) d\tau ,
\nonumber
\end{eqnarray}
where all the stationary dynamics is completely contained by the second line expression on the right-hand side. For a better exposition, we shall disregard the non-stochastic force $V_{j}^{\beta}$ from the following discussion by doing a local unitary transformation,
\begin{eqnarray}
\hat Q_{i}^{\alpha}(t)=\hat q_{i}^{\alpha}(t)+\sum_{j=1}^{n}\int_{t_{0}}^{t}\left( \vect G_{R}\right)_{ij}^{\alpha\beta}(t-\tau) V_{j}^{\beta}d\tau.
\nonumber
\end{eqnarray}

From the expressions (\ref{HEF}) and (\ref{GLES}) one may see that, for the system particles reach a stationary state independent of an initially prepared configuration, the retarded Green function $\vect G_{R}(t)$ must completely decay (either algebraically or exponentially) at long time. As previously show in Sec.\ref{PSNDR}, this information is completely encoded by its (time) Fourier transform (\ref{GRF}). It is known from previous studies related to the generalized Langevin equation for the conventional damped harmonic oscillator \cite{pagel20131,alamoudi19981,alamoudi19991,dhar20071,joichi19971} that the existence of a well-defined stationary solution entails that $\tilde{\vect G}_{R}(\omega)$ has no an isolated pole $\omega_{b}$ lying outside of the environment dense spectrum, or equivalently, it does not blow up at some value $\omega_{b}$ with $0<\omega_{b}^2< \kappa^2$ (one way to see this is by evaluating its Fourier transform by means of the Riemann-Lebesgue lemma). This immediately implies that the determinant of $\tilde{\vect G}_{R}^{-1}(\omega)$ can only have roots $\omega_{r}$ obeying $\kappa^2<\omega_{r}^2$. By substituting the frequency-domain expression of the retarded self-energy in (\ref{GRF}), the search of these roots may be rephrased in terms of the eigenvalue problem of the real part of $\tilde{\vect G}_{R}^{-1}(\omega)$, i.e. $\text{Re} \ \tilde{\vect G}_{R}^{-1}(\lambda)=m\lambda^2\vect I_{2n}$ with
\begin{eqnarray}
\text{Re} \ \left( \tilde{\vect G}_{R}^{-1}\right)^{\alpha\beta}_{ij} (\lambda)&=&V_{ij}^{\alpha\beta}-\frac{\text{sgn}\left(\lambda \right)}{2\pi}\text{Im}\mathcal{J}_{\alpha\beta}\left(  \big|\lambda\big|,\Delta\bar{q}_{ij}\right) \nonumber \\
&-&\frac{1}{2\pi}\mathcal{H}\Big[\text{Re}\mathcal{J}_{\alpha\beta}\big(|\omega|,\Delta\bar{q}_{ij}\big)\Big]\left( \lambda\right)  \label{STC0} , 
\end{eqnarray}
where $\vect I_{2n}$ denotes the $2n\times 2n$ identity matrix and $\text{sgn}$ stands for the usual sign function. Using results borrowed from the matrix theory \cite{petersen20121}, the condition under which $\vect G_{R}(t)$ asymptotically vanishes could be then expressed in the following compact form 
\begin{equation}
m\kappa^2\vect I_{2n}<\text{Re}\ \tilde{\vect G}_{R}^{-1}(\omega_{r}),
\label{STC}
\end{equation}
that is, $\text{Re}\ \tilde{\vect G}_{R}^{-1}(\omega_{r})-m\kappa^2\vect I_{2n}$ is a positive-definite matrix for $\omega_{r}$.
In the particular case of vanishing Chern-Simons action, the condition (\ref{STC}) returns the known result for the conventional damped harmonic oscillator \cite{alamoudi19981,alamoudi19991,joichi19971}. When (\ref{STC}) is obeyed, the dissipative Hamiltonian (\ref{HMCSF2}) is prevented to have a bound, estable normal mode with frequency $\omega_{b}$. Intuitively speaking, the condition (\ref{STC}) ensures that the spectrum of the gauge field will well accommodate the bare frequencies of the harmonic $n$-particle system (i.e. $\kappa<\omega_{i}<1/\sqrt{2\sigma}$ for $i\in \{1,n\}$), and, consequently, it makes possible an irreversible energy transfer from the system to the environment, at least in a finite time sufficiently larger than the natural time scale of the system particles.

As similarly occurs for an intricate non-Markovian Langevin dynamics in the conventional Brownian motion, it may be rather difficult to obtain the analytic structure of $\tilde{\vect G}_{R}^{-1}(\omega) $ for most cases as manifested by the sophisticated spectral density (\ref{SPD}) and expression (\ref{IFSE}) for the retarded self-energy. Concretely, the real part of the retarded self-energy, which is obtained via the Kramers-Kronig relation (\ref{KKR}), regards a Bessel Hilbert transform that has no analytic expression  in general, and thus, one has to resort to numerical methods \cite{xu20141}. Nonetheless, this treatment substantially simplifies for the previously discussed situation when all the system particles are very close to each other and the Chern-Simons action is weak (i.e. $\sigma\kappa^2\ll 1$ and $\frac{|\Delta\bar{q}_{ij}|}{\sqrt{2\sigma}}\ll 1$ for arbitrary $i$ and $j$). Recall that the positive constraint upon the microscopic Hamiltonian (\ref{DHPCE}) discussed in Sec.\ref{SQD} imposes the consistency condition (\ref{PCREII}) as well. As shown previously, the dissipation produced by the MCS environment is then characterized by a sort of super-Ohmic spectral density (\ref{SPDA}). Replacing this in (\ref{STC0}) and computing the corresponding Hilbert transform, it may be verified that there is no (non-trivial) solution $\omega_{r}$ which violates (\ref{STC}) provided the particle frequencies lies in the MCS environment spectrum (i.e. $\kappa<\omega_{i}$ for $i\in \{1,n\}$), and further, the subsidiary condition (\ref{PCREII}) is obeyed. To see this we must realize that the off-diagonal elements of $\tilde{\vect G}_{R}^{-1}(\omega_{r})$ represent a perturbative contribution to $\det\tilde{\vect G}_{R}^{-1}(\omega)$ by virtue of (\ref{PCREII}). Hence, we expect that the harmonic oscillators relax towards a thermal equilibrium state in the closed-particle and weak-Chern-Simons-action picture. In more general scenarios, the condition (\ref{STC}) will be satisfied depending mainly on the values of the oscillator bare frequencies $\omega_{i}$, width of the particle charged distribution $\sigma$, Chern-Simons constant $\kappa$, and environmental coupling $e$.

Once the stationary solution is guaranteed (and the condition (\ref{STC}) is fully satisfied), it is convenient to take the limit $t-t_{0}\rightarrow \infty$ in Eq.(\ref{GLES}) and rewrite the stationary solution of the position operator in terms of its Fourier transform, i.e. $\hat Q_{i,s}^{\alpha}(\omega)=\sum_{l=1}^{n}\left( \tilde{\vect G}_{R}\right)_{il}^{\alpha\gamma}(\omega)\tilde{\xi}_{l}^{\gamma}(\omega)$. After replacing this into both the spectral (\ref{SPE}) and statistical (\ref{SCE}) correlators, and averaging over the initial state $\hat\rho_{\beta}$, we arrive at their Fourier transform $\tilde{\Lambda}^{\alpha\beta}_{ij}(\omega,\omega')$ and $\tilde{\Delta}^{\alpha\beta}_{ij}(\omega,\omega')$, respectively. These are directly obtained by taking into account the  anti-commutator (\ref{IFSE}) and commutator (\ref{CFFE}) expressions of the fluctuating force in the frequency domain. As a result, we find for the spectral correlator
\begin{eqnarray}
&&\tilde{\Lambda}^{\alpha\beta}_{ij}(\omega,\omega')= \label{KMSR1} \\
&=&2i\delta(\omega-\omega')\sum_{l,l'=1}^{n}\bigg[\left( \tilde{\vect G}_{R}\right)_{il}^{\alpha\gamma}\ \mathcal{I}\tilde{\Sigma}_{ll'}^{\gamma\gamma'}\left( \tilde{\vect G}_{R}^{\dagger}\right)_{l'j}^{\gamma'\beta}\bigg] (\omega), \nonumber
\end{eqnarray}
which is absent of the non-stochastic fluctuations $\Upsilon_{ij}^{\alpha\beta}(t,t',t_{0}) $ and $\Xi_{ij}^{\alpha\beta}(t,t',t_{0})$. In contrast, the Fourier transform of the latter functions are involved in the expression for $\tilde{\Delta}^{\alpha\beta}_{ij}(\omega,\omega')$. Once again we may evaluate the time asymptotic limit of the statistical correlator by using the aforementioned Riemann-Lebesgue lemma after interchanging it by the integral in the frequency domain (as similarly discussed in the previous section). Recall that $\tilde{\Upsilon}_{ij}^{\alpha\beta}(\omega,\omega',t_{0}) $ and $\tilde{\Xi}_{ij}^{\alpha\beta}(\omega,\omega',t_{0})$ explicitly inherit the fast oscillatory behavior exhibited by (\ref{GGE}) and (\ref{FFE}), such that we may drop the non-stochastic fluctuation contribution in the asymptotic limit $t-t_{0}\rightarrow \infty$ once we appeal to the fact that the retarded Green's function is integrable according to the condition (\ref{STC}). Thus it can be shown that this yields
\begin{eqnarray}
\tilde{\Delta}_{ij}^{\alpha\beta}(\omega,\omega') &=&-\frac{i}{2} \left( 1+2 n\left( \omega, \beta^{-1}\right) \right) \tilde{\Lambda}_{ij}^{\alpha\beta}(\omega,\omega'). \label{KMSR2} 
\end{eqnarray}
This result is in agreement with the previous finding about the asymptotic vanishing of the non-stochastic fluctuations in Sec\ref{PSNDR}. Now, it is simple to verify that the KMS relation (\ref{KMSC}) is recovered from (\ref{KMSR1}) and (\ref{KMSR2}) via their inverse Fourier transform. Consequently, this certifies that the harmonic $n$-particle system asymptotically reaches a thermal equilibrium state regardless its initial configuration, as we wanted to show. On the other hand, the KMS relation (\ref{KMSC}) reveals that the system particle undergoes a stationary Gaussian process in the long-time limit completely determined by the statistical propagator $\Delta_{ij}^{\alpha\beta}(t'+\tau,t')=\Delta_{ij}^{\alpha\beta}(\tau)$ and its time derivatives. The latter probes that the microscopic Hamiltonian (\ref{HMCSF2}) in the asymptotic limit provides us with a (numerically) solvable model for the dissipative dynamics of harmonic planar systems. 

In the following section, we shall focus the attention in a non-equilibrium situation of physical interest which arises when the dissipative dynamics is well described by time-local damping, or equivalently, the retarded Green's function in the time domain is exclusively dominated by an exponential decaying behavior at long time. This is commonly recognized as the Markovian regime \cite{hanggi20051}, and has been extensively studied for the damped harmonic oscillator based on the independent-oscillator model \cite{caldeira19831,weiss20121,haake19851,ford19881,hu19921}, or alternatively, by phenomenological stochastic models \cite{grabert19841}. We shall illustrate how this description emerges in the present treatment and briefly discuss its validity.

\section{Example: Markovian Langevin dynamics}\label{MLDS}

Since the specific form of the spectral density (\ref{SPE}) consists of an algebraic combination of Bessel functions evaluated in square roots and weighted by a Gaussian function, we may expect that the retarded Green's function (endowed with the condition (\ref{STC})) may display an intricate mixture of "particle" poles and brunch cut singularities in the complex $\omega$-plane \cite{gautier20121,joichi19971,alamoudi19981,alamoudi19991}. It is well known that the latter singularities contribute to the dynamics with a power-law decaying evolution, whilst the former render the previously mentioned exponential behavior characteristic of the Markovian dynamics. Now we shall assume that the dissipative dynamics of the harmonic $n$-particle system is mainly dominated by particle poles. Since the latter mathematically represent (simple) complex poles, the Markovian case corresponds to focus the attention when $(\tilde{\vect G}_{R})_{ij}^{\alpha\beta}(\omega)$ takes the form of a rational function \cite{spiegel19931}. Formally, this is equivalent to approximate $\tilde{\vect G}_{R}(\omega)$ by a Breit-Wigner resonance shape \cite{alamoudi19981,alamoudi19991,gautier20121,joichi19971,anisimov20091}, i.e. $\tilde{\vect G}_{R}(\omega)\simeq  \tilde{\vect G}_{BW}(\omega+i0^{+})$ with 
\begin{equation}
\left(\tilde{\vect G}^{-1}_{BW}\right)_{ij}^{\alpha\beta}(\omega)= \frac{-\delta_{ij}\delta_{\alpha\beta}\omega^2-i 2\omega \Gamma_{ij}^{\alpha\beta}+ \left( \Omega^{\circ2}\right)_{ij}^{\alpha\beta}}{(2\pi m)^{-1}Z_{ij}^{\alpha\beta}} ,
\label{BWRE}
\end{equation}
where the matrix entries $\left( \Omega^{\circ2}\right)_{ij}^{\alpha\beta}$ and $\Gamma_{ij}^{\alpha\beta}$  are given by,
\begin{eqnarray}
m\left( \Omega^{\circ2}\right)_{ij}^{\alpha\beta}&=&V_{ij}^{\alpha\beta} \label{BWCI}  \\
&-&\frac{1}{2\pi}\text{sgn}\left(\Omega_{ij}^{\alpha\beta} \right)\text{Im}\mathcal{J}_{\alpha\beta}\left( \big|\Omega_{ij}^{\alpha\beta}\big|,\Delta\bar{q}_{ij}\right) \nonumber\\
&-&\frac{1}{2\pi}\mathcal{H}\Big[\text{Re}\mathcal{J}_{\alpha\beta}\big(|x|,\Delta\bar{q}_{ij}\big)\Big]\left(\Omega_{ij}^{\alpha\beta} \right) ,  \nonumber \\
\Gamma^{\alpha\beta}_{ij}&=&\frac{m^{-1}Z_{ij}^{\alpha\beta}}{2(2\pi)\Omega_{ij}^{\alpha\beta}}\bigg(\frac{1}{2\pi}\text{Re}\mathcal{J}_{\alpha\beta}\left(\big|\Omega^{\alpha\beta}_{ij}\big|,\Delta\bar{q}_{ij}\right)  \label{BWCII}\\
&-&\frac{1}{2\pi}\mathcal{H}\Big[\text{sgn}(x)\text{Im}\mathcal{J}_{\alpha\beta}\big(|x|,\Delta\bar{q}_{ij}\big)\Big]\left( \Omega^{\alpha\beta}_{ij}\right)\bigg) ,
\nonumber
\end{eqnarray}
with $Z_{ij}^{\alpha\beta}$ being a renormalization factor,
\begin{eqnarray}
Z_{ij}^{\alpha\beta}&=& 2\pi\Bigg[1 \label{ZRF} \\
&+&\frac{1}{2\pi m}\frac{\partial}{\partial\omega^2}\bigg(\mathcal{H}\Big[\text{Re}\mathcal{J}_{\alpha\beta}\big(|x|,\Delta\bar{q}_{ij}\big)\Big](\Omega^{\alpha\beta}_{ij}) \nonumber \\
&+& \text{sgn}\left(\Omega_{ij}^{\alpha\beta} \right)\text{Im}\mathcal{J}_{\alpha\beta}\left( \big|\Omega_{ij}^{\alpha\beta}\big|,\Delta\bar{q}_{ij}\right)\bigg)\Bigg]^{-1}.
\nonumber
\end{eqnarray}
The matrix elements $\Omega_{ij}^{\alpha\beta}$ are obtained from the Hadamard (entrywise) power $\left( \Omega^{\circ2}\right)_{ij}^{\alpha\beta}= \Omega_{ij}^{\alpha\beta}\Omega_{ij}^{\alpha\beta}$, and further, they satisfy $\epsilon_{\alpha\beta}(\Omega^{\circ2})_{ij}^{\alpha\beta}=\epsilon_{\beta\alpha}(\Omega^{\circ2})_{ij}^{\beta\alpha}$ in order (\ref{BWRE}) preserves the Hermitian property of the equations. Additionally, to be the Breit-Wigner approximation (\ref{BWRE}) physically consistent with a Markovian treatment of dissipative harmonic systems, we demand
\begin{eqnarray}
\frac{\Omega_{ij}^{\alpha\beta}}{\Omega_{ij}^{\alpha\alpha}}<\frac{\Gamma^{\alpha\alpha}_{ij} }{ \Omega_{ij}^{\alpha\alpha}}&\ll & 1 \ \ \text{with}  \ \alpha\neq \beta, \label{BWREC}  
\end{eqnarray}
as well as $0< \Gamma^{\alpha\alpha}_{ij},\Omega_{ij}^{\alpha\alpha}$ for $i,j\in\left\lbrace1,n \right\rbrace $ and $\alpha,\beta=1,2$. Expression (\ref{BWREC}) reflects nothing else but the fact that the Makovian dynamics emerges when the system-environment coupling is weak in comparison with the particle bare  frequencies \cite{riseborough19851,haake19851}. On the other hand, the left-hand side of (\ref{BWREC}) is motivated by the subsidiary condition (\ref{PCREII}) discussed in Sec.\ref{SQD}. This point will be cleared in the next section when we deal with a concrete example, e.g. the single harmonic oscillator case. 

The ansatz (\ref{BWRE}) is inspired by the Breit-Wigner resonance shape for the scenario of a two-dimensional system composed of $n$ independent damped harmonic oscillators. Following the prescription from Refs.\cite{gautier20121,anisimov20091}, this may be obtained by doing a Taylor expansion of the real and imaginary parts of the retarded self-energy around the particle pole. Recalling that $\text{Im}\tilde{\Sigma}_{ij}^{\alpha\beta}(\omega)$ and $\text{Re}\tilde{\Sigma}_{ij}^{\alpha\beta}(\omega)$ are respectively odd and even functions in the frequency domain (according to the Kramers-Kronig relation (\ref{KKR})), such prescription may be extended to our treatment by considering the following Taylor expansions near $\Omega_{ij}^{\alpha\beta}$,
\begin{eqnarray}
\text{Im}\tilde{\Sigma}_{ij}^{\alpha\beta}(\omega)&\simeq &\text{Im}\tilde{\Sigma}_{ij}^{\alpha\beta}\left( \Omega_{ij}^{\alpha\beta}\right) +\frac{4\pi m  \Gamma_{ij}^{\alpha\beta}}{Z_{ij}^{\alpha\beta}} \left( \omega -\Omega^{\alpha\beta}_{ij}\right), \label{BWFI} \\
\text{Re}\tilde{\Sigma}_{ij}^{\alpha\beta}(\omega)&\simeq &\text{Re}\tilde{\Sigma}_{ij}^{\alpha\beta}\left( \Omega_{ij}^{\alpha\beta}\right) \label{BWFII} \\
&-& m\left(1-\frac{2\pi}{Z_{ij}^{\alpha\beta}} \right) \left(\delta_{ij}\delta_{\alpha\beta}\omega^2- (\Omega^{\circ2})_{ij}^{\alpha\beta}\right),  \nonumber
\end{eqnarray}
where we have defined without loss of generality,
\begin{eqnarray}
\frac{\partial \text{Im} \tilde{\Sigma}_{ij}^{\alpha\beta}(\omega)}{\partial \omega}\Bigg|_{\omega=\Omega_{ij}^{\alpha\beta}}&=&\frac{4\pi m\Gamma_{ij}^{\alpha\beta}}{Z_{ij}^{\alpha\beta}}, \nonumber \\
\frac{\partial \text{Re} \tilde{\Sigma}_{ij}^{\alpha\beta}(\omega)}{\partial \omega^2}\Bigg|_{\omega=\Omega_{ij}^{\alpha\beta}}&=&-m\left(1-\frac{2\pi}{Z_{ij}^{\alpha\beta}} \right) \nonumber. 
\end{eqnarray}
By replacing equations (\ref{BWFI}) and (\ref{BWFII}) in the inverse of the retarded Green's function (\ref{GRF}), we are lead to the constraints (\ref{BWCI}) and (\ref{BWCII}) by requiring this takes the form of the ansatz (\ref{BWRE}). Notice that the final expression of these constraints is obtained after substituting the retarded self-energy by the spectral density, where the factor $2\pi$ essentially appears because the definition of the latter consider here. Similarly, (\ref{ZRF}) follows directly from (\ref{BWFII}), and corresponds to the so-called wave function renormalization in the particular case of independent damped harmonic oscillators. One may verify that Eqs. (\ref{BWFI}) and (\ref{BWFII}) retrieve the Breit-Wigner resonance shape for $n$ two-dimensional oscillator following the conventional Brownian motion, as desired. 

By considering the ansatz (\ref{BWRE}), we obtain a Markovian dynamics in agreement with the dissipative properties encoded by the spectral density. Indeed, it can be shown that the inverse Fourier transform of (\ref{BWRE}) returns the following Markovian Langevin equation for our harmonic $n$-particle system,
\begin{eqnarray}
\frac{2\pi m}{Z_{ij}^{\alpha\alpha}}\ddot{\hat{Q}}_{i}^{\alpha}(t)&+&2\pi m\sum_{j=1}^{n}\sum_{\beta=1,2}\frac{2\Gamma_{ij}^{\alpha\beta}}{Z_{ij}^{\alpha\beta}}\dot{\hat{Q}}_{j}^{\beta}(t) \label{MLGE}\\
&+&2\pi m\sum_{j=1}^{n}\sum_{\beta=1,2}\frac{\left( \Omega^{\circ2}\right)_{ij}^{\alpha\beta} }{Z_{ij}^{\alpha\beta}}\hat{Q}_{j}^{\beta}(t)=\hat{\xi}_{i,BW}^{\alpha}(t),
\nonumber 
\end{eqnarray}
where $\hat{\xi}_{i,BW}^{\alpha}(t)$ is the associated thermal noise which is determined from (\ref{TNE}). Aside the contribution from the quadratic potential $V_{ij}^{\alpha\alpha}$ contained in $\Omega_{ij}^{\alpha\alpha}$ (notice that $V_{ij}^{\alpha\beta}=0$ for $\alpha\neq\beta$), one may realize from (\ref{MLGE}) that the MSC environment in the Markovian limit may mediate an effective interaction between transversal spatial components that is characteristic of a linear drift force \cite{chun20181}. More concretely, the latter is equivalent to an array of non-conservative rotational forces acting on the harmonic oscillators
\begin{equation}
\hat{\vect F}_{R,i}=-2\pi m\sum_{j=1}^{n}\Omega^{2}_{ij} ( \hat{\vect Q}_{j}\times \hat{\vect e}_{3}), \ \text{with} \ i\in \{1,n\} \label{RFCS}
\end{equation}
where $\left( \Omega^{\circ2}\right)_{ij}^{\alpha\beta}=\epsilon^{\alpha\beta}Z_{ij}^{\alpha\beta}\Omega^{2}_{ij}$ and $\hat{\vect e}_{3}$ denotes the (unit) normal vector to the plane defined by the system. Upon close inspection of equation (\ref{BWCI}), one may see that $\hat F^{\alpha}_{R,i}$ must exclusively arise from the imaginary part of the spectral density (see Eqs. (\ref{MKEII}) and (\ref{SPD})). The latter pinpoints the Chern-Simons electric field $\hat{\vect E}_{CS}$ as the responsible for (\ref{RFCS}), which is primarily induced by the particle-attached Chern-Simons flux discussed in Secs. \ref{SGID} and \ref{SQD}. Interesting enough, this situation is common in static MCS electrodynamics \cite{moura20011}, and invokes a physical picture for our dissipative system that recalls an intricate ensemble of dissipative coupled magnetic-like vorteces: each oscillator follows a rotational symmetric motion carrying certain Chern-Simons flux that simultaneously interacts between each other with strength determined by $\Omega_{ij}$. A similar vortex-like dynamics was also found in a disipative microscopic description of type-II superconductors \cite{ao19991}. To be in agreement with an asymptotic stationary picture, there must exist a complex interplay between these driving forces and the energy dissipated by the corresponding environmental noise. Indeed, we show below that the noise associated to (\ref{RFCS}) via the fluctuation-dissipation relation (\ref{FDTI}) resemblances to a magnetic flux noise, which reinforces the aforementioned vision of flux-carrying particles. Accordingly, these rotational forces constitute a perturvative repulsive interaction (which may globally drift away the system particles) that counteracts the particle confining potential.

Alternatively, the ansatz (\ref{BWRE}) can be though of as the retarded Green's function resulting from the generalized Langevin equation after assuming certain form for the spectral density, namely $\mathcal{J}_{\alpha\beta}^{BW}$, that fulfills Eqs. (\ref{BWCI}) and (\ref{BWCII}). In other words, we should be able to deduce a Markovian Langevin equation identical to the one associated to the Breit-Wigner ansatz (\ref{MLGE}) if we replace such spectral density $\mathcal{J}_{\alpha\beta}^{BW}$ in the expression (\ref{MKEII}) for the retarded self-energy. In this way, we could follow an inverse line of thinking to figure out the specific form of $\mathcal{J}_{\alpha\beta}^{BW}$ by requiring the expression (\ref{MKEII}) reproduces a time-local memory kernel which agrees with (\ref{MLGE}), i.e. $\Sigma_{ij}^{\alpha\beta}(t-t')\propto \delta(t-t')$. On one hand, since the off-diagonal elements of the spectral density are just contained in the Fourier cosine transform of (\ref{MKEII}), the transformation rules for the latter entails (for arbitrary $i,j$): $\Gamma_{ij}^{\alpha\beta}=0$ and $\mathcal{J}_{\alpha\beta}^{BW}(\omega,\Delta\bar{q}_{ij})\propto i \Omega_{ij}^2$ for $\alpha\neq\beta$. On the other hand, $\mathcal{J}_{\alpha\beta}^{BW}$ must be Hermitian to provide a sensible description as explained in Sec.\ref{SGLE}. Introducing together these results in (\ref{BWCI}) and (\ref{BWCII}), we are led to the following spectral density associated to the Breit-Wigner approximation (\ref{BWRE}),
\begin{equation}
\mathcal{J}_{\alpha\beta}^{BW}(\omega,\Delta\bar{q}_{ij})=\delta_{\alpha\beta}\frac{2m\pi\Gamma_{ij}^{\alpha\alpha}}{Z_{ij}^{\alpha\alpha}}\omega-i\epsilon_{\alpha\beta}\frac{2\pi m\Omega^{2}_{ij}}{2}.
\label{SPEBW}
\end{equation}
Using the Fourier sine and cosine transformation rules, it is simple to verify that expression (\ref{SPEBW}) matches with the Markovian Langevin equation (\ref{MLGE}). Interestingly, the diagonal elements of $\mathcal{J}_{\alpha\beta}^{BW}$ takes the form of an Ohmic spectral density (which is characteristic of the strict Markovian limit \cite{weiss20121}), whilst the off-diagonal contribution exhibits a flat or white spectrum. 

Additionally, we may obtain the fluctuation-dissipation relation determining the corresponding fluctuating force $\hat{\xi}_{i,BW}^{\alpha}(t)$ by replacing (\ref{SPEBW}) in the expression (\ref{TNE}). In the high temperature limit $\Gamma_{ij}^{\alpha\alpha}\beta \ll 1$, this takes the particular form
\begin{eqnarray}
&&\left\langle \left\lbrace \hat \xi_{i,BW}^{\alpha} (t'+t), \hat \xi^{\beta\dagger}_{j,BW} ( t')\right\rbrace  \right\rangle  =\delta_{\alpha\beta}\frac{2m(2\pi)\Gamma_{ij}^{\alpha\alpha}}{Z_{ij}^{\alpha\alpha}\beta}\delta(t-t') \nonumber \\
&+&\epsilon_{\alpha\beta}\frac{2\pi m\Omega^{2}_{ij}}{2\beta}\text{sgn}(t-t'),  \label{MWN}
\end{eqnarray}
where again we have made use of the well-known properties of the Dirac delta function and the standard tables of Fourier cosine and sine transforms \cite{gradshteyn20141}. 

The statistical correlator (\ref{MWN}) reveals intriguing features of the fluctuating force acting on the harmonic $n$-particle system at high temperatures, specially for the off-diagonal correlations arising from the Chern-Simons effects. We may clearly recognize the first line in (\ref{MWN}) as the well-known fluctuation-dissipation relation due to the white noise presented in the conventional Brownian motion \cite{hanggi20051}. Remarkably, the second line closely resemblances to the ordinary Hall response of two-dimensional particles found in the dissipative Hofstadter model \cite{callan19921,novais20051}, such that the fluctuations of the transversal spatial components may be interpreted as the Hall effect counterpart in the present context, occurring here because the Chern-Simons flux. Moreover, the second line in the frequency domain identically coincides with the fluctuations of an antisymmetric 1/f noise of power spectrum $S(\omega)=(i\omega)^{-1}$. The latter may be realized by paying attention to (\ref{TNE}): the hyperbolic cotangent renders an effective inverse scaling for the power spectrum of the transversal correlations, whereas the spectral density contributes with a constant factor owning to the form (\ref{SPEBW}). Interestingly, $1/f$ power law is characteristic of the low-frequency magnetic flux noise in superconducting circuits \cite{anton20131,lee20081}, which suggests that the environmental noise acting upon transversal spatial degrees of freedom originates from the fluctuations of the Chern-Simons flux carried by the harmonic oscillators. The time-reversal asymmetry and parity violation of the present microscopic description can be clearly appreciated from Eqs. (\ref{MLGE}) and (\ref{MWN}). In principle, the latter can be thought of as an extension to the classical Einstein relation \cite{hanggi20051}.

A final remark in this section is that the spectral density (\ref{SPEBW}) cannot provide a fully physical description of the dissipative dynamics (and thus neither (\ref{MLGE})), as similarly occurs for the Ohmic spectral density in the standard microscopic model (i.e. it gives rise to the so-called ultraviolet catastrophe). Here, it is important to emphasize that the Markovian Langevin equation (\ref{MLGE}) will be valid when the coupling between the harmonic $n$-particle system and gauge field is weak according to (\ref{BWREC}). Physically, this could well correspond to the scenario of closed particles and weak Chern-Simons action in the weak-damping regime, i.e. $e/\omega_{i}\ll 1$ and $\omega_{i}\sqrt{2\sigma}\ll1$.

\subsection{Single harmonic oscillator}\label{SHOS}
So far we have followed a very general treatment of the harmonic $n$-particle system. Aiming to provide a clear comparison with the conventional (two-dimensional) isotropic damped harmonic oscillator \cite{hanggi20051}, we now address the asymptotic dissipative dynamics of the single harmonic oscillator case (i.e. $n=1$) within the previous Markovian framework. For a better exposition, we take the parameters involved in Eqs. (\ref{BWRE}) and (\ref{SPEBW}) as follows:  $\Omega^{12}=-\Omega^{21}=(2\pi)^{-\frac{1}{2}}\Omega_{CS}$ , while for the diagonal elements $\Omega^{11}=\Omega^{22}=\Omega_{0}$ and $\Gamma^{11}=\Gamma^{22}=\Gamma_{0}$. Furthermore, substituting (\ref{MWN}) in (\ref{ZRF}) yields $Z^{\alpha\alpha}=2\pi$ for $\alpha,\beta=1,2$. Note that we have chosen an isotropic damping rate since the apparent anisotropy of the spectral density cancels out for the single oscillator case as discussed in Sec.\ref{SGLE}, and further, $\Omega_{CS}$ determines the strength of the rotational forces discussed in the previous section.

Owing to the rational form of the Breit-Wigner approximation, the retarded Green's function $ \tilde{\vect G}_{SO}(\omega)$ is found to decay algebraically as faster as $\sim\omega^{-3}$ and has no brunch cut. Instead it possesses four complex-conjugate simple poles, denoted by $-i\lambda_{\pm}$, in the $\omega$-plane, i.e.
\begin{eqnarray}
\lambda_{\pm}=\Gamma_{0}\pm \eta \ \ \text{with} \ \ \eta=\left( \Gamma_{0}^2-\Omega_{0}^2+i\Omega_{CS}^2\right)^{\frac{1}{2}},
\end{eqnarray}
aside its complex conjugate. Hence we may use contour integration methods \cite{spiegel19931} to obtain its inverse Fourier transform. Note that the contours at infinity do not contribute to the final integral due to $\tilde{\vect G}_{SO}(\omega)$ vanishes rapidly at large frequency. Hence, it is simple to verify that, 
\begin{eqnarray}
\vect G_{SO}(\tau)=\Theta(\tau)\left( \begin{array}{cc}
f_{+}(\eta \tau) & -if_{-}(\eta \tau) \\
 i f_{-}(\eta \tau) & f_{+}(\eta \tau) \\  
\end{array}\right) ,
\label{RGFSO}
\end{eqnarray}
where $\tau=t-t_{0}$ and we have introduced the following auxiliary functions
\begin{eqnarray}
f_{\pm}(tz)=\frac{e^{-\Gamma_{0} t}}{2m}\bigg(\frac{\sinh(tz^{\dagger})}{z^{\dagger}}\pm\frac{\sinh(tz)}{z}\bigg),
\nonumber
\end{eqnarray}
which returns the familiar solution of the Markovian damped harmonic oscillator when $\Omega_{CS}\rightarrow 0$. At first sight, the retarded Green's function (\ref{RGFSO}) does not seem to decay at late times, preventing the single harmonic oscillator to relax towards an equilibrium state. However, one may show that the consistency condition (\ref{BWREC}) guarantees (\ref{RGFSO}) asymptotically vanishes. Concretely, it is immediate to see that (\ref{RGFSO}) will exhibit an exponential decayment if the following inequality holds,
\begin{equation}
\text{Re}\left\lbrace \sqrt{\Gamma_{0}^2-\Omega_{0}^2+i\Omega_{CS}^2}\right\rbrace  <\Gamma_{0}.
\nonumber
\end{equation}
By means of identities for the complex square root, this inequality is equivalent to $\Omega_{CS}^4<4\Gamma_{0}^2\Omega_{0}^2$, which is clearly satisfied by recalling (\ref{BWREC}), as we wanted to show. Note this result is in complete agreement with the subsidiary condition (\ref{PCREII}) derived in Sec.\ref{SQD}. 

Now we study the position autocorrelation function denoted by $\Delta^{\alpha\alpha}_{SO}(t)$ and defined in (\ref{SCE}). As before, this may be obtained via equations (\ref{KMSR1}) and (\ref{KMSR2}) by using standard contour integration techniques to compute the corresponding inverse Fourier transform, once we have replaced the single-oscillator expressions for both the spectral density (\ref{MWN}) and retarded Green's function $\tilde{\vect G}_{SO}(\omega)$. Specifically, we arrive at the following identities
\begin{eqnarray}
\Delta^{11}_{SO}(t) &=&\Delta^{22}_{SO}(t) \nonumber \\
&=& \text{Re}\left\lbrace S_{0}(t)+S_{CS}(t)\right\rbrace  +\frac{4\Gamma_{0}}{m\beta} \dot{S}_{0}^{(q)}(t), \label{AFSOI}
\end{eqnarray}
where we have introduced,
\begin{eqnarray}
S_{0}(t)&=&\frac{i\mu_{+}}{2m\mu_{-}(\lambda_{+}-\lambda_{-})}\bigg(\coth\left(\frac{i\beta\lambda_{-}}{2}\right)e^{-\lambda_{-}t} \nonumber \\
&-&\coth\left(\frac{i\beta\lambda_{+}}{2}\right)e^{-\lambda_{+}t} \bigg),\nonumber \\
S_{CS}(t)&=&\frac{2\Gamma_{0}\Omega_{CS}^2}{m\mu_{-}\left(\lambda_{+}-\lambda_{-}\right) }\bigg(\lambda_{+}\coth\left(\frac{i\beta\lambda_{-}}{2}\right)e^{-\lambda_{-}t} \nonumber \\
&-&\lambda_{-}\coth\left(\frac{i\beta\lambda_{+}}{2}\right)e^{-\lambda_{+}t}\bigg) , \label{CSRE} 
\end{eqnarray}
with $\mu_{\pm}=4\Gamma_{0}^{2}\Omega_{0}^2\pm \Omega_{CS}^{4}$, and the quantum corrections,
\begin{eqnarray}
S_{0}^{(q)}(t)&=&\sum_{n=1}^{\infty}\frac{e^{-\nu_{n}t}\left(\left(\nu_{n}^2+\Omega_{0}^2\right)^2+3\Omega_{CS}^4-4\Gamma_{0}^2 \nu_{n}^2\right)}{(\nu_{n}^2-\lambda_{+}^2)(\nu_{n}^2-\lambda_{-}^2)(\nu_{n}^2-\lambda_{+}^{\dagger 2})(\nu_{n}^2-\lambda_{-}^{\dagger 2})},\nonumber \\
S_{CS}^{(q)}(t)&=&\sum_{n=1}^{\infty}\frac{e^{-\nu_{n}t}\left(\left(\nu_{n}^2+\Omega_{0}^2\right)^2+\Omega_{CS}^4-12\Gamma_{0}^2 \nu_{n}^2\right)}{(\nu_{n}^2-\lambda_{+}^2)(\nu_{n}^2-\lambda_{-}^2)(\nu_{n}^2-\lambda_{+}^{\dagger 2})(\nu_{n}^2-\lambda_{-}^{\dagger 2})},\nonumber \\
  \label{QSRE} 
\end{eqnarray}
with $\nu_{n}$ being the positive bosonic Matsubara frequencies, i.e. $\nu_{n}=2\pi n/\beta $. It is straightforward to show that Eq.(\ref{AFSOI}) returns the well-known results of the Markovian damped harmonic oscillator \cite{grabert19841} for zero Chern-Simons action (i.e. $\Omega_{CS}\rightarrow 0$). From expression (\ref{AFSOI}) is immediate to get the mean-square dispersion of the harmonic oscillator position along one dimension, i.e. $\left\langle \hat Q^2\right\rangle_{\beta}=\Delta^{11}_{SO}(0)=\Delta^{22}_{SO}(0)$. To evaluate the classical and quantum limit, it is convenient to rewrite (\ref{AFSOI}) in terms of the psi function $\psi(x)$, which is the logarithmic derivative of the gamma function. In particular the quantum corrections $\dot{S}_{0}^{(q)}(0)$ could be written as a combination of $\psi(x)$ by following a similar procedure to Ref.\cite{grabert19841}. 

In the high temperature limit ($\Gamma_{0}\beta\ll 1$), the quantum corrections vanishes as can be clearly seen from equation (\ref{QSRE}). After some straightforward manipulation we then arrive at,
\begin{eqnarray}
\left\langle \hat Q^2\right\rangle_{\beta}&=&\frac{\Omega_{CS}^4\Omega_{0}^2+4\Gamma_{0}^2\left(\Omega_{0}^4+2\Omega_{CS}^4 \right) }{m\beta \left(4\Gamma_{0}^2\Omega_{0}^2-\Omega_{CS}^4 \right)\left(\Omega_{0}^4+\Omega_{CS}^4 \right) }\label{MSDC} \\
&= &\frac{1}{m\beta\Omega_{0}^2}\left(1+\left(1+\frac{\Omega_{0}^2}{2\Gamma_{0}^2}\right) \left( \frac{\Omega_{CS}}{\Omega_{0}}\right)^4  \right) \nonumber\\
&+&\mathcal{O}\left( \left( \frac{\Omega_{CS}}{\Omega_{0}}\right)^8\right),
\nonumber
\end{eqnarray}
where the first term coincides identically with the classical correlation function of the damped harmonic oscillator \cite{grabert19841}. Clearly, the above expression shows that the Chern-Simons action in the Markovian limit induces an effective broadening and shifting of the harmonic oscillator spectrum, so that the two-dimensional particle is expected to be in a thermal equilibrium distribution incorporating these effects. 

\begin{figure}[ht]
\includegraphics[width=0.95\columnwidth]{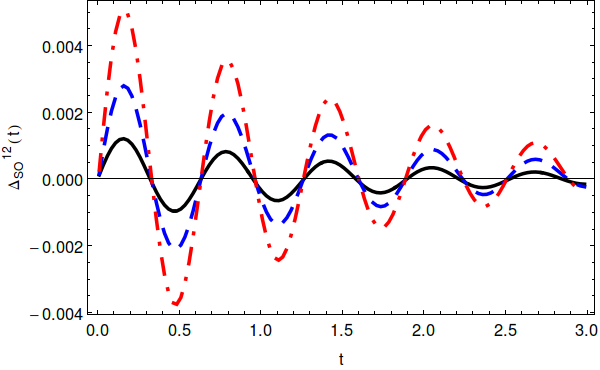}
\caption{(color online). The position cross-correlation as a function of time. The solid balck, dashed blue and dot-dashed red lines correspond to the temperatures $\beta^{-1}=0.01\Omega_{0}$, $\beta^{-1}=\Omega_{0}/2$, and $\beta^{-1}=\Omega_{0}$, respectively. We have fixed the rest of parameters as follows: $\Omega_{0}=10$, $\Gamma_{0}=0.1 \Omega_{0}$, $\Omega_{CS}=\Gamma_{0}/2$, and $m=1$. \label{Fig2}} 
\end{figure}

Although (\ref{MSDC}) manifests that the Chern-Simons effects may be eventually negligible in the classical dynamics, this result may significantly diverse from the quantum scenario. In the zero temperature limit $\beta^{-1}=0$, we consider the asymptotic expression of the psi function for large arguments, i.e. $ \psi(1+z)\approx \log z$ for $1\ll z$. It can be shown that this leads to
\begin{eqnarray}
\left\langle \hat Q^2\right\rangle_{\beta}
&=&\frac{\log\left( \frac{\Gamma_0+\sqrt{\Gamma_0^2-\Omega_0^2}}{\Gamma_0-\sqrt{\Gamma_0^2-\Omega_0^2}}\right)}{2\pi m\sqrt{\Gamma_0^2-\Omega_0^2}} +\frac{1}{2m\Gamma_{0}}\left(\frac{\Omega_{CS}}{\Omega_{0}} \right)^2\nonumber \\
&+&\mathcal{O}\left(\left(\frac{\Omega_{CS}}{\Omega_{0}} \right)^4 \right), \label{MSDQ}
\end{eqnarray}
where again the first term of the right-hand side identifies with the quantum result for the mean-square dispersion found in the damped harmonic oscillator \cite{weiss20121,grabert19841,caldeira19831,caldeira20141}. Unlike the previous classical limit, this result underlines that the Chern-Simons effects may substantially influence the Brownian motion, and could be interpreted as an "hyperfine" structure of the two-dimensional dissipative dynamics.

Finally, we evaluate the position cross-correlation $\Delta^{12}_{SO}(t)$ between the transversal spatial degrees of freedom. We can follow an identical procedure as to compute (\ref{AFSOI}). We find for $0<t$
\begin{eqnarray}
&&\Delta^{12}_{SO}(t) =\text{Im}\left\lbrace  S_{0}(t)+S_{CS}(t)\right\rbrace +\frac{2\Omega_{CS}^2}{m\beta}S_{CS}^{(q)}(t)\nonumber \\
&+&\frac{2\Omega_{CS}^2}{\pi m\beta}\int_{0}^{\infty}d\omega\frac{\sin(\omega t)\left(\left(\omega^2-\Omega_{0}^2\right)^2+\Omega_{CS}^4+12\Gamma_{0}^2 \omega^2\right)}{\omega(\omega^2+\lambda_{+}^2)(\omega^2+\lambda_{-}^2)(\omega^2+\lambda_{+}^{\dagger 2})(\omega^2+\lambda_{-}^{\dagger 2})},\nonumber \\\label{AFSOII}
\end{eqnarray}
which vanishes for zero Chern-Simons action. Figure\ref{Fig2} illustrates the short-time behavior of the position cross-correlation at different temperatures. 

From Eq.(\ref{AFSOII}) we may also evaluate the long-time behavior of $\Delta^{12}_{SO}(t)$. Paying attention to Eqs. (\ref{CSRE}) and (\ref{QSRE}), it is readily to see that in the high temperature limit $\Gamma_{0}\beta \ll 1$ the position cross-correlation (\ref{AFSOII}) will decrease exponentially with a ratio given by $\Gamma_{0}$ at long time, since the quantum corrections decay as $e^{-\beta^{-1} t}$. However, this scenario drastically changes in the low temperature regime where the quantum corrections dominate the long time behavior. In particular, in the zero temperature limit the quantum corrections sum up to an algebraic long-time behavior, i.e
\begin{eqnarray}
\Delta^{12}_{SO}(t) &\sim& \frac{\Omega_{CS}^2}{m\pi(\Omega_{0}^4+\Omega_{CS}^4)}\frac{1}{t}= \frac{1}{m \pi \Omega_{0}^2}\frac{1}{t} \left( \frac{\Omega_{CS}}{\Omega_{0}}\right)^2 \nonumber \\
&+&\mathcal{O}\left( \left( \frac{\Omega_{CS}}{\Omega_{0}}\right)^6\right), \ \ \ \text{for} \ \  t \rightarrow \infty \label{AFSOIITL}.
\end{eqnarray}
To obtain the above result we evaluate the infinite sum involved in the quantum corrections $S_{CS}^{(q)}(t)$  after expanding the rational part of its arguments about the origin $\nu_{n}\rightarrow 0$, and then taking the strict zero-temperature limit. Interestingly, expression (\ref{AFSOIITL}) shows that the transversal spatial components exhibit long-time correlations in the quantum regimen. Let us stress that this feature is a consequence of the quantum Hall response previously discussed in Sec.\ref{MLDS} and has no counterpart in the conventional Brownian motion, rather the latter  generates a similar time algebraic decayment just for the position autocorrelation functions \cite{grabert19841} (i.e. $\Delta^{\alpha\alpha}_{SO}(t)\backsim t^{-2}$ for $\alpha=1,2$). Thus we see from Eqs. (\ref{MSDQ}) and (\ref{AFSOIITL}) that the dissipative dynamics arising from the Chern-Simons action in the Markovian limit constitutes a second-order correction to the damped harmonic oscillator in the quantum regime.

\section{Outlook and concluding remarks}\label{OCR}
Adopting a combined gauge field and open-system theoretical view, we have shown that the non-relativistic Maxwell-Chern-Simons electrodynamics leads to a novel microscopic model for the non-equilibrium dynamics of planar harmonic systems that fulfills the essential ingredients of a (linear) non-anomalous dissipative description: local U(1) gauge invariance and relaxation towards a thermal equilibrium state, as well as it encompasses the conventional Brownian motion as a particular instance (i.e. the deduced microscopic Hamiltonian exactly maps on the independent-oscillator model in the limit of zero Chern-Simons constant). Specifically, we have shown that the symmetry structure provided by the Chern-Simons action has two main effects in the long-wavelength regime: a gap environmental spectrum, and a particle-attached magnetic-like flux which yields an ordinary Hall response of the harmonic oscillators. As a counterpart, these come along with a backreaction on the environment and renormalized potential interaction of the system particles. Importantly, if we disregard the latter effects, the conceived dissipative model provides a legitimate description for the Brownian motion of free particles as well, owing to the fact that the microscopic Hamiltonian turns identically into a minimal-coupling theory with a gauge field bearing the basic dissipative mechanism. It is also worthwhile to remark that, though we have considered a Gaussian particle charge distribution, this condition could be substantially relaxed to contain a broad class of form factors (e.g. we could chose a well-behaved function which falls to zero at sufficiently large distances) without changing the overall properties of the dissipative dynamics, since such choice mainly affects the cutoff factor of the spectral density.

In the Markovian regime, we have found that the explicit influence of the Chern-Simons action at the level of the Langevin equation is twofold: an additional rotational force and magnetic-like flux noise acting upon the harmonic oscillators; such that the system could be regarded as an ensemble of interacting dissipative vortex particles, unlike the conventional Brownian motion. Moreover, for the single harmonic oscillator case we have provided concise expressions for the mean-square dispersion in the high- and zero- temperature limit. Interestingly, from the latter follows that the Chern-Simons action can be though of as a second-order correction to the well-known damped harmonic oscillator model, from an open-system theory perspective.

From an experimental point of view, many results of planar condensed matter systems predicted by the Maxwell-Chern-Simons electrodynamics are still challenging to be experimentally tested. Nonetheless, there exist several examples, for instance in the context of cold Rydberg atoms \cite{caruso20131}, where this theory reproduces the true electromagnetic interaction. In particular, it was probed in the Ref.\cite{marino19931} that an effective Maxwell-Chern-Simons description naturally emerges in the study of systems composed of charged particles constrained to move on an infinite plane and subjected to an ordinary electromagnetic interaction, like in the quantum Hall effect. Concretely, it was shown that the Chern-Simons kinetic term may arise from the underlying topological structure of the (3+1)-dimensional electrodynamics by dimensional reduction \cite{marino19931}, which emphasize the genuine topological nature of the Chern-Simons action \cite{jackiw19902}. Together with the previous discussion, we could draw the conclusion that the proposed description could be tested in near-future two-dimensional experiments addressing the dissipative dynamics induced by electric and magnetic fields.

In the last decade the quantum physics of two-dimensional systems have attracted a renewed interest (e.g. in quantum computation theory), for which the Maxwell-Chern-Simons theory proved to be useful in the description of a great diversity of phenomena connected to condensed matter systems. Similarly, the proposed microscopic description could provide a theoretical testing ground for new ideas in two-dimensional quantum thermodynamics, optics or information theory (e.g. as occurs for the independent-oscillator model). In particular, it is appealing to further investigate which are the nonequilibrium thermodynamic properties of the rising flux-carrying Brownian particles.

\acknowledgments
The author is grateful to L.A. Correa, S. Kohler, I. de Vega and D. Alonso for fruitful discussions. The author warmly thanks J.J. Garc\' ia-Ripoll and M.S. Kim for their support through the development of the present work. The author acknowledges financial support from Physical Science Research Council (project EP/K034480/1) and the Air Force Office of Scientific Research (project FA2386-18-1-4019).

\appendix

\section{Canonical Hamiltonian in the Coulomb gauge}\label{app1}
In this Appendix we briefly illustrate the derivation of the canonical Hamiltonian (\ref{HMCS2}) presented in Sec.\ref{SGID}. First, let us indicate that we enforce $A_{0}$ as a Lagrange multiplier in \ref{CHE} \cite{dunne19991,jackiw19901}, in order to satisfy the Coulomb law constraint (\ref{FCCII}). It is easy to verify that equation (\ref{CII}) is a solution of the latter. Now, we replace this in (\ref{CHE}) once we have rewritten the field canonical variables in terms of their longitudinal and transversal parts. Doing this, we obtain two terms coming from the dot and cross products, respectively,
\begin{equation}
I_{1}=\int d \vect x \ \vect \Pi^{\parallel}\cdot \vect \Pi^{\parallel}, \ \ \ I_{2}=-\kappa \int d \vect x \ \vect \Pi^{\parallel}\times \vect A^{\perp}.
\nonumber
\end{equation}
Owing to the boundary properties of the Coulomb Green's function, gauge field, and charged distribution function (i.e. $\nabla G_{c}(\vect x)\rightarrow 0 $, $\partial_{\alpha}A^{\perp}_{\beta}(\vect x)\rightarrow 0 $, and $\rho(\vect x)\rightarrow 0$ for $|\vect x| \rightarrow \infty$), both terms may be manipulated as follows, 
\begin{widetext}
\begin{eqnarray}
I_{1}&=&\int d \vect y  \int d \vect y' \ \left(\rho(\vect y)-\frac{\kappa}{2} \nabla \times\vect{A}^{\perp}(\vect y) \right)\left(\rho(\vect y')-\frac{\kappa}{2} \nabla \times\vect{A}^{\perp}(\vect y') \right) \int d \vect x \ \partial_{\alpha} G_{c}(\vect x-\vect y)\partial_{\alpha} G_{c}(\vect x-\vect y') \nonumber \\
&=& -\int d \vect x \int d \vect y   \ \left(\rho(\vect y)-\frac{\kappa}{2} \nabla \times\vect{A}^{\perp}(\vect y) \right) G_{c}(\vect x-\vect y)\int d \vect y'\nabla^{2}_{\vect x} G_{c}(\vect x-\vect y')\left(\rho(\vect y')-\frac{\kappa}{2} \nabla \times\vect{A}^{\perp}(\vect y') \right)  \nonumber \\
&=& -\int d \vect x \int d \vect y   \ \left(\rho(\vect y)-\kappa \nabla \times\vect{A}^{\perp}(\vect y) \right) G_{c}(\vect x-\vect y)\rho(\vect x) -\frac{\kappa^{2}}{4}\int d \vect x \int d \vect y   \ \nabla \times\vect{A}^{\perp}(\vect x) G_{c}(\vect x-\vect y) \nabla \times \vect{A}^{\perp}(\vect y) \nonumber \\
&=&-\int d \vect x \int d \vect y   \ \left(\rho(\vect y)-\kappa \nabla \times\vect{A}^{\perp}(\vect y) \right) G_{c}(\vect x-\vect y)\rho(\vect x) +\frac{\kappa^{2}}{4}\int d \vect x \int d \vect y'  \ G(\vect x') \epsilon_{\alpha\beta}\epsilon_{\alpha'\beta'}A^{\perp}_{\beta'}(\vect x)\partial^{\vect x}_{\alpha}\partial^{\vect x}_{\alpha'}A^{\perp}_{\beta}(\vect x+\vect y'), \nonumber
\end{eqnarray}
as well as 
\begin{eqnarray}
I_{2}&=&-\kappa \int d \vect x\int d \vect y' \ G_{c}(\vect y')\epsilon_{\alpha\beta} \vect{A}_{\beta}^{\perp}(\vect x) \partial_{\alpha}^{\vect x}\left(\rho(\vect x+\vect y')-\frac{\kappa}{2} \nabla \times\vect{A}^{\perp}(\vect x+\vect y') \right)   \nonumber \\
&=&\kappa\int d \vect x\int d \vect y \ \nabla\times \vect{A}^{\perp}(\vect x) G(\vect x-\vect y)\rho(\vect y)+\frac{\kappa^{2}}{2}\int d \vect x\int d \vect y' \ G_{c}(\vect y') \epsilon_{\alpha\beta}\epsilon_{\alpha'\beta'}A^{\perp}_{\beta'}(\vect x)\partial^{\vect x}_{\alpha}\partial^{\vect x}_{\alpha'}A^{\perp}_{\beta}(\vect x+\vect y'). \nonumber 
\end{eqnarray}
Now we may go further by gathering together the above results,
\begin{eqnarray}
I_{1}+I_{2}&=&-\int d \vect x \int d \vect y   \ \left(\rho(\vect y)-2\kappa \nabla \times\vect{A}^{\perp}(\vect y) \right) G_{c}(\vect x-\vect y)\rho(\vect x) \nonumber \\ 
&+&\frac{3\kappa^{2}}{4}\int d \vect x \int d \vect y   \  G_{c}(\vect x-\vect y) (\delta_{\alpha \alpha'}\delta_{\beta \beta'}-\delta_{\alpha \beta'}\delta_{ \alpha'\beta})A^{\perp}_{\beta}(\vect x)\partial^{\vect y}_{\alpha'}\partial^{\vect y}_{\alpha}A^{\perp}_{\beta'}(\vect y)  \nonumber \\
&=&-\int d \vect x \int d \vect y   \ \left(\rho(\vect y)-2\kappa \nabla \times\vect{A}^{\perp}(\vect y) \right) G_{c}(\vect x-\vect y)\rho(\vect x) +\frac{3\kappa^{2}}{4}\int d \vect x \int d \vect y   \  \nabla^{2}_{\vect y}G_{c}(\vect x-\vect y) A^{\perp}_{\alpha}(\vect x)A^{\perp}_{\alpha}(\vect y)  \nonumber \\
&=&-\int d \vect x \int d \vect y   \ \left(\rho(\vect y)-2\kappa \nabla \times\vect{A}^{\perp}(\vect y) \right) G_{c}(\vect x-\vect y)\rho(\vect x) +\frac{3\kappa^{2}}{4}\int d \vect x \ \vect{A}^{\perp}\cdot\vect{A}^{\perp}, \nonumber 
\end{eqnarray}
\end{widetext}
where we have made use of the fact $\partial^{\vect y}_{\alpha'}\vect{A}^{\perp}_{\alpha'}(\vect y)=0 $ according to the Coulomb gauge fixing. By substituting this result in the Hamiltonian expressed in terms of the longitudinal and transversal parts, we arrive at 
\begin{eqnarray}
\hat \H&=&\sum_{i=1}^{n}\bigg(\frac{1}{2m}\left(\hat{\vect p}_{i}-e \int d^2\vect x \varphi(\vect x- \hat{\vect q_{i}})\hat{\vect A}^{\perp}(\vect x)\right)^{2}+ V(\hat{\vect q}_{i})\bigg) \nonumber  \\ 
&-&\frac{1}{2}\int d^2\vect xd^2\vect y \ \rho(\vect x)G_{c}(\vect x-\vect y)\left(\rho(\vect y) -2\kappa \nabla \times \hat{\vect{A}}^{\perp}(\vect y)\right) \nonumber \\
&+& \hat \H_{MCS}  \label{HMCS1},
\end{eqnarray}
where $\hat \H_{MCS} $ is defined in Sec.\ref{SGID}. It is immediate to see that the canonical Hamiltonian (\ref{HMCS2}) is obtained after replacing the Coulomb Green's function and the charge density in (\ref{HMCS1}).

\section{Derivation of the positive condition}\label{app3}
Here we compute the positive constraint (\ref{DHPCE}) presented in Sec.\ref{SQD}, as well as we provide explicit expressions for the potentials (\ref{QPE}) and (\ref{VSPE}). First, we illustrate the basic procedure by starting to compute the quadratic potential $V_{ij}^{\alpha\beta}$. To carry out the discrete sum in $\vect k$, it is convenient to take the dense spectrum limit after switching the integral to polar-coordinate variables,
\begin{eqnarray}
V_{ij}^{\alpha\beta}&=&\delta_{ij}\delta_{\alpha\beta}m\omega_{i}^2+  \frac{e^2}{8\pi^2}\int_{0}^{\infty} dk \ k \ e^{-2\sigma k^2}  \nonumber \\
&\times & \bigg(\int_{-\pi}^{\pi} d\theta e^{i k|\bar{\vect q}_{i}-\bar{\vect q}_{j}|\cos\theta} \Big(\delta_{1\alpha }\delta_{1\beta } \sin^2\theta+\delta_{2\alpha }\delta_{2\beta }\cos^2\theta \nonumber \\
&-& (1-\delta_{\alpha\beta})\cos\theta\sin\theta  \Big)+c.c.\bigg),  \nonumber 
\end{eqnarray}
where we have rewritten $|\vect k|=k$, and $\theta$ is chosen to be the azimuthal angle of the vector $\vect k$ and the axis defined by $\Delta\bar{\vect q}_{ij}$. The above equation is further simplified by introducing the definition of the Bessel function of first kind, i.e.  
\begin{eqnarray}
V_{ij}^{\alpha\beta}&=&\delta_{ij}\delta_{\alpha\beta}m\omega_{i}^2+\frac{e^2}{2\pi}\int_{0}^{\infty} dk \ k \ e^{-2\sigma k^2} \label{AQPE}\\
&\times &\bigg( \delta_{\alpha\beta}\frac{J_{1}(k|\bar{\vect q}_{i}-\bar{\vect q}_{j}|)}{k|\bar{\vect q}_{i}-\bar{\vect q}_{j}|} - \delta_{2\alpha }\delta_{2\beta}J_{2}(k|\bar{\vect q}_{i}-\bar{\vect q}_{j}|)\bigg). \nonumber 
\end{eqnarray}
Fortunately, the integral involving the Bessel functions can be exactly obtained by making use of the integration tables (see \S 6.618 and \S 6.631 in Ref\cite{gradshteyn20141}), 
\begin{eqnarray}
V_{ij}^{\alpha\beta}&=&\delta_{ij}\delta_{\alpha\beta}m\omega_{i}^2\nonumber \\
&+&\frac{e^2e^{-\frac{|\bar{\vect q}_{i}-\bar{\vect q}_{j}|^2}{16\sigma}}}{4\pi\sqrt{2\sigma}|\bar{\vect q}_{i}-\bar{\vect q}_{j}|}\bigg(\delta_{\alpha\beta}\sqrt{\pi}I_{\frac{1}{2}}\left(\frac{|\bar{\vect q}_{i}-\bar{\vect q}_{j}|^2}{16\sigma} \right) \nonumber \\
&-&\delta_{2\sigma}\delta_{2\beta}M_{\frac{1}{2},1}\left(\frac{|\bar{\vect q}_{i}-\bar{\vect q}_{j}|^2}{8\sigma} \right)\bigg),
\label{QPV}
\end{eqnarray}
where $I_{n}$ denotes the $n$-order modified Bessel function of the first kind and $M_{\frac{1}{2},1}$ is the Whitakker function \cite{gradshteyn20141}. Note that the above expression is valid for $0< |\bar{\vect q}_{i}-\bar{\vect q}_{j}|$ and $0 <\sigma$. 

Now, paying attention to the Hamiltonian (\ref{HMCSF2}), it is clear that the latter will be a positive-definite operator as long as the following quadratic expression is satisfied,
\begin{eqnarray}
\frac{1}{2}\sum_{j,i=1}^{n}\Big(V_{ij}^{\alpha\beta}-\sum_{\vect k}\omega(\vect k)(h_{\alpha}(\vect k,\bar{\vect q}_{i})h_{\beta}^{\dagger}(\vect k,\bar{\vect q}_{j})+c.c.)\Big)\hat{q}_{i}^{\alpha}\hat{q}_{j}^{\beta} \nonumber \\
+\sum_{i=1}^{n}V_{i}^{\alpha}\hat{q}_{i}^{\alpha}-\sum_{i,j=1}^{n}\sum_{\vect k}\frac{\mathcal{P}_{\alpha\beta}(\vect k)}{\omega(\vect k)}g_{\alpha}^{\dagger}(\vect k,\bar{\vect q}_{i})g_{\beta}(\vect k,\bar{\vect q}_{j})\geq 0 \nonumber . \\
\label{APDCI}
\end{eqnarray}
Let us address the second term in the first line of equation (\ref{APDCI}). Firstly, by replacing the expression of the system-environment coupling coefficient (\ref{EIT}), we obtain after some algebra 
\begin{eqnarray}
&&h_{\alpha}(\vect k,\bar{\vect q}_{i})h_{\beta}^{\dagger}(\vect k,\bar{\vect q}_{j})=\frac{e^2 \ e^{i\vect k\cdot(\bar{\vect q}_{i}-\bar{\vect q}_{j})}e^{-2\sigma|\vect k|^2}}{8\pi^2 L^2\omega(\vect k)}\bigg( \mathcal{P}_{\alpha\beta}(\vect k) \nonumber \\
&+&\frac{\kappa^{2}k_{\alpha}k_{\beta}}{|\vect k|^4\omega^2(\vect k)}\left(|\vect k|^{2}-k_{\gamma}k_{\delta}\mathcal{P}_{\gamma\delta}(\vect k)  \right)   \nonumber \\
&+& i\frac{\kappa\epsilon_{\gamma\delta}k_{\gamma}}{|\vect k|^2\omega(\vect k)}\left(k_{\beta}\mathcal{P}_{\alpha\delta}(\vect k) -k_{\alpha}\mathcal{P}_{\beta\delta}(\vect k)\right) \bigg) . \nonumber  
\end{eqnarray}
Inserting this result and the dispersion relation (\ref{DRB}) yields
\begin{eqnarray}
&&\sum_{\vect k}\omega(\vect k)h_{\alpha}(\vect k,\bar{\vect q}_{i})h_{\beta}^{\dagger}(\vect k,\bar{\vect q}_{j})= \frac{e^2 }{4\pi}\int_{0}^{\infty} dk \ k e^{-2\sigma k^2}\nonumber \\
&\times & \bigg[ \bigg(\bigg(\delta_{\alpha\beta}- \delta_{1\alpha}\delta_{1\beta}\bigg(1-\frac{ \kappa^{2}}{\left( k^{2}+\kappa^2\right) }\bigg)\bigg)J_{0}(k|\bar{\vect q}_{i}-\bar{\vect q}_{j}|)     \nonumber \\
&+&(\delta_{\alpha\beta}-2\delta_{2\alpha}\delta_{2\beta})\frac{k}{\left( k^{2}+\kappa^2\right) }\frac{J_{1}(k|\bar{\vect q}_{i}-\bar{\vect q}_{j}|)}{|\bar{\vect q}_{i}-\bar{\vect q}_{j}|} \bigg) \nonumber  \\
&-&i\epsilon_{\alpha\beta}\frac{\kappa J_{0}(k|\bar{\vect q}_{i}-\bar{\vect q}_{j}|)}{\sqrt{k^{2}+\kappa^2} }\bigg].
\nonumber
\end{eqnarray}
We can analytically computed the above integral via standard contour integration techniques by writing the Bessel function as a combination of the Hankel functions and noting that the functions within the integral has simple poles $\pm i \kappa$ \cite{schwartz19821}, i.e.
\begin{eqnarray}
&&\sum_{\vect k}\omega(\vect k)(h_{\alpha}(\vect k,\bar{\vect q}_{i})h_{\beta}^{\dagger}(\vect k,\bar{\vect q}_{j})+c.c.)=    \label{AINT1}\\
&=&\delta_{2\alpha}\delta_{2\beta}\frac{e^2}{8\pi\sigma}e^{-\frac{|\bar{\vect q}_{i}-\bar{\vect q}_{j}|^2}{8\sigma}}
\nonumber \\
& +&\frac{e^2\ e^{2\sigma \kappa^2} }{8}  \bigg[\frac{\kappa(\delta_{\alpha\beta}-2\delta_{1\alpha}\delta_{1\beta})}{|\bar{\vect q}_{i}-\bar{\vect q}_{j}|} \bigg(H_{1}^{(2)}\left(-|\bar{\vect q}_{i}-\bar{\vect q}_{j}|i\kappa\right)\nonumber \\
&+&H_{1}^{(1)}\left(|\bar{\vect q}_{i}-\bar{\vect q}_{j}|i\kappa\right)\bigg) +\delta_{1\alpha}\delta_{1\beta}i\kappa^2 \bigg(H_{0}^{(1)}\left(|\bar{\vect q}_{i}-\bar{\vect q}_{j}|i\kappa\right)\nonumber \\
&-&H_{0}^{(2)}\left(-|\bar{\vect q}_{i}-\bar{\vect q}_{j}|i\kappa\right)\bigg)\bigg]. \nonumber 
\end{eqnarray}

Similarly, the previous procedure can be replicated to obtain the potential (\ref{SPE}). That is,
\begin{eqnarray}
&&\sum_{\vect k}h_{\alpha}(\vect k,\bar{\vect q}_{i})g_{\beta}^{\dagger}(\vect k,\bar{\vect q}_{j}) \varepsilon_{\beta}^{\dagger}(\vect k)=\frac{e^2\kappa}{8\pi^2}\int_{0}^{\infty} dk \ k e^{-2\sigma k^2} \nonumber \\
&\times &\int_{-\pi}^{\pi} d \theta \frac{e^{i k|\bar{\vect q}_{i}-\bar{\vect q}_{j}|\cos\theta }}{k^2\sqrt{ k^{2}+\kappa^2}}\bigg(\epsilon_{\gamma\alpha}k_{\gamma}-i\frac{\kappa k_{\alpha}}{\sqrt{ k^{2}+\kappa^2}}    \bigg) \nonumber \\
&=&\frac{e^2\kappa}{4\pi}\int_{0}^{\infty} dk \frac{ e^{-2\sigma k^2}J_{1}(k|\bar{\vect q}_{i}-\bar{\vect q}_{j}|)}{\sqrt{ k^{2}+\kappa^2}}\Bigg(-\frac{\kappa \delta_{1\alpha}}{\sqrt{k^{2}+\kappa^{2}}}+i\delta_{2\alpha}\Bigg). \nonumber
\end{eqnarray}
and thus,
\begin{eqnarray}
V_{i}^{\alpha}&=&\sum_{j=1}^{n}\Bigg( 2\left( V_{c}'(\bar{\vect q}_{i}-\bar{\vect q}_{j})\right)_{\alpha}  \nonumber \\
&-&\delta_{1\alpha}\frac{e^2\kappa^{2}}{2\pi}\int_{0}^{\infty} dk \ \frac{e^{-2\sigma k^2}}{k^{2}+\kappa^{2}} J_{1}(k|\bar{\vect q}_{i}-\bar{\vect q}_{j}|)\Bigg) 
\label{ASPE}
\end{eqnarray}
On the other hand, we find for the independent term of the quadratic form (\ref{APDCI}),
\begin{eqnarray}
&&\sum_{\vect k}\frac{\mathcal{P}_{\alpha\beta}(\vect k)}{\omega(\vect k)}g_{\alpha}(\vect k,\bar{\vect q}_{i})g_{\beta}(\vect k,\bar{\vect q}_{j})=   \nonumber \\
&=&\frac{e^{2}\kappa^{2}}{4\pi} \int_{0}^{\infty} dk \ \frac{e^{-2\sigma k^2} }{k\left(k^{2}+\kappa^2 \right)}J_{0}(k|\bar{\vect q}_{i}-\bar{\vect q}_{j}|) 
\label{AINT2}
\end{eqnarray}

Finally, the positive constraint (\ref{DHPCE}) is directly obtained by replacing (\ref{AINT1}), (\ref{ASPE}), and (\ref{AINT2}) in (\ref{APDCI}).


\section{Retarded self-energy and spectral density}\label{app4}
In this appendix we illustrate the derivation of the retarded self-energy (\ref{MKEII}) and (field) spectral density (\ref{SPE}), and the Kramers-Kronig relation (\ref{KKR}) and (\ref{IFSE}) presented in Sec.\ref{SGLE}. 

We start from computing the anti-commutator appearing in the definition of the retarded self-energy (\ref{MKE}) by using the following expression of the pseudo-electric field which follows from (\ref{SFFE}),
\begin{equation}
\hat{\mathcal{E}}_{i}^{\alpha} (t)=\sum_{\vect k}\omega(\vect k)\Big(h_{\alpha}(\vect k,\bar{\vect q}_{i})e^{-i\omega(\vect k)(t-t_{0})}\hat a_{d} (\vect k, t_{0})+c.c.\Big),
\label{AEPEF}
\end{equation}
where $\hat a_{d} (\vect k, t_{0})$ are the environmental quasiparticle operators at the initial time after introducing the displacement produced by the backreaction effects discussed in Sec.\ref{SQD}. By considering an equilibrium canonical initial state $\hat \rho_{\beta}$, we obtain the following statistics for the displaced quasiparticle operators
\begin{eqnarray}
&&\left\langle \left\lbrace \hat a_{d}(\vect k,t_{0} ), \hat a^{\dagger}_{d}(\vect k', t_{0}) \right\rbrace  \right\rangle_{\hat \rho_{\beta}}= \nonumber \\
&=&\delta(\vect k-\vect k')(1+2n(\omega(\vect k),\beta^{-1})) \nonumber\\  
&+&2 \sum_{i,j=1}^{n}\frac{\varepsilon_{\alpha}(\vect k)\varepsilon_{\beta}^{\dagger} (\vect k')}{\omega(\vect k)\omega(\vect k')}g_{\alpha}(\vect k,\bar{\vect q}_{i})g_{\beta}^{\dagger}(\vect k',\bar{\vect q}_{j}), \nonumber \\
&&\left\langle \left\lbrace \hat a_{d}(\vect k,t_{0} ), \hat a_{d}(\vect k', t_{0})  \right\rbrace\right\rangle_{\hat \rho_{\beta}} = \nonumber \\
&=&2\sum_{i,j=1}^{n}\frac{\varepsilon_{\alpha}(\vect k)\varepsilon_{\beta} (\vect k')}{\omega(\vect k)\omega(\vect k')}g_{\alpha}(\vect k,\bar{\vect q}_{i})g_{\beta}(\vect k',\bar{\vect q}_{j}), \label{ACOV}
\end{eqnarray}
with $n(\omega(\vect k),\beta^{-1})$ denoting the average quasiparticle number of the free MCS field at temperature $\beta^{-1}$,
\begin{equation}
n(\omega(\vect k), \beta^{-1})=\frac{1}{e^{\beta\omega(\vect k)}-1}.
\label{AQNH}
\end{equation}

Now from (\ref{AEPEF}) we obtain
\begin{widetext}
\begin{eqnarray}
\left\langle \left[  \hat{\mathcal{E}}_{i}^{\alpha} (t),\hat{\mathcal{E}}_{j}^{\beta\dagger} (t')\right]_{\pm}   \right\rangle_{\hat \rho_{\beta}} &=&\sum_{\vect k,\vect k'}\omega(\vect k)\omega(\vect k')\Bigg(h_{\alpha}(\vect k,\bar{\vect q}_{i})h_{\beta}^{\dagger}(\vect k',\bar{\vect q}_{j})e^{-i\left(\omega(\vect k)t-\omega(\vect k')t'-\left(\omega(\vect k)-\omega(\vect k') \right)t_{0}  \right)} \left\langle \left[   a_{d} (\vect k,t_{0}), a_{d}^{\dagger} (\vect k', t_{0})\right]_{\pm}  \right\rangle_{\hat \rho_{0}} \nonumber \\
&+&h_{\alpha}(\vect k,\bar{\vect q}_{i})h_{\beta}(\vect k',\bar{\vect q}_{j})e^{-i\left(\omega(\vect k)t+\omega(\vect k')t'-\left(\omega(\vect k)+\omega(\vect k') \right)t_{0}  \right)} \left\langle \left[  a_{d} (\vect k,t_{0}), a_{d} (\vect k', t_{0})\right]_{\pm}  \right\rangle_{\hat \rho_{0}} \nonumber \\
&+& h_{\alpha}^{\dagger}(\vect k,\bar{\vect q}_{i})h_{\beta}^{\dagger}(\vect k',\bar{\vect q}_{j})e^{i\left(\omega(\vect k)t+\omega(\vect k')t'-\left(\omega(\vect k)+\omega(\vect k') \right)t_{0}  \right)} \left\langle \left[    a_{d}^{\dagger} (\vect k,t_{0}), a_{d}^{\dagger} (\vect k', t_{0})\right]_{\pm}  \right\rangle_{\hat \rho_{0}} \nonumber \\
&+&h_{\alpha}^{\dagger}(\vect k,\bar{\vect q}_{i})h_{\beta}(\vect k',\bar{\vect q}_{j})e^{i\left(\omega(\vect k)t-\omega(\vect k')t'-\left(\omega(\vect k)-\omega(\vect k') \right)t_{0}  \right)} \left\langle \left[   a_{d}^{\dagger} (\vect k,t_{0}), a_{d} (\vect k', t_{0})\right]_{\pm}  \right\rangle_{\hat \rho_{0}}\Bigg). 
\label{ACESH}
\end{eqnarray}
where $\left[\bullet, \bullet \right]_{\pm} $ represents the anti-commutator and commutator in a short-hand notation, respectively. By taking the dense spectrum limit after some straightforward manipulation in the anti-commutator expression (\ref{ACESH}), we arrive at 
\begin{eqnarray}
&& \left\langle\left[  \hat{\mathcal{E}}_{i}^{\alpha} (t), \hat{\mathcal{E}}_{j}^{\beta\dagger} (t')\right] \right\rangle_{\hat \rho_{\beta}}   =\sum_{\vect k, \vect k'}\omega(\vect k)\omega(\vect k')\Big(e^{-i(\omega(\vect k)t-\omega(\vect k')t'-t_{0}(\omega(\vect k)-\omega(\vect k')))} h_{\alpha}(\vect k,\bar{\vect q}_{i})h_{\beta}^{\dagger}(\vect k',\bar{\vect q}_{j})\left[a_{d}(\vect k,t_{0}),a^{\dagger}_{d}(\vect k',t_{0}) \right]_{-}  \nonumber \\
&+&e^{i(\omega(\vect k)t-\omega(\vect k')t'-t_{0}(\omega(\vect k)-\omega(\vect k')))}h_{\alpha}^{\dagger}(\vect k,\bar{\vect q}_{i})h_{\beta}(\vect k',\bar{\vect q}_{j}) \left[a_{d}^{\dagger}(\vect k,t_{0}),a_{d}(\vect k',t_{0}) \right]_{-} \Big) \nonumber \\
&=& - 2i\sum_{\vect k}\omega^{2}(\vect k)\bigg(\text{Re}\Big\{h_{\alpha}(\vect k,\bar{\vect q}_{i})h_{\beta}^{\dagger}(\vect k,\bar{\vect q}_{j})\Big\} \sin(\omega(\vect k)(t-t'))  -\text{Im}\Big\{h_{\alpha}(\vect k,\bar{\vect q}_{i})h_{\beta}^{\dagger}(\vect k,\bar{\vect q}_{j})\Big\} \cos(\omega(\vect k)(t-t'))\bigg)                                             \nonumber \\
&=&- \frac{2i}{\pi}\int_{0}^{\infty}d\omega\bigg(\text{Re}\big\{\mathcal{J}_{\alpha\beta}(\Delta\bar{q}_{ij},\omega)\big\} \sin(\omega(t-t')) +\text{Im}\big\{\mathcal{J}_{\alpha\beta}(\Delta\bar{q}_{ij},\omega)\big\} \cos(\omega(t-t'))\bigg)  ,                                          
\label{MKEA}
\end{eqnarray}
where $\mathcal{J}_{\alpha\beta}$ represents the spectral density. It is immediate to see that we get expression (\ref{MKEII}) after substituting (\ref{MKEA}) in the definition (\ref{MKE}). In the above equation, we introduced the following expression for the spectral density, 
\begin{eqnarray}
\mathcal{J}_{\alpha\beta}(\Delta\bar{q}_{ij},\omega)&=&\pi\sum_{\vect k}\omega^{2}(\vect k)\Big(\text{Re}\Big\{h_{\alpha}(\vect k,\bar{\vect q}_{i})h_{\beta}^{\dagger}(\vect k,\bar{\vect q}_{j})\Big\}-i\text{Im}\Big\{h_{\alpha}(\vect k,\bar{\vect q}_{i})h_{\beta}^{\dagger}(\vect k,\bar{\vect q}_{j})\Big\}\Big)\delta(\omega(\vect k)-\omega) \nonumber\\
&=&\frac{e^2\omega}{8\pi L^2}\sum_{\vect k}e^{i\vect k\cdot\Delta\bar{q}_{ij}}e^{-2\sigma|\vect k|^2}\bigg(  \delta_{\alpha\beta}-\frac{k_{\alpha}k_{\beta}}{|\vect k|^2} +\frac{\kappa^{2}k_{\alpha}k_{\beta}}{|\vect k|^2\omega^2}-i \frac{\kappa}{|\vect k|^2\omega}(\epsilon_{\gamma\alpha}k_{\gamma}k_{\beta} -\epsilon_{\gamma\beta}k_{\gamma}k_{\alpha})  \bigg) \delta(\omega(\vect k)-\omega)     \nonumber \\
&=& \frac{e^2\omega^2}{8\pi} \int_{0}^{\infty} dk  \  k e^{-2\sigma k^2} \frac{\delta\left( k-\sqrt{\omega^{2}-\kappa^{2}}\right) }{\sqrt{\omega^{2}-\kappa^{2}}}\int_{-\pi}^{\pi}d \theta \  e^{ik|\Delta\bar{q}_{ij}|\cos\theta}\mathcal{W}_{\alpha\beta}(k,\theta),
\label{SPDA1}
\end{eqnarray}
where we have replaced the expression for the system-environment coupling coefficients (\ref{EIT}). In the last line of (\ref{SPDA1}), the integral is expressed in terms of the polar-coordinate variables $(k=|\vect k|,\theta)$ (where $\theta$ is the azimuthal angle as before), and the following matrix,
\begin{eqnarray}
\mathcal{W}_{\alpha\beta}(k,\theta) =\left( \begin{array}{cc}
1+\left(-1+\frac{\kappa^{2}}{\omega^{2}} \right)\cos^{2}\theta   & \left(-1+\frac{\kappa^{2}}{\omega^{2}} \right)\cos\theta  \sin\theta+i\frac{\kappa }{\omega}\\
\left(-1+\frac{\kappa^{2}}{\omega^{2}} \right)\cos\theta  \sin\theta-i\frac{\kappa }{\omega}  &   1+\left(-1+\frac{\kappa^{2}}{\omega^{2}} \right)\sin^{2}\theta
\end{array}\right). \nonumber
\end{eqnarray}
To get (\ref{SPDA1}) (after taking the dense spectrum limit) we employed the dispersion relation (\ref{DRB}) of the Maxwell-Chern-Simons gauge field combined with the properties from the Dirac delta function\cite{landau19711}, i.e. $\delta(\omega(\vect k)-\omega')=\omega'(\omega'^{2}-\kappa^{2})^{-\frac{1}{2}}\delta\left( k-\sqrt{\omega'^{2}-\kappa^{2}}\right) $. Now, by paying attention to the trigonometric integral in (\ref{SPDA1}), we may identify the definition of the zero- and first- order Bessel functions of first kind \cite{abramowitz19641}. Once we have rewritten the integral in terms of the latter, one may readily verify that the integral in the variable $k$ directly leads to the desired expression for the spectral density (\ref{SPD}).

On the other hand, from expression (\ref{MKEA}) we may compute the commutator of the Maxwell-Chern-Simons electric field defined by Eq.(\ref{MCSEF}) in Sec.\ref{SQD}. Since the backreaction displacement on the environmental quasiparticle operators must leave invariant the commutator of the pseudo-electric field force, from (\ref{MKEA}) follows that 
\begin{eqnarray}
\left[  \hat{E}_{MCS}^{\alpha}(\bar{\vect q}_{i}), \hat{E}_{MCS}^{\beta}(\bar{\vect q}_{j})\right]&=&\frac{1}{e^2}\left[  \hat{\mathcal{E}}_{i}^{\alpha} (t), \hat{\mathcal{E}}_{j}^{\beta\dagger} (t)\right]=-\frac{i\kappa \epsilon_{\alpha\beta}}{2\pi}\int_{0}^{\infty} dk \ k \ e^{-2\sigma k^2} J_{0}(k|\Delta\bar{q}_{ij}|),
\label{MCSEFCCR}
\end{eqnarray}
which clearly manifests that the Maxwell-Chern-Simons electric field has non-commutative components.

Let us turn the attention to the Kramers-Kronig relation (\ref{KKR}) and expression (\ref{IFSE}). This may be obtained directly by carrying out the inverse Fourier transform on the previously deduced Eq.(\ref{MKEII}), i.e.
\begin{eqnarray}
&&\tilde{\Sigma}_{ij}^{\alpha\beta}(\omega)=\frac{1}{2\pi}\int_{-\infty}^{\infty}dt  \ e^{i\omega t}\Sigma_{ij}^{\alpha\beta}(t)= \nonumber \\                          \nonumber \\
&=&\frac{1}{\pi} \sum_{\vect k}\omega^{2}(\vect k)\bigg(\text{Re}\Big\{h_{\alpha}(\vect k,\bar{\vect q}_{i})h_{\beta}^{\dagger}(\vect k,\bar{\vect q}_{j})\Big\}e^{i\omega|\Delta \bar{\vect q}_{ij}|}\int_{-\infty}^{\infty}d\tau  \ e^{i\omega \tau}     \Theta\left(\tau\right)\sin(\omega(\vect k)(\tau+|\Delta \bar{\vect q}_{ij}|))    \nonumber \\
&-&\text{Im}\Big\{h_{\alpha}(\vect k,\bar{\vect q}_{i})h_{\beta}^{\dagger}(\vect k,\bar{\vect q}_{j})\Big\}e^{i\omega|\Delta \bar{\vect q}_{ij}|}\int_{-\infty}^{\infty}d\tau  \ e^{i\omega \tau}   \Theta\left(\tau\right)  \cos(\omega(\vect k)(\tau+|\Delta \bar{\vect q}_{ij}|))\bigg)                            \nonumber \\                              \nonumber \\
&=&  \frac{i}{2\pi}\sum_{\vect k}\omega^{2}(\vect k)\bigg(\text{Re}\Big\{h_{\alpha}(\vect k,\bar{\vect q}_{i})h_{\beta}^{\dagger}(\vect k,\bar{\vect q}_{j})\Big\}\lim_{\epsilon\rightarrow 0^{+}}\int_{0}^{\infty}d\tau \ \Big( e^{-i(\omega(\vect k)-\omega)(\tau-\epsilon\tau+|\Delta \bar{\vect q}_{ij}|)}-e^{i(\omega(\vect k)+\omega)(\tau-\epsilon\tau+|\Delta \bar{\vect q}_{ij}|)}\Big)  \nonumber \\
&+&i\text{Im}\Big\{h_{\alpha}(\vect k,\bar{\vect q}_{i})h_{\beta}^{\dagger}(\vect k,\bar{\vect q}_{j})\Big\}\lim_{\epsilon\rightarrow 0^{+}}\int_{0}^{\infty}d\tau \ \Big(e^{-i(\omega(\vect k)-\omega)(\tau-\epsilon\tau+|\Delta \bar{\vect q}_{ij}|)}+e^{i(\omega(\vect k)+\omega)(\tau-\epsilon\tau+|\Delta \bar{\vect q}_{ij}|)}\Big) \bigg) \nonumber \\
&=&\frac{i}{2}\sum_{\vect k}\omega^{2}(\vect k)\bigg(\Big(-\text{Re}\Big\{h_{\alpha}(\vect k,\bar{\vect q}_{i})h_{\beta}^{\dagger}(\vect k,\bar{\vect q}_{j})\Big\}+i\text{Im}\Big\{h_{\alpha}(\vect k,\bar{\vect q}_{i})h_{\beta}^{\dagger}(\vect k,\bar{\vect q}_{j})\Big\}\Big)e^{-i(\omega(\vect k)+\omega)|\Delta \bar{\vect q}_{ij}|}\delta(\omega(\vect k)+\omega)  \nonumber \\
&+&\Big(\text{Re}\Big\{h_{\alpha}(\vect k,\bar{\vect q}_{i})h_{\beta}^{\dagger}(\vect k,\bar{\vect q}_{j})\Big\}+i\text{Im}\Big\{h_{\alpha}(\vect k,\bar{\vect q}_{i})h_{\beta}^{\dagger}(\vect k,\bar{\vect q}_{j})\Big\}\Big)e^{-i(\omega(\vect k)-\omega)|\Delta \bar{\vect q}_{ij}|}\delta(\omega(\vect k)-\omega)\bigg) \nonumber \\
&+&\frac{1}{2\pi}\sum_{\vect k}\omega^{2}(\vect k)\bigg(P\int_{-\infty}^{\infty}\frac{d\omega'}{\omega'-\omega}\Big(-\text{Re}\Big\{h_{\alpha}(\vect k,\bar{\vect q}_{i})h_{\beta}^{\dagger}(\vect k,\bar{\vect q}_{j})\Big\}+i\text{Im}\Big\{h_{\alpha}(\vect k,\bar{\vect q}_{i})h_{\beta}^{\dagger}(\vect k,\bar{\vect q}_{j})\Big\}\Big)e^{-i(\omega(\vect k)+\omega)|\Delta \bar{\vect q}_{ij}|}\delta(\omega(\vect k)+\omega') \nonumber \\
&+&P\int_{-\infty}^{\infty}\frac{d\omega'}{\omega'-\omega}\Big(\text{Re}\Big\{h_{\alpha}(\vect k,\bar{\vect q}_{i})h_{\beta}^{\dagger}(\vect k,\bar{\vect q}_{j})\Big\}+i\text{Im}\Big\{h_{\alpha}(\vect k,\bar{\vect q}_{i})h_{\beta}^{\dagger}(\vect k,\bar{\vect q}_{j})\Big\}\Big)e^{-i(\omega(\vect k)-\omega)|\Delta \bar{\vect q}_{ij}|}\delta(\omega(\vect k)-\omega')\bigg).
\label{SEFT}
\end{eqnarray}
Recalling that the self-energy must be analytic in the upper-half complex $\omega$ plane (which is feature by the Heaviside step function), in the above Fourier transform we have introduced a positive infinitesimal quantity $0^{+}$ that provides the correct pole prescriptions in the frequency domain. The latter gives rise to the principal-value terms. Paying attention to the real and imaginary parts in the last few lines of the right-hand side of Eq.(\ref{SEFT}), we may easily identify the spectral density definition (\ref{SPDA1}). While the imaginary terms directly lead to (\ref{IFSE}), the principal-value terms only contribute to the real part of the self-energy Fourier transform, and thus, they retrieve the extended Kramers-Kronig relation (\ref{KKR}).

\section{Derivation of the fluctuation-dissipation relation}\label{app5}
This appendix is devoted to the derivation of the fluctuation-dissipation relation (\ref{CFFE}) of Sec.\ref{SGLE}, and further, the non-equilibrium spectral functions (\ref{GGE}) and (\ref{FFE}). We start from the commutator expression computed from Eq.(\ref{ACESH}) in the Appendix\ref{app4}. By inserting the average values for the environmental displaced quasiparticle operators (\ref{ACOV}), we find
\begin{eqnarray}
\frac{1}{2}\left\langle \left\lbrace \hat{\mathcal{E}}_{i}^{\alpha} (t), \hat{\mathcal{E}}_{j}^{\beta\dagger} (t')\right\rbrace  \right\rangle_{\hat \rho_{0}}  &=&\sum_{\vect k,\vect k'}\Big(\delta(\vect k-\vect k')\omega^{2}(\vect k)(1+2n(\omega(\vect k),\beta^{-1}))\text{Re}\Big\{h_{\alpha}(\vect k,\bar{\vect q}_{i})h_{\beta}^{\dagger}(\vect k,\bar{\vect q}_{j})e^{-i\omega(\vect k)(t-t')}\Big\} \nonumber \\
&+&2\sum_{l,m=1}^{n}\text{Re}\Big\{e^{-i\left(\omega(\vect k)t-\omega(\vect k')t'-\left(\omega(\vect k)-\omega(\vect k') \right)t_{0}  \right)}\varepsilon_{\gamma}(\vect k)\varepsilon_{\delta}^{\dagger} (\vect k')g_{\gamma}(\vect k,\bar{\vect q}_{l})g_{\delta}^{\dagger}(\vect k',\bar{\vect q}_{m}) h_{\alpha}(\vect k,\bar{\vect q}_{i})h_{\beta}^{\dagger}(\vect k',\bar{\vect q}_{j})\Big\} \nonumber \\
&+&2\sum_{l,m=1}^{n}\text{Re}\Big\{e^{-i\left(\omega(\vect k)t+\omega(\vect k')t'-\left(\omega(\vect k)+\omega(\vect k') \right)t_{0}  \right)}\varepsilon_{\gamma}(\vect k)\varepsilon_{\delta}(\vect k')g_{\gamma}(\vect k,\bar{\vect q}_{l})g_{\delta}(\vect k',\bar{\vect q}_{m})h_{\alpha}(\vect k,\bar{\vect q}_{i})h_{\beta}(\vect k',\bar{\vect q}_{j}) \Big\}\Big).
\nonumber
\end{eqnarray}
Now we compute the Fourier transform of the above anti-commutator by following a similar procedure as to compute the commutator expression (\ref{MKEA}) in the appendix\ref{app4}. Specifically, considering $t'< t$ one obtains
\begin{eqnarray}
&&\frac{1}{2}\int_{-\infty}^{\infty}\int_{-\infty}^{\infty}dt dt' e^{i\omega t}e^{-i\omega' t'}\left\langle \left\lbrace \hat{\mathcal{E}}_{i}^{\alpha}(t), \hat{\mathcal{E}}_{j}^{\beta\dagger} (t')\right\rbrace  \right\rangle = \nonumber \\
&=&\sum_{\vect k, \vect k'}\bigg[\delta(\vect k-\vect k')\omega^{2}(\vect k)(1+2n(\omega(\vect k),\beta^{-1}))\frac{(2\pi)^2}{2}\bigg(\text{Re}\Big\{h_{\alpha}(\vect k,\bar{\vect q}_{i})h_{\beta}^{\dagger}(\vect k,\bar{\vect q}_{j})\Big\}(\delta(\omega-\omega')(\delta(\omega(\vect k)-\omega)+\delta(\omega(\vect k)+\omega)) ) \nonumber \\
&-&i\text{Im}\Big\{h_{\alpha}(\vect k,\bar{\vect q}_{i})h_{\beta}^{\dagger}(\vect k,\bar{\vect q}_{j})\Big\}(\delta(\omega-\omega')(\delta(\omega(\vect k)+\omega)-\delta(\omega(\vect k)-\omega)) )\bigg) \nonumber \\
&+&\frac{(2\pi)^2}{2}\sum_{l,m=1}^{n}\bigg(\text{Re}\Big\{e^{i\left( \omega(\vect k)-\omega(\vect k')\right) t_{0}}\varepsilon_{\gamma}(\vect k)\varepsilon_{\delta}^{\dagger} (\vect k')g_{\gamma}(\vect k,\bar{\vect q}_{l})g_{\delta}^{\dagger}(\vect k',\bar{\vect q}_{m}) h_{\alpha}(\vect k,\bar{\vect q}_{i})h_{\beta}^{\dagger}(\vect k',\bar{\vect q}_{j})\Big\}\Big(\delta(\omega(\vect k)-\omega)\delta(\omega(\vect k')-\omega') \nonumber \\
&+&\delta(\omega(\vect k)+\omega)\delta(\omega(\vect k')+\omega') \Big) \nonumber \\
&-&i \text{Im}\Big\{e^{i\left( \omega(\vect k)-\omega(\vect k')\right)t_{0} }\varepsilon_{\gamma}(\vect k)\varepsilon_{\delta}^{\dagger} (\vect k') g_{\gamma}(\vect k,\bar{\vect q}_{l})g_{\delta}^{\dagger}(\vect k',\bar{\vect q}_{m})h_{\alpha}(\vect k,\bar{\vect q}_{i})h_{\beta}^{\dagger}(\vect k',\bar{\vect q}_{j})\Big\} \Big(\delta(\omega(\vect k)-\omega)\delta(\omega(\vect k')-\omega') \nonumber \\
&-&\delta(\omega(\vect k)+\omega)\delta(\omega(\vect k')+\omega') \bigg) \nonumber \\
&+&\frac{(2\pi)^2}{2}\sum_{l,m=1}^{n}\bigg(\text{Re}\Big\{e^{i\left( \omega(\vect k)+\omega(\vect k')\right)t_{0} }\varepsilon_{\gamma}(\vect k)\varepsilon_{\delta} (\vect k') g_{\gamma}(\vect k,\bar{\vect q}_{l})g_{\delta}(\vect k',\bar{\vect q}_{m})h_{\alpha}(\vect k,\bar{\vect q}_{i})h_{\beta}(\vect k',\bar{\vect q}_{j})\Big\}\Big(\delta(\omega(\vect k)-\omega)\delta(\omega(\vect k)'+\omega') \nonumber \\
&+&\delta(\omega(\vect k)+\omega)\delta(\omega(\vect k')-\omega') \Big) \nonumber \\
&-&i \text{Im}\Big\{e^{i\left( \omega(\vect k)+\omega(\vect k')\right) t_{0}}\varepsilon_{\gamma}(\vect k)\varepsilon_{\delta} (\vect k')g_{\gamma}(\vect k,\bar{\vect q}_{l})g_{\delta}(\vect k',\bar{\vect q}_{m})h_{\alpha}(\vect k,\bar{\vect q}_{i})h_{\beta}(\vect k',\bar{\vect q}_{j}) \Big\} \Big(\delta(\omega(\vect k)-\omega)\delta(\omega(\vect k')+\omega') \nonumber \\
&-&\delta(\omega(\vect k)+\omega)\delta(\omega(\vect k')-\omega')\bigg)\bigg].
\label{ACFFE}
\end{eqnarray}
One may bring the bracket appearing in the first and second lines of (\ref{ACFFE}) into the form of the stationary fluctuations in (\ref{CFFE}) after using the properties of the Dirac delta function (e.g. $f(\omega(\vect k))\delta(\omega-\omega(\vect k))=f(\omega)\delta(\omega-\omega(\vect k))$ for a given function $f(\omega)$) as well as $1+2n(\omega,\beta^{-1})$ is an odd function in the variable $\omega$, and then, substituting the definition of the spectral density given by (\ref{SPDA1}). Similarly, by directly comparing  the first and second brackets on the variables $l,m$ in (\ref{ACFFE}) with the second and third lines of equation (\ref{CFFE}), it is easy to recognize the following expressions for the non-stationary spectral functions, 
\begin{eqnarray}
\tilde{\mathcal{G}}_{ij}^{\alpha\beta}(\omega,\omega',t_{0})=\frac{(2\pi)^{2}}{2}e^{i\left( \omega-\omega'\right)t_{0} }\sum_{l,m=1}^{n}\sum_{\vect k,\vect k'}\varepsilon_{\gamma} (\vect k)\varepsilon_{\delta}^{\dagger}(\vect k')g_{\gamma}(\vect k,\bar{\vect q}_{l})g_{\delta}^{\dagger}(\vect k',\bar{\vect q}_{m})h_{\alpha}(\vect k,\bar{\vect q}_{i})h_{\beta}^{\dagger}(\vect k',\bar{\vect q}_{j})\delta(\omega(\vect k)-\omega)\delta(\omega(\vect k')-\omega'), \nonumber \\
\label{GGEA}
\end{eqnarray}
as well as
\begin{eqnarray}
\tilde{\mathcal{F}}_{ij}^{\alpha\beta}(\omega,\omega',t_{0})=\frac{(2\pi)^{2}}{2}e^{i\left( \omega+\omega'\right)t_{0} }\sum_{l,m=1}^{n}\sum_{\vect k,\vect k'}\varepsilon_{\gamma}(\vect k)\varepsilon_{\delta}(\vect k')g_{\gamma}(\vect k,\bar{\vect q}_{l})g_{\delta}(\vect k',\bar{\vect q}_{m})h_{\alpha}(\vect k,\bar{\vect q}_{i})h_{\beta}(\vect k',\bar{\vect q}_{j})\delta(\omega(\vect k)-\omega)\delta(\omega(\vect k')-\omega'). \nonumber \\  \label{FFEA} 
\end{eqnarray}

Let us next to obtain equations (\ref{GGE}) and (\ref{FFE}) starting from (\ref{GGEA}) and (\ref{FFEA}), respectively. We show first the derivation for $\tilde{\mathcal{G}}_{ij}^{\alpha\beta}$, and an identical procedure can be carried out to obtain $\tilde{\mathcal{F}}_{ij}^{\alpha\beta}$. Substituting the corresponding expressions (\ref{PVP}), (\ref{GEIT}) and (\ref{EIT}) in (\ref{GGEA}), we find
\begin{eqnarray}
\frac{2\tilde{\mathcal{G}}_{ij}^{\alpha\beta}(\omega,\omega',t_{0})}{(2\pi)^{2}e^{it_{0}\left( \omega-\omega'\right) }}
&=&\frac{e^4\kappa^2}{4(2\pi)^4 L^{4}\omega\omega'}\sum_{l,m=1}^{n}\sum_{\vect k,\vect k'}e^{-2\sigma\left( |\vect k|^{2}+|\vect k'|^{2}\right) }e^{i(\vect k'\cdot(\bar{\vect q}_{j}+\bar{\vect q}_{m})-\vect k\cdot(\bar{\vect q}_{i}+\bar{\vect q}_{l}))} \nonumber \\
&\times &\Bigg(\Bigg( \frac{\kappa^{2}k_{\alpha}k_{\beta}'\epsilon_{\iota\nu}\epsilon_{\nu\gamma}\epsilon_{\iota'\nu'}\epsilon_{\nu'\gamma'}k_{\iota}k_{\iota'}'k_{\gamma}k_{\gamma'}' e^{i\frac{\kappa}{|\kappa|}(\theta(\vect k')-\theta(\vect k))}}{|\vect k'|^{2}|\vect k|^{2}\omega(\vect k)\omega(\vect k')} \nonumber \\
&-&i \kappa e^{-i\frac{\kappa}{|\kappa|}(\theta(\vect k')-\theta(\vect k))} \bigg(\frac{\epsilon_{\iota\nu}\epsilon_{\nu\gamma}\epsilon_{\beta\gamma'}k_{\alpha}k_{\iota}k_{\gamma}k_{\gamma'}'}{|\vect k|^{2}\omega(\vect k)}-\frac{\epsilon_{\iota'\nu'}\epsilon_{\nu'\gamma}\epsilon_{\alpha\gamma'}k_{\beta}'k_{\iota'}'k_{\gamma}'k_{\gamma'}}{|\vect k'|^{2}\omega(\vect k')}\bigg) \nonumber \\
&+& \epsilon_{\alpha\gamma}\epsilon_{\beta\gamma'}k_{\gamma}k_{\gamma'}'e^{-i\frac{\kappa}{|\kappa|}(\theta(\vect k')-\theta(\vect k))}\Bigg)\frac{(k_{1}^{2}+k_{2}^{2})(k_{1}'^{2}+k_{2}'^{2})}{|\vect k|^{3}|\vect k'|^{3}}e^{-i\frac{\kappa}{|\kappa|}(\theta(\vect k')-\theta(\vect k))}\Bigg)\delta(\omega(\vect k)-\omega)\delta(\omega(\vect k')-\omega') \nonumber \\
&=&\frac{e^4\kappa^2}{4(2\pi)^4 L^{4}\omega\omega'}\sum_{l,m=1}^{n}\sum_{\vect k,\vect k'}e^{-2\sigma\left( |\vect k|^{2}+|\vect k'|^{2}\right) }e^{-i(\vect k'\cdot(\bar{\vect q}_{j}+\bar{\vect q}_{m})-\vect k\cdot(\bar{\vect q}_{i}+\bar{\vect q}_{l}))}e^{-i\frac{2\kappa}{|\kappa|}(\theta(\vect k')-\theta(\vect k))} \nonumber \\
&\times&\Bigg( \frac{\kappa^{2}k_{\alpha}k_{\beta}'}{\omega(\vect k)\omega(\vect k')} +i\kappa\Big(\frac{\epsilon_{\beta\gamma'}k_{\alpha}k_{\gamma'}'}{\omega(\vect k)}-\frac{\epsilon_{\alpha\gamma'}k_{\beta}'k_{\gamma'}}{\omega(\vect k')}\Big)\nonumber \\
&+& \delta_{\alpha\beta}(k_{1}k'_{1}+k_{2}k'_{2})-k_{\beta}k_{\alpha}'\Bigg)\frac{1}{|\vect k||\vect k'|}\delta(\omega(\vect k)-\omega)\delta(\omega(\vect k')-\omega') \nonumber \\
&=&\frac{e^4\kappa^2}{4(2\pi)^4}\sum_{l,m=1}^{n}\int_{0}^{\infty}  dk dk'\ kk'e^{-2\sigma\left( k^2+k'^{2}\right) } \ \frac{\delta\left( k-\sqrt{\omega^{2}-\kappa^{2}}\right) }{\sqrt{\omega^{2}-\kappa^{2}}}\frac{\delta\left( k'-\sqrt{\omega'^{2}-\kappa^{2}}\right) }{\sqrt{\omega'^{2}-\kappa^{2}}} 
\nonumber \\
&\times &\mathcal{R}_{\alpha\beta}(k,k',|\bar{\vect q}_{i}+\bar{\vect q}_{l}|,|\bar{\vect q}_{j}+\bar{\vect q}_{m}|), \label{GGEAII}
\end{eqnarray}
where we have rewrite $\theta(\vect k')=\theta'$ and $\theta(\vect k)=\theta$, and introduced the following matrix
\begin{eqnarray}
\mathcal{R}_{\alpha\beta}(k,k',|\bar{\vect q}_{i}+\bar{\vect q}_{l}|,|\bar{\vect q}_{j}+\bar{\vect q}_{m}|)=\int_{-\pi}^{\pi}d\theta d\theta' e^{-i\frac{2\kappa}{|\kappa|}(\theta'-\theta)}e^{-i(k'|\bar{\vect q}_{j}+\bar{\vect q}_{m}|\cos\theta-k|\bar{\vect q}_{i}+\bar{\vect q}_{l}|\cos\theta')}\mathcal{D}_{\alpha\beta},
\label{BF1}
\end{eqnarray}
with 
\begin{eqnarray}
\mathcal{D}_{11}&=&\frac{\kappa^2}{\omega\omega'}\cos\theta\cos\theta'+i\kappa\left( \frac{1}{\omega}\cos\theta\sin\theta'-\frac{1}{\omega'}\cos\theta'\sin\theta\right)  +\sin\theta\sin\theta', \nonumber \\
\mathcal{D}_{12}&=& \frac{\kappa^2}{\omega\omega'}\cos\theta\sin\theta'-\cos\theta'\sin\theta-i\kappa\left(\frac{1}{\omega}\cos\theta\cos\theta'+\frac{1}{\omega'}\sin\theta\sin\theta'\right) , \nonumber \\
\mathcal{D}_{21}&=&\frac{\kappa^2}{\omega\omega'}\cos\theta'\sin\theta-\cos\theta\sin\theta'+i\kappa\left( \frac{1}{\omega'}\cos\theta\cos\theta'+\frac{1}{\omega}\sin\theta\sin\theta'\right)  ,     \nonumber \\
\mathcal{D}_{22}&=& \frac{\kappa^2}{\omega\omega'}\sin\theta\sin\theta'+i\kappa\left(\frac{1}{\omega'}\sin\theta'\cos\theta-\frac{1}{\omega}\sin\theta\cos\theta'\right) +\cos\theta\cos\theta' .    \nonumber
\end{eqnarray}
As we have previously done in appendix\ref{app4} to obtain the expression for the spectral density, we now replace the trigonometric integrals appearing in (\ref{BF1}) by the corresponding definition of the first- and second- order Bessel functions of first kind, i.e.
\begin{eqnarray}
\mathcal{R}_{11}(k,k',a,b)&=& \frac{4\pi^2}{abkk'\omega\omega'}\bigg(\kappa (ak J_{1}(ak)-2J_{2}(ak))+2\frac{\kappa\omega}{|\kappa|}J_{2}(ak)\bigg)\bigg(\kappa (bk'J_{1}(bk')-2J_{2}(bk'))+\omega'\frac{2\kappa}{|\kappa|}J_{2}(bk')\bigg) , \nonumber\\
\mathcal{R}_{12}(k,k',a,b)&=&\frac{-4i\pi^2}{abkk'\omega\omega'}\bigg(\kappa (ak J_{1}(ak)-2J_{2}(ak))-2\frac{\kappa \omega}{|\kappa|} J_{2}(ak)\bigg)\bigg(bk'\omega' J_{1}(bk')-\bigg(2\omega'-\frac{\kappa^2k'}{|\kappa|}\bigg)J_{2}(bk')\bigg), \nonumber \\
\mathcal{R}_{21}(k,k',a,b)&=&\frac{4i\pi^2}{abkk'\omega\omega'}\bigg(\kappa (bk' J_{1}(bk')-2J_{2}(bk'))-2\frac{\kappa\omega'}{|\kappa|}J_{2}(bk')\bigg)\bigg(ak\omega J_{1}(ak)-\bigg(2\omega-\frac{\kappa^2k}{|\kappa|}\bigg)J_{2}(ak)\bigg),  \nonumber \\
\mathcal{R}_{22}(k,k',a,b)&=& \frac{4\pi^2}{abkk'\omega\omega'}\bigg(ak\omega J_{1}(ak)-\bigg(2\omega-\frac{\kappa^2k}{|\kappa|}\bigg)J_{2}(ak)\bigg) \bigg(bk'\omega' J_{1}(bk')-\bigg(2\omega'-\frac{\kappa^2k'}{|\kappa|}\bigg)J_{2}(bk')\bigg).
\label{GGEAM}
\end{eqnarray}
Finally, Eq.(\ref{GGE}) is obtained by firstly inserting (\ref{GGEAM}) into (\ref{GGEAII}), and then, carrying out the integral on the real variables $k$ and $k'$, which is immediately obtained by using the properties of the Dirac delta function. Although we obtain a formidable expression, the particular shape of (\ref{GGEAM}) permits to cast the non-equilibrium spectral density in the form (\ref{GGE}) after some simple manipulation.

Applying a similar procedure to (\ref{FFEA}) as before yields  
\begin{eqnarray}
\frac{2\tilde{\mathcal{F}}_{ij}^{\alpha\beta}(\omega,\omega',t_{0})}{(2\pi)^{2}e^{-it_{0}\left( \omega+\omega'\right) }}
&=&\frac{e^4\kappa^2}{4(2\pi)^4 L^{4}\omega\omega'}\sum_{l,m=1}^{n}\sum_{\vect k,\vect k'}e^{-2\sigma(|\vect k|^{2}+|\vect k'|^{2})}e^{i(\vect k'\cdot(\bar{\vect q}_{j}+\bar{\vect q}_{m})+\vect k\cdot(\bar{\vect q}_{i}+\bar{\vect q}_{l}))}\bigg(\left( \frac{\kappa\epsilon_{\iota\nu}k_{\iota}k_{\alpha}\varepsilon_{\nu}(\vect k)}{|\vect k|^2\omega(\vect k)}+i\varepsilon_{\alpha}(\vect k)\right) \label{FFEAII}\\
&\times&\left( \frac{\kappa\epsilon_{\iota'\nu'}k_{\iota'}'k_{\beta}'\varepsilon_{\nu'}(\vect k')}{|\vect k'|^2\omega(\vect k')}+i\varepsilon_{\beta}(\vect k')\right) \frac{\epsilon_{\lambda\gamma}\epsilon_{\lambda'\delta}k_{\lambda}k_{\lambda'}'}{|\vect k|^2|\vect k'|^2}\frac{\epsilon_{\gamma\mu}\epsilon_{\delta\mu'}k_{\mu}k_{\mu'}'}{|\vect k||\vect k'|}e^{i\frac{\kappa}{|\kappa|}(\theta(\vect k')+\theta(\vect k))}\bigg)\delta(\omega(\vect k)-\omega)\delta(\omega(\vect k')-\omega')  \nonumber \\
&=&\frac{e^4\kappa^2}{4(2\pi)^4 L^{4}\omega\omega'}\sum_{l,m=1}^{n}\sum_{\vect k,\vect k'}e^{-2\sigma(|\vect k|^{2}+|\vect k'|^{2})}e^{i(\vect k'\cdot(\bar{\vect q}_{j}+\bar{\vect q}_{m})+\vect k\cdot(\bar{\vect q}_{i}+\bar{\vect q}_{l}))}e^{i\frac{2\kappa}{|\kappa|}(\theta(\vect k')+\theta(\vect k))} \nonumber \\
&\times &\Bigg( -\frac{\kappa^{2}k_{\alpha}k_{\beta}'}{\omega(\vect k)\omega(\vect k')} -i\kappa\Big(\frac{\epsilon_{\beta\gamma'}k_{\alpha}k_{\gamma'}'}{\omega(\vect k)}+\frac{\epsilon_{\alpha\gamma'}k_{\beta}'k_{\gamma'}}{\omega(\vect k')}\Big) \nonumber \\
&-& (\delta_{\alpha\beta}(k_{1}k'_{1}+k_{2}k'_{2})-k_{\beta}k_{\alpha}')\Bigg)\frac{1}{|\vect k||\vect k'|}\delta(\omega(\vect k)-\omega)\delta(\omega(\vect k')-\omega')  \nonumber \\
&=&-\frac{e^4\kappa^2}{4(2\pi)^4}\sum_{l,m=1}^{n}\int_{0}^{\infty}  dk dk' \ kk' e^{-2\sigma(|\vect k|^{2}+|\vect k'|^{2})}\frac{\delta\left( k-\sqrt{\omega^{2}-\kappa^{2}}\right) }{\sqrt{\omega^{2}-\kappa^{2}}}\frac{\delta\left( k'-\sqrt{\omega'^{2}-\kappa^{2}}\right) }{\sqrt{\omega'^{2}-\kappa^{2}}} \nonumber \\
&\times &\mathcal{T}_{\alpha\beta}(k,k',|\bar{\vect q}_{j}+\bar{\vect q}_{m}|,|\bar{\vect q}_{i}+\bar{\vect q}_{l}|) , \nonumber 
\end{eqnarray}
where we have introduced the following matrix after doing some tedious algebra as before, 
\begin{eqnarray}
\mathcal{T}_{11}(k,k',a,b)&=&   \frac{4\pi^{2}}{abkk'\omega\omega'}\bigg(ak\kappa J_{1}(ak)\bigg(-bk'\kappa  J_{1}(bk')+2\bigg(\kappa-\frac{\kappa\omega'}{|\kappa|}\bigg)J_{2}(bk')  \bigg)\label{AEFTM} \\
&+& 2 J_{2}(ak)\bigg(bk'\kappa\bigg(\kappa-\frac{\kappa\omega}{|\kappa|}\bigg)J_{1}(bk')+2\bigg(\frac{\kappa^{2}}{|\kappa|}(\omega'+\omega)+\omega\omega'-\kappa^2\bigg) J_{2}(bk') \bigg) \bigg)                      , \nonumber \\
\mathcal{T}_{12}(k,k',a,b)&=&\frac{-4i\pi^{2}}{abkk'\omega\omega'}\bigg(a^2k^{2}\kappa J_{1}(ak)\bigg(bk'\omega' J_{1}(bk')-2\bigg(\omega'-\frac{\kappa^{2}}{|\kappa|}\bigg)J_{2}(bk')  \bigg)\nonumber \\
&+& 2akJ_{2}(ak)\bigg(bk'\omega'\bigg(\kappa+\frac{\kappa\omega}{|\kappa|}\bigg)J_{1}(bk')+2\bigg((\omega'+\omega)\kappa+\kappa^2+\frac{\kappa}{|\kappa|}\omega\omega'\bigg) J_{2}(bk') \bigg) \bigg),   \nonumber \\
\mathcal{T}_{21}(k,k',a,b)&=&\frac{-4i\pi^{2}}{abkk'\omega\omega'}\bigg(a^2k^2\omega J_{1}(ak)\bigg(bk'^{2}\kappa  J_{1}(bk')-2\bigg(\kappa+\frac{\kappa\omega'}{|\kappa|}\bigg)J_{2}(bk')  \bigg)\nonumber \\
&-&2 ak J_{2}(ak)\bigg(bk'\kappa\bigg( \kappa+\frac{\kappa\omega}{|\kappa|}\bigg)J_{1}(bk')-2\bigg(\kappa(\omega'+\omega)-\kappa^2+\frac{\kappa}{|\kappa|}\omega\omega'\bigg) J_{2}(bk') \bigg) \bigg)   ,                      \nonumber \\
\mathcal{T}_{22}(k,k',a,b)&=&\frac{4\pi^{2}}{abkk'\omega\omega'}\bigg(-ak\omega J_{1}(ak)\bigg(bk'\omega'\kappa  J_{1}(bk')+2\bigg(\omega'+\frac{\kappa^{2}}{|\kappa|}k'\bigg)J_{2}(bk')  \bigg)\nonumber \\
&+&2 J_{2}(ak)\bigg( bk'\omega'\kappa\bigg(2\omega+\frac{\kappa^{2}}{|\kappa|} k\bigg)J_{1}(bk')-2\bigg(\kappa^2+\frac{\kappa^{2}}{|\kappa|}(\omega'+\omega)-\omega\omega')\bigg) J_{2}(bk') \bigg) \bigg).                         \nonumber
\end{eqnarray} 
Once again, the integrals appearing in the last line of (\ref{FFEAII}) are directly obtained by making use of the Dirac delta properties, once we have replaced \ref{AEFTM}. In this way, we finally arrive to the expression (\ref{FFE}) for the non-stochastic spectral function presented in Sec.\ref{SGLE}. 

\section{Non-stochastic fluctuations, thermal noise and retarded self-energy for closed particles and weak Chern-Simons action}\label{app6}
In this section, we extend the discussion, presented in Sec.\ref{PSNDR}, about the evaluation of the non-stochastic fluctuations appearing in (\ref{ENE}) when it is taking the strict limit $t-t_{0}\rightarrow \infty$, and we provide explicit expressions for the self-energy and the thermal correlations (\ref{TNE}) in the zero-temperature limit as well.

By paying attention to the expressions obtained for (\ref{GGE}) and (\ref{FFE}), it is readily to see that both are integrable functions for $\omega,\omega'$ into the domain of the MSC environment bandwidth (i.e. $\kappa\leq \omega\leq (2\sigma)^{-\frac{1}{2}}$), as well as they decay as fast as an exponential function for large frequencies since the Bessel functions decrease algebraically. As mentioned in Sec.\ref{SGLE}, these properties permit to use the Riemann-Lebesgue lemma to evaluate the non-equilibrium fluctuations $\Upsilon_{ij}^{\alpha\beta}(t,t',t_{0}) $ and $\Xi_{ij}^{\alpha\beta}(t,t',t_{0})$ involved in expression (\ref{ENE}), which are respectively obtained from the second and third line of (\ref{CFFE}) via the inverse Fourier transform. Specifically, these non-stochastic fluctuations takes the following form in the time domain,
\begin{equation}
\Upsilon_{ij}^{\alpha\beta}(t,t',t_{0})=\frac{4}{\pi^2}\int d \omega d\omega' \Big(\text{Re}\left\lbrace \tilde{\mathcal{G}}_{ij}^{\alpha\beta}(\omega,\omega',t_{0})\right\rbrace \cos(\omega t-\omega' t')+\text{Im}\left\lbrace \tilde{\mathcal{G}}_{ij}^{\alpha\beta}(\omega,\omega',t_{0})\right\rbrace \sin(\omega t-\omega' t')\Big),  \label{TNFI}
\end{equation}
and 
\begin{equation}
\Xi_{ij}^{\alpha\beta}(t,t',t_{0})=\frac{4}{\pi^2}\int d\omega d\omega' \Big( \text{Re}\left\lbrace\tilde{\mathcal{F}}_{ij}^{\alpha\beta}(\omega, \omega',t_{0})\right\rbrace \cos(\omega t+\omega t')  +\text{Im}\left\lbrace\tilde{\mathcal{F}}_{ij}^{\alpha\beta}(\omega, \omega',t_{0})\right\rbrace \sin(\omega t+\omega t') \Big).\label{TNFII}
\end{equation}
where we recall that $\tilde{\mathcal{G}}_{ij}^{\alpha\beta}(\omega,\omega',t_{0})$ and $\tilde{\mathcal{F}}_{ij}^{\alpha\beta}(\omega, \omega',t_{0})$ are given by (\ref{GGE}) and (\ref{FFE}), respectively. Formally, the aforementioned lemma can be enunciated as follows: Let $f(\omega)$ be an complex-valued function that is absolutely integrable on $\mathbb{R}$. Then the Riemann-Lebesgue lemma states that \cite{chandrasekharan20121},
\begin{equation}
\smash{\displaystyle\lim_{|t|\to \infty}\int_{-\infty}^{\infty}f(\omega)e^{i\omega t}d\omega\rightarrow 0}.
\end{equation} 
One says $f(\omega)$ is an absolutely integrable function on $\mathbb{R}$ if it is fulfill $\smash{\displaystyle\int_{-\infty}^{\infty}|f(\omega)|d\omega< 0}$,  or equivalently, if it belongs to the class of $L^{1}(\mathbb{R})$-functions. Clearly, the Riemann-Lebesgue lemma has an intuitive interpretation: the integral becomes so highly oscillatory that everything cancels out.

To see more clearly how such lemma applied to  (\ref{TNFI}) and (\ref{TNFII}), it is convenient to rewrite them as the integrals of two independent dispersion functions in the variables $\omega$ and $\omega'$. Let first pay attention to the non-stochastic fluctuation encoded by $\Upsilon_{ij}^{\alpha\beta}(t,t',t_{0})$. By using trigonometric identities once replaced equation (\ref{GGE}) into (\ref{TNFI}), we can bring this into the following form 
\begin{eqnarray}
\Upsilon_{ij}^{\alpha\beta}(t,t',t_{0})&=& \int_{0}^{\infty}  d \omega d\omega'  \text{Re}\left\lbrace\rho_{i}^{\alpha\beta}(\omega)\right\rbrace \text{Re}\left\lbrace\rho_{j}^{\alpha\beta}(\omega')\right\rbrace  c(\omega,\omega',t-t_{0})   \nonumber \\
&+&\int_{0}^{\infty}  d \omega d\omega' \text{Im}\left\lbrace\rho_{i}^{\alpha\beta}(\omega)\right\rbrace \text{Im}\left\lbrace\rho_{j}^{\alpha\beta}(\omega')\right\rbrace     s(\omega,\omega',t-t_{0}) \label{LNEF}
\end{eqnarray}
where
\begin{eqnarray}
c(\omega,\omega',\tau)&=&\cos\left(\omega\tau \right) \cos\left(\omega'\tau \right)  +\sin\left(\omega\tau \right) \sin\left(\omega'\tau\right)  \nonumber \\
s(\omega,\omega',\tau)&=&\sin\left(\omega\tau \right) \cos\left(\omega'\tau \right)  -\sin\left(\omega\tau \right) \cos\left(\omega'\tau\right) \nonumber
\end{eqnarray}
and with the dispersion function given by,
\begin{eqnarray}
\rho^{\alpha\beta}_{i/j}(\omega)&=&\frac{e^2\kappa e^{-2\sigma\left( \omega^{2}-\kappa^{2}\right) }}{\sqrt{128}\omega} \sum_{l=1}^{n}r_{i/j}^{\alpha\beta}\left( \sqrt{\omega^{2}-\kappa^{2}},|\bar{\vect q}_{i/j}+\bar{\vect q}_{l}|\right), \ \ \ \kappa \leq \omega \label{WEF}
\end{eqnarray}
where the elements of the matrix function $\vect r_{i}\left( \sqrt{\omega^{2}-\kappa^{2}},|\bar{\vect q}_{i}+\bar{\vect q}_{l}|\right)$ ($\vect r_{j}\left( \sqrt{\omega'^{2}-\kappa^{2}},|\bar{\vect q}_{j}+\bar{\vect q}_{l}|\right)$) are directly obtained from the elements $\vect R_{\alpha\beta}$, determined by (\ref{SDGM}), after removing the dependence on the variables $\omega'$ ($\omega$) and $\bar{\vect q}_{j}$ ($\bar{\vect q}_{i}$). It is found that the elements $r_{i/j}^{\alpha\beta}$ are basically convergent combinations of first- and second- order Bessel functions of the first kind. 

From (\ref{WEF}) it is now clear that both dispersion functions $ \rho^{\alpha\beta}_{i/j}(x)$ (for arbitrary $i$ and $j$) are continuously differentiable for $\kappa\leq x< \infty$, as well as they exponentially decay for values $x$ larger than $(2\sigma)^{-\frac{1}{2}}$. As a result, it follows from the above Riemann-Lebesgue lemma that all integrals in (\ref{LNEF}) asymptotically vanishes in the long time limit $t-t_{0}\rightarrow \infty$. On the other hand, starting from the expression for $\tilde{\mathcal{F}}_{ij}^{\alpha\beta}(\omega,\omega', t_{0})$ (see equation (\ref{AEFTM}) in Appendix\ref{app5}), we can follow the same procedure to show that the corresponding dispersion functions characterizing the non-stochastic fluctuation $\Xi_{ij}^{\alpha\beta}(t,t',t_{0})$ share these features with $ \rho^{\alpha\beta}_{i/j}(x)$, so we could conclude an identical statement for (\ref{TNFII}) after applying the Riemann-Lebesgue lemma.

Now we address the retarded self-energy and thermal fluctuations in the realistic situation when the system particles are sufficiently closed (i.e $\frac{|\Delta\bar{q}_{ij}|}{\sqrt{2\sigma}}\ll 1$) and the Chern-Simons action is considered to be weak (i.e. $\sqrt{2\sigma}\kappa\ll 1$). As a result of these approximations, the spectral density substantially simplifies to (\ref{SPDA1}), which is illustrated in Sec.\ref{PSNDR}. Replacing this in the expression for the retarded self-energy (\ref{MKEII}) yields upon integration, 
\begin{eqnarray}
\Sigma_{ij}^{11}(t)&\simeq&\frac{e^2}{16\pi\sqrt{2(4\sigma)^5}}\bigg( \Big(32 \sigma (1 + 2 \kappa^2 \sigma) - |\Delta\bar{q}_{ij}|^2 (5 +18 \kappa^2 \sigma)\Big)x + 2 |\Delta\bar{q}_{ij}|^2(1 + 2 \kappa^2 \sigma)x^3 \label{SEAI}  \\
&-&\frac{\sqrt{\pi}}{2}e^{-x^2}\text{erfi}(x) \Big( 3 |\Delta\bar{q}_{ij}|^2 -32 \sigma + 14 |\Delta\bar{q}_{ij}|^2 \kappa^2 \sigma - 192\kappa^2 \sigma^2  \nonumber \\
&+&\Big(-12 |\Delta\bar{q}_{ij}|^2 + 64 \sigma - 40 |\Delta\bar{q}_{ij}|^2 \kappa^2 \sigma + 
 128 \kappa^2 \sigma^2\Big)x^2+ \Big(4|\Delta\bar{q}_{ij}|^2 + 8 |\Delta\bar{q}_{ij}|^2 \kappa^2 \sigma \Big)x^4 \Big) \bigg), \nonumber
\end{eqnarray}
and 
\begin{eqnarray}
\Sigma_{ij}^{22}(t)&\simeq& \frac{e^2}{16\pi\sqrt{2(4\sigma)^5}}\bigg(\Big(32 \sigma (1 + 2 \kappa^2 \sigma) - |\Delta\bar{q}_{ij}|^2 (15 +22 \kappa^2 \sigma)\Big)x + 6|\Delta\bar{q}_{ij}|^2(1 + 2 \kappa^2 \sigma)x^3    \label{SEAII}\\
&-&\frac{\sqrt{\pi}}{2}e^{-x^2}\text{erfi}(x) \Big( 9 |\Delta\bar{q}_{ij}|^2 - 32 \sigma + 10 |\Delta\bar{q}_{ij}|^2 \kappa^2 \sigma - 192 \kappa^2 \sigma^2  \nonumber \\
&+&\Big(-32 |\Delta\bar{q}_{ij}|^2 +64 \sigma - 56 |\Delta\bar{q}_{ij}|^2 \kappa^2 \sigma + 
128 \kappa^2 \sigma^2\Big)x^2+ \Big(12 |\Delta\bar{q}_{ij}|^2 + 24 |\Delta\bar{q}_{ij}|^2 \kappa^2 \sigma \Big)x^4\Big)\bigg),   \nonumber
\end{eqnarray}
as well as the off-diagonal element given by (\ref{SEAIII}) in Sec.\ref{PSNDR}. The behavior of (\ref{SEAI}) and (\ref{SEAII}) in the time domain is illustrated in the figure\ref{Fig1}. Moreover, we may follow a similar procedure to get the zero-temperature limit of the thermal fluctuations (\ref{TNE}) once we have replaced the spectral density by (\ref{SPDA1}). In this way, we consider the zero-temperature limit where $n\left(  \omega, \beta^{-1}\right)\rightarrow 0$. Using the standard tables of integration, we find after some tedious manipulation 
\begin{eqnarray}
\left\langle \left\lbrace \hat \xi_{i}^{1} (t'+t), \hat \xi^{1\dagger}_{j} ( t')\right\rbrace  \right\rangle &\simeq &\frac{e^2\ e^{-x^2}}{2\sqrt{2\pi(16\sigma)^{5}}}\bigg(-3 |\Delta\bar{q}_{ij}|^2 +32 \sigma - 14 |\Delta\bar{q}_{ij}|^2 \kappa^2 \sigma +  192 \kappa^2 \sigma^2 \nonumber \\
&+&\Big(12 |\Delta\bar{q}_{ij}|^2 - 64 \sigma + 40 |\Delta\bar{q}_{ij}|^2 \kappa^2 \sigma -  128 \kappa^2 \sigma^2\Big)x^2  \nonumber\\
&-&4\Big(|\Delta\bar{q}_{ij}|^2 + 2 |\Delta\bar{q}_{ij}|^2 \kappa^2 \sigma\Big) x^4   \bigg),
\label{TFAI}
\end{eqnarray}
and 
\begin{eqnarray}
\left\langle \left\lbrace \hat \xi_{i}^{2} (t'+t), \hat \xi^{2\dagger}_{j} ( t')\right\rbrace  \right\rangle &\simeq &\frac{e^2\ e^{-x^2}}{2\sqrt{2\pi(16\sigma)^{5}}}\bigg(-9 |\Delta\bar{q}_{ij}|^2 + 32 \sigma - 10|\Delta\bar{q}_{ij}|^2 \kappa^2 \sigma +  192 \kappa^2 \sigma^2  \nonumber \\
&+&\Big(36 |\Delta\bar{q}_{ij}|^2 - 64 \sigma + 56 |\Delta\bar{q}_{ij}|^2 \kappa^2 \sigma -  128 \kappa^2 \sigma^2\Big)x^2  \nonumber \\
&-& 12\Big(|\Delta\bar{q}_{ij}|^2 +2|\Delta\bar{q}_{ij}|^2 \kappa^2 \sigma\Big) x^4 \bigg),
\label{TFAII}
\end{eqnarray}
whereas for the off-diagonal term,
\begin{eqnarray}
\left\langle \left\lbrace \hat \xi_{i}^{1} (t'+t), \hat \xi^{2\dagger}_{j} ( t')\right\rbrace  \right\rangle&\simeq &\frac{e^2 \kappa}{256\sqrt{\pi}\sigma^2}xe^{-x^2}\bigg( 16\sigma+|\Delta\bar{q}_{ij}|^2(-3+x^2) \bigg),
\label{TFAIII}
\end{eqnarray}
with $x=\frac{t}{\sqrt{8\sigma}}$.The above expressions are illustrated in figure\ref{Fig1} as functions of time.

\end{widetext}


\begin{thebibliography}{85}%
\makeatletter
\providecommand \@ifxundefined [1]{%
 \@ifx{#1\undefined}
}%
\providecommand \@ifnum [1]{%
 \ifnum #1\expandafter \@firstoftwo
 \else \expandafter \@secondoftwo
 \fi
}%
\providecommand \@ifx [1]{%
 \ifx #1\expandafter \@firstoftwo
 \else \expandafter \@secondoftwo
 \fi
}%
\providecommand \natexlab [1]{#1}%
\providecommand \enquote  [1]{``#1''}%
\providecommand \bibnamefont  [1]{#1}%
\providecommand \bibfnamefont [1]{#1}%
\providecommand \citenamefont [1]{#1}%
\providecommand \href@noop [0]{\@secondoftwo}%
\providecommand \href [0]{\begingroup \@sanitize@url \@href}%
\providecommand \@href[1]{\@@startlink{#1}\@@href}%
\providecommand \@@href[1]{\endgroup#1\@@endlink}%
\providecommand \@sanitize@url [0]{\catcode `\\12\catcode `\$12\catcode
  `\&12\catcode `\#12\catcode `\^12\catcode `\_12\catcode `\%12\relax}%
\providecommand \@@startlink[1]{}%
\providecommand \@@endlink[0]{}%
\providecommand \url  [0]{\begingroup\@sanitize@url \@url }%
\providecommand \@url [1]{\endgroup\@href {#1}{\urlprefix }}%
\providecommand \urlprefix  [0]{URL }%
\providecommand \Eprint [0]{\href }%
\providecommand \doibase [0]{http://dx.doi.org/}%
\providecommand \selectlanguage [0]{\@gobble}%
\providecommand \bibinfo  [0]{\@secondoftwo}%
\providecommand \bibfield  [0]{\@secondoftwo}%
\providecommand \translation [1]{[#1]}%
\providecommand \BibitemOpen [0]{}%
\providecommand \bibitemStop [0]{}%
\providecommand \bibitemNoStop [0]{.\EOS\space}%
\providecommand \EOS [0]{\spacefactor3000\relax}%
\providecommand \BibitemShut  [1]{\csname bibitem#1\endcsname}%
\let\auto@bib@innerbib\@empty
\bibitem [{\citenamefont {Weiss}(2012)}]{weiss20121}%
  \BibitemOpen
  \bibfield  {author} {\bibinfo {author} {\bibfnamefont {U.}~\bibnamefont
  {Weiss}},\ }\href@noop {} {\emph {\bibinfo {title} {Quantum dissipative
  systems}}},\ Vol.~\bibinfo {volume} {13}\ (\bibinfo  {publisher} {World
  scientific},\ \bibinfo {year} {2012})\BibitemShut {NoStop}%
\bibitem [{\citenamefont {Garc{\'{i}}a-Ripoll}\ \emph
  {et~al.}(2009)\citenamefont {Garc{\'{i}}a-Ripoll}, \citenamefont
  {D{\"{u}}rr}, \citenamefont {Syassen}, \citenamefont {Bauer}, \citenamefont
  {Lettner}, \citenamefont {Rempe},\ and\ \citenamefont
  {Cirac}}]{garciaripoll20091}%
  \BibitemOpen
  \bibfield  {author} {\bibinfo {author} {\bibfnamefont {J.~J.}\ \bibnamefont
  {Garc{\'{i}}a-Ripoll}}, \bibinfo {author} {\bibfnamefont {S.}~\bibnamefont
  {D{\"{u}}rr}}, \bibinfo {author} {\bibfnamefont {N.}~\bibnamefont {Syassen}},
  \bibinfo {author} {\bibfnamefont {D.~M.}\ \bibnamefont {Bauer}}, \bibinfo
  {author} {\bibfnamefont {M.}~\bibnamefont {Lettner}}, \bibinfo {author}
  {\bibfnamefont {G.}~\bibnamefont {Rempe}}, \ and\ \bibinfo {author}
  {\bibfnamefont {J.~I.}\ \bibnamefont {Cirac}},\ }\href {\doibase
  10.1088/1367-2630/11/1/013053} {\bibfield  {journal} {\bibinfo  {journal}
  {New J. Phys.}\ }\textbf {\bibinfo {volume} {11}},\ \bibinfo {pages} {013053}
  (\bibinfo {year} {2009})}\BibitemShut {NoStop}%
\bibitem [{\citenamefont {Sieberer}\ \emph {et~al.}(2016)\citenamefont
  {Sieberer}, \citenamefont {Buchhold},\ and\ \citenamefont
  {Diehl}}]{sieberer20161}%
  \BibitemOpen
  \bibfield  {author} {\bibinfo {author} {\bibfnamefont {L.~M.}\ \bibnamefont
  {Sieberer}}, \bibinfo {author} {\bibfnamefont {M.}~\bibnamefont {Buchhold}},
  \ and\ \bibinfo {author} {\bibfnamefont {S.}~\bibnamefont {Diehl}},\ }\href
  {\doibase 10.1088/0034-4885/79/9/096001} {\bibfield  {journal} {\bibinfo
  {journal} {Rep. Prog. Phys.}\ }\textbf {\bibinfo {volume} {79}},\ \bibinfo
  {pages} {096001} (\bibinfo {year} {2016})}\BibitemShut {NoStop}%
\bibitem [{\citenamefont {Unruh}\ and\ \citenamefont
  {Zurek}(1989)}]{unruh19891}%
  \BibitemOpen
  \bibfield  {author} {\bibinfo {author} {\bibfnamefont {W.~G.}\ \bibnamefont
  {Unruh}}\ and\ \bibinfo {author} {\bibfnamefont {W.~H.}\ \bibnamefont
  {Zurek}},\ }\href {\doibase 10.1103/PhysRevD.40.1071} {\bibfield  {journal}
  {\bibinfo  {journal} {Phys. Rev. D}\ }\textbf {\bibinfo {volume} {40}},\
  \bibinfo {pages} {1071} (\bibinfo {year} {1989})}\BibitemShut {NoStop}%
\bibitem [{\citenamefont {Gautier}\ and\ \citenamefont
  {Serreau}(2012)}]{gautier20121}%
  \BibitemOpen
  \bibfield  {author} {\bibinfo {author} {\bibfnamefont {F.}~\bibnamefont
  {Gautier}}\ and\ \bibinfo {author} {\bibfnamefont {J.}~\bibnamefont
  {Serreau}},\ }\href {\doibase 10.1103/PhysRevD.86.125002} {\bibfield
  {journal} {\bibinfo  {journal} {Phys. Rev. D}\ }\textbf {\bibinfo {volume}
  {86}},\ \bibinfo {pages} {125002} (\bibinfo {year} {2012})}\BibitemShut
  {NoStop}%
\bibitem [{\citenamefont {Calzetta}\ and\ \citenamefont
  {Hu}(2008)}]{calzetta20081}%
  \BibitemOpen
  \bibfield  {author} {\bibinfo {author} {\bibfnamefont {E.~A.}\ \bibnamefont
  {Calzetta}}\ and\ \bibinfo {author} {\bibfnamefont {B.-L.}\ \bibnamefont
  {Hu}},\ }\href@noop {} {\emph {\bibinfo {title} {Nonequilibrium quantum field
  theory}}}\ (\bibinfo  {publisher} {Cambridge University Press},\ \bibinfo
  {year} {2008})\BibitemShut {NoStop}%
\bibitem [{\citenamefont {Anisimov}\ \emph {et~al.}(2009)\citenamefont
  {Anisimov}, \citenamefont {Buchm{\"{u}}ller}, \citenamefont {Drewes},\ and\
  \citenamefont {Mendizabal}}]{anisimov20091}%
  \BibitemOpen
  \bibfield  {author} {\bibinfo {author} {\bibfnamefont {A.}~\bibnamefont
  {Anisimov}}, \bibinfo {author} {\bibfnamefont {W.}~\bibnamefont
  {Buchm{\"{u}}ller}}, \bibinfo {author} {\bibfnamefont {M.}~\bibnamefont
  {Drewes}}, \ and\ \bibinfo {author} {\bibfnamefont {S.}~\bibnamefont
  {Mendizabal}},\ }\href {\doibase 10.1016/j.aop.2009.01.001} {\bibfield
  {journal} {\bibinfo  {journal} {Ann. Phys.}\ }\textbf {\bibinfo {volume}
  {324}},\ \bibinfo {pages} {1234} (\bibinfo {year} {2009})}\BibitemShut
  {NoStop}%
\bibitem [{\citenamefont {Boyanovsky}(2015)}]{boyanovsky20151}%
  \BibitemOpen
  \bibfield  {author} {\bibinfo {author} {\bibfnamefont {D.}~\bibnamefont
  {Boyanovsky}},\ }\href
  {http://iopscience.iop.org/article/10.1088/1367-2630/17/6/063017} {\bibfield
  {journal} {\bibinfo  {journal} {New J. Phys.}\ }\textbf {\bibinfo {volume}
  {17}} (\bibinfo {year} {2015})}\BibitemShut {NoStop}%
\bibitem [{\citenamefont {de~Vega}\ and\ \citenamefont
  {Alonso}(2017)}]{devega20171}%
  \BibitemOpen
  \bibfield  {author} {\bibinfo {author} {\bibfnamefont {I.}~\bibnamefont
  {de~Vega}}\ and\ \bibinfo {author} {\bibfnamefont {D.}~\bibnamefont
  {Alonso}},\ }\href {\doibase 10.1103/RevModPhys.89.015001} {\bibfield
  {journal} {\bibinfo  {journal} {Rev. Mod. Phys.}\ }\textbf {\bibinfo {volume}
  {89}},\ \bibinfo {pages} {015001} (\bibinfo {year} {2017})}\BibitemShut
  {NoStop}%
\bibitem [{\citenamefont {Breuer}\ and\ \citenamefont
  {Petruccione}()}]{breuer20021}%
  \BibitemOpen
  \bibfield  {author} {\bibinfo {author} {\bibfnamefont {H.-P.}\ \bibnamefont
  {Breuer}}\ and\ \bibinfo {author} {\bibfnamefont {F.}\ \bibfnamefont {Petruccione}},\ }\href@noop {} {\emph {\bibinfo
  {title} {The theory of open quantum systems}}}\BibitemShut {NoStop}%
\bibitem [{\citenamefont {Grabert}\ \emph {et~al.}(1988)\citenamefont
  {Grabert}, \citenamefont {Schramm},\ and\ \citenamefont
  {Ingold}}]{grabert19881}%
  \BibitemOpen
  \bibfield  {author} {\bibinfo {author} {\bibfnamefont {H.}~\bibnamefont
  {Grabert}}, \bibinfo {author} {\bibfnamefont {P.}~\bibnamefont {Schramm}}, \
  and\ \bibinfo {author} {\bibfnamefont {G.~L.}\ \bibnamefont {Ingold}},\
  }\href {\doibase 10.1016/0370-1573(88)90023-3} {\bibfield  {journal}
  {\bibinfo  {journal} {Physics Reports}\ }\textbf {\bibinfo {volume} {168}},\
  \bibinfo {pages} {115} (\bibinfo {year} {1988})}\BibitemShut {NoStop}%
\bibitem [{\citenamefont {Rivas}\ and\ \citenamefont
  {Huelga}(2012)}]{rivas20121}%
  \BibitemOpen
  \bibfield  {author} {\bibinfo {author} {\bibfnamefont {A.}~\bibnamefont
  {Rivas}}\ and\ \bibinfo {author} {\bibfnamefont {S.~F.}\ \bibnamefont
  {Huelga}},\ }\href@noop {} {\emph {\bibinfo {title} {Open quantum systems}}}\
  (\bibinfo  {publisher} {Springer},\ \bibinfo {year} {2012})\BibitemShut
  {NoStop}%
\bibitem [{\citenamefont {Caldeira}(2014)}]{caldeira20141}%
  \BibitemOpen
  \bibfield  {author} {\bibinfo {author} {\bibfnamefont {A.~O.}\ \bibnamefont
  {Caldeira}},\ }\href@noop {} {\emph {\bibinfo {title} {An Introduction to
  Macroscopic Quantum Phenomena and Quantum Dissipation}}}\ (\bibinfo
  {publisher} {Cambridge University Press},\ \bibinfo {year}
  {2014})\BibitemShut {NoStop}%
\bibitem [{\citenamefont {Grabert}\ \emph {et~al.}(1984)\citenamefont
  {Grabert}, \citenamefont {Weiss},\ and\ \citenamefont
  {Talkner}}]{grabert19841}%
  \BibitemOpen
  \bibfield  {author} {\bibinfo {author} {\bibfnamefont {H.}~\bibnamefont
  {Grabert}}, \bibinfo {author} {\bibfnamefont {U.}~\bibnamefont {Weiss}}, \
  and\ \bibinfo {author} {\bibfnamefont {P.}~\bibnamefont {Talkner}},\ }\href
  {\doibase 10.1007/BF01307505} {\bibfield  {journal} {\bibinfo  {journal} {Z.
  Phys. B}\ }\textbf {\bibinfo {volume} {55}},\ \bibinfo {pages} {87} (\bibinfo
  {year} {1984})}\BibitemShut {NoStop}%
\bibitem [{\citenamefont {Riseborough}\ \emph {et~al.}(1985)\citenamefont
  {Riseborough}, \citenamefont {H{\"{a}}nggi},\ and\ \citenamefont
  {Weiss}}]{riseborough19851}%
  \BibitemOpen
  \bibfield  {author} {\bibinfo {author} {\bibfnamefont {P.~S.}\ \bibnamefont
  {Riseborough}}, \bibinfo {author} {\bibfnamefont {P.}~\bibnamefont
  {H{\"{a}}nggi}}, \ and\ \bibinfo {author} {\bibfnamefont {U.}~\bibnamefont
  {Weiss}},\ }\href {\doibase 10.1103/PhysRevA.31.471} {\bibfield  {journal}
  {\bibinfo  {journal} {Phys. Rev. A}\ }\textbf {\bibinfo {volume} {31}},\
  \bibinfo {pages} {471} (\bibinfo {year} {1985})}\BibitemShut {NoStop}%
\bibitem [{\citenamefont {Haake}\ and\ \citenamefont
  {Reibold}(1985)}]{haake19851}%
  \BibitemOpen
  \bibfield  {author} {\bibinfo {author} {\bibfnamefont {F.}~\bibnamefont
  {Haake}}\ and\ \bibinfo {author} {\bibfnamefont {R.}~\bibnamefont
  {Reibold}},\ }\href {\doibase 10.1103/PhysRevA.32.2462} {\bibfield  {journal}
  {\bibinfo  {journal} {Phys. Rev. A}\ }\textbf {\bibinfo {volume} {32}},\
  \bibinfo {pages} {2462} (\bibinfo {year} {1985})}\BibitemShut {NoStop}%
\bibitem [{\citenamefont {Boyanovsky}\ and\ \citenamefont
  {Jasnow}(2017{\natexlab{a}})}]{boyanovsky20171}%
  \BibitemOpen
  \bibfield  {author} {\bibinfo {author} {\bibfnamefont {D.}~\bibnamefont
  {Boyanovsky}}\ and\ \bibinfo {author} {\bibfnamefont {D.}~\bibnamefont
  {Jasnow}},\ }\href {\doibase 10.1103/PhysRevA.96.062108} {\bibfield
  {journal} {\bibinfo  {journal} {Phys. Rev. A}\ }\textbf {\bibinfo {volume}
  {96}},\ \bibinfo {pages} {062108} (\bibinfo {year}
  {2017}{\natexlab{a}})}\BibitemShut {NoStop}%
\bibitem [{\citenamefont {Hu}\ \emph {et~al.}(1992)\citenamefont {Hu},
  \citenamefont {Paz},\ and\ \citenamefont {Zhang}}]{hu19921}%
  \BibitemOpen
  \bibfield  {author} {\bibinfo {author} {\bibfnamefont {B.~L.}\ \bibnamefont
  {Hu}}, \bibinfo {author} {\bibfnamefont {J.~P.}\ \bibnamefont {Paz}}, \ and\
  \bibinfo {author} {\bibfnamefont {Y.}~\bibnamefont {Zhang}},\ }\href
  {\doibase 10.1103/PhysRevD.45.2843} {\bibfield  {journal} {\bibinfo
  {journal} {Phys. Rev. D}\ }\textbf {\bibinfo {volume} {45}},\ \bibinfo
  {pages} {2843} (\bibinfo {year} {1992})}\BibitemShut {NoStop}%
\bibitem [{\citenamefont {Alamoudi}\ \emph {et~al.}(1998)\citenamefont
  {Alamoudi}, \citenamefont {Boyanovsky}, \citenamefont {Vega},\ and\
  \citenamefont {Holman}}]{alamoudi19981}%
  \BibitemOpen
  \bibfield  {author} {\bibinfo {author} {\bibfnamefont {S.~M.}\ \bibnamefont
  {Alamoudi}}, \bibinfo {author} {\bibfnamefont {D.}~\bibnamefont
  {Boyanovsky}}, \bibinfo {author} {\bibfnamefont {H.~J.~D.}\ \bibnamefont
  {Vega}}, \ and\ \bibinfo {author} {\bibfnamefont {R.}~\bibnamefont
  {Holman}},\ }\href {\doibase 10.1103/PhysRevD.59.025003} {\bibfield
  {journal} {\bibinfo  {journal} {Phys. Rev. D}\ }\textbf {\bibinfo {volume}
  {59}},\ \bibinfo {pages} {025003} (\bibinfo {year} {1998})}\BibitemShut
  {NoStop}%
\bibitem [{\citenamefont {Alamoudi}\ \emph {et~al.}(1999)\citenamefont
  {Alamoudi}, \citenamefont {Boyanovsky},\ and\ \citenamefont
  {Vega}}]{alamoudi19991}%
  \BibitemOpen
  \bibfield  {author} {\bibinfo {author} {\bibfnamefont {S.~M.}\ \bibnamefont
  {Alamoudi}}, \bibinfo {author} {\bibfnamefont {D.}~\bibnamefont
  {Boyanovsky}}, \ and\ \bibinfo {author} {\bibfnamefont {H.~J.~D.}\
  \bibnamefont {Vega}},\ }\href {\doibase 10.1103/PhysRevE.60.94} {\bibfield
  {journal} {\bibinfo  {journal} {Phys. Rev. E}\ }\textbf {\bibinfo {volume}
  {60}},\ \bibinfo {pages} {94} (\bibinfo {year} {1999})}\BibitemShut {NoStop}%
\bibitem [{\citenamefont {Ford}\ \emph {et~al.}(1988)\citenamefont {Ford},
  \citenamefont {Lewis},\ and\ \citenamefont {O'Connell}}]{ford19881}%
  \BibitemOpen
  \bibfield  {author} {\bibinfo {author} {\bibfnamefont {W.}~\bibnamefont
  {Ford}}, \bibinfo {author} {\bibfnamefont {J.}~\bibnamefont {Lewis}}, \ and\
  \bibinfo {author} {\bibfnamefont {R.}~\bibnamefont {O'Connell}},\ }\href
  {\doibase 10.1103/PhysRevA.37.4419} {\bibfield  {journal} {\bibinfo
  {journal} {Phys. Rev. A}\ }\textbf {\bibinfo {volume} {37}},\ \bibinfo
  {pages} {4419} (\bibinfo {year} {1988})}\BibitemShut {NoStop}%
\bibitem [{\citenamefont {H{\"{a}}nggi}\ and\ \citenamefont
  {Ingold}(2005)}]{hanggi20051}%
  \BibitemOpen
  \bibfield  {author} {\bibinfo {author} {\bibfnamefont {P.}~\bibnamefont
  {H{\"{a}}nggi}}\ and\ \bibinfo {author} {\bibfnamefont {G.-l.}\ \bibnamefont
  {Ingold}},\ }\href {\doibase 10.1063/1.1853631} {\bibfield  {journal}
  {\bibinfo  {journal} {Chaos}\ }\textbf {\bibinfo {volume} {15}},\ \bibinfo
  {pages} {026105} (\bibinfo {year} {2005})}\BibitemShut {NoStop}%
\bibitem [{\citenamefont {Philbin}(2012)}]{philbin20121}%
  \BibitemOpen
  \bibfield  {author} {\bibinfo {author} {\bibfnamefont {T.~G.}\ \bibnamefont
  {Philbin}},\ }\href {\doibase 10.1088/1367-2630/14/8/083043} {\bibfield
  {journal} {\bibinfo  {journal} {New J. Phys.}\ }\textbf {\bibinfo {volume}
  {14}},\ \bibinfo {pages} {083043} (\bibinfo {year} {2012})}\BibitemShut
  {NoStop}%
\bibitem [{\citenamefont {Feynman}\ and\ \citenamefont
  {Vernon~Jr}(1963)}]{feynman19631}%
  \BibitemOpen
  \bibfield  {author} {\bibinfo {author} {\bibfnamefont {R.~P.}\ \bibnamefont
  {Feynman}}\ and\ \bibinfo {author} {\bibfnamefont {F.}~\bibnamefont
  {Vernon~Jr}},\ }\href@noop {} {\bibfield  {journal} {\bibinfo  {journal}
  {Ann. Phys.}\ }\textbf {\bibinfo {volume} {24}},\ \bibinfo {pages} {118}
  (\bibinfo {year} {1963})}\BibitemShut {NoStop}%
\bibitem [{\citenamefont {Caldeira}\ and\ \citenamefont
  {Leggett}(1983)}]{caldeira19831}%
  \BibitemOpen
  \bibfield  {author} {\bibinfo {author} {\bibfnamefont {A.}~\bibnamefont
  {Caldeira}}\ and\ \bibinfo {author} {\bibfnamefont {A.~J.}\ \bibnamefont
  {Leggett}},\ }\href@noop {} {\bibfield  {journal} {\bibinfo  {journal}
  {Annals of physics}\ }\textbf {\bibinfo {volume} {149}},\ \bibinfo {pages}
  {374} (\bibinfo {year} {1983})}\BibitemShut {NoStop}%
\bibitem [{\citenamefont {G.~W.~Ford}\ and\ \citenamefont
  {Mazur}(1965)}]{ford19651}%
  \BibitemOpen
  \bibfield  {author} {\bibinfo {author} {\bibfnamefont {M.~K.}\ \bibnamefont
  {G.~W.~Ford}}\ and\ \bibinfo {author} {\bibfnamefont {P.}~\bibnamefont
  {Mazur}},\ }\href {\doibase 10.1063/1.1704304} {\bibfield  {journal}
  {\bibinfo  {journal} {J. Math. Phys.}\ }\textbf {\bibinfo {volume} {6}},\
  \bibinfo {pages} {504} (\bibinfo {year} {1965})}\BibitemShut {NoStop}%
\bibitem [{\citenamefont {Correa}\ \emph {et~al.}(2015)\citenamefont {Correa},
  \citenamefont {Mehboudi}, \citenamefont {Adesso},\ and\ \citenamefont
  {Sanpera}}]{correa20151}%
  \BibitemOpen
  \bibfield  {author} {\bibinfo {author} {\bibfnamefont {L.~A.}\ \bibnamefont
  {Correa}}, \bibinfo {author} {\bibfnamefont {M.}~\bibnamefont {Mehboudi}},
  \bibinfo {author} {\bibfnamefont {G.}~\bibnamefont {Adesso}}, \ and\ \bibinfo
  {author} {\bibfnamefont {A.}~\bibnamefont {Sanpera}},\ }\href {\doibase
  10.1103/PhysRevLett.114.220405} {\bibfield  {journal} {\bibinfo  {journal}
  {Phys. Rev. Lett.}\ }\textbf {\bibinfo {volume} {114}},\ \bibinfo {pages}
  {220405} (\bibinfo {year} {2015})}\BibitemShut {NoStop}%
\bibitem [{\citenamefont {Hovhannisyan}\ and\ \citenamefont
  {Correa}(2018)}]{hovhannisyan20181}%
  \BibitemOpen
  \bibfield  {author} {\bibinfo {author} {\bibfnamefont {K.~V.}\ \bibnamefont
  {Hovhannisyan}}\ and\ \bibinfo {author} {\bibfnamefont {L.~A.}\ \bibnamefont
  {Correa}},\ }\href {\doibase 10.1103/PhysRevB.98.045101} {\bibfield
  {journal} {\bibinfo  {journal} {Phys. Rev. B}\ }\textbf {\bibinfo {volume}
  {98}},\ \bibinfo {pages} {045101} (\bibinfo {year} {2018})}\BibitemShut
  {NoStop}%
\bibitem [{\citenamefont {Correa}\ \emph {et~al.}(2017)\citenamefont {Correa},
  \citenamefont {Perarnau-Llobet}, \citenamefont {Hovhannisyan}, \citenamefont
  {Hern\'andez-Santana}, \citenamefont {Mehboudi},\ and\ \citenamefont
  {Sanpera}}]{correa20171}%
  \BibitemOpen
  \bibfield  {author} {\bibinfo {author} {\bibfnamefont {L.~A.}\ \bibnamefont
  {Correa}}, \bibinfo {author} {\bibfnamefont {M.}~\bibnamefont
  {Perarnau-Llobet}}, \bibinfo {author} {\bibfnamefont {K.~V.}\ \bibnamefont
  {Hovhannisyan}}, \bibinfo {author} {\bibfnamefont {S.}~\bibnamefont
  {Hern\'andez-Santana}}, \bibinfo {author} {\bibfnamefont {M.}~\bibnamefont
  {Mehboudi}}, \ and\ \bibinfo {author} {\bibfnamefont {A.}~\bibnamefont
  {Sanpera}},\ }\href {\doibase 10.1103/PhysRevA.96.062103} {\bibfield
  {journal} {\bibinfo  {journal} {Phys. Rev. A}\ }\textbf {\bibinfo {volume}
  {96}},\ \bibinfo {pages} {062103} (\bibinfo {year} {2017})}\BibitemShut
  {NoStop}%
\bibitem [{\citenamefont {Valido}\ \emph {et~al.}(2015)\citenamefont {Valido},
  \citenamefont {Ruiz},\ and\ \citenamefont {Alonso}}]{valido20151}%
  \BibitemOpen
  \bibfield  {author} {\bibinfo {author} {\bibfnamefont {A.}~\bibnamefont
  {Valido}}, \bibinfo {author} {\bibfnamefont {A.}~\bibnamefont {Ruiz}}, \ and\
  \bibinfo {author} {\bibfnamefont {D.}~\bibnamefont {Alonso}},\ }\href
  {\doibase 10.1103/PhysRevE.91.062123} {\bibfield  {journal} {\bibinfo
  {journal} {Phys. Rev. E}\ }\textbf {\bibinfo {volume} {91}} (\bibinfo {year}
  {2015}),\ 10.1103/PhysRevE.91.062123}\BibitemShut {NoStop}%
\bibitem [{\citenamefont {Valido}\ \emph
  {et~al.}(2013{\natexlab{a}})\citenamefont {Valido}, \citenamefont {Correa},\
  and\ \citenamefont {Alonso}}]{valido20132}%
  \BibitemOpen
  \bibfield  {author} {\bibinfo {author} {\bibfnamefont {A.~A.}\ \bibnamefont
  {Valido}}, \bibinfo {author} {\bibfnamefont {L.~A.}\ \bibnamefont {Correa}},
  \ and\ \bibinfo {author} {\bibfnamefont {D.}~\bibnamefont {Alonso}},\ }\href
  {\doibase 10.1103/PhysRevA.88.012309} {\bibfield  {journal} {\bibinfo
  {journal} {Phys. Rev. A}\ }\textbf {\bibinfo {volume} {88}},\ \bibinfo
  {pages} {012309} (\bibinfo {year} {2013}{\natexlab{a}})}\BibitemShut
  {NoStop}%
\bibitem [{\citenamefont {Boyanovsky}\ and\ \citenamefont
  {Jasnow}(2017{\natexlab{b}})}]{boyanovsky20172}%
  \BibitemOpen
  \bibfield  {author} {\bibinfo {author} {\bibfnamefont {D.}~\bibnamefont
  {Boyanovsky}}\ and\ \bibinfo {author} {\bibfnamefont {D.}~\bibnamefont
  {Jasnow}},\ }\href {\doibase 10.1103/PhysRevA.96.012103} {\bibfield
  {journal} {\bibinfo  {journal} {Phys. Rev. A}\ }\textbf {\bibinfo {volume}
  {96}},\ \bibinfo {pages} {012103} (\bibinfo {year}
  {2017}{\natexlab{b}})}\BibitemShut {NoStop}%
\bibitem [{\citenamefont {Hsiang}\ \emph {et~al.}(2018)\citenamefont {Hsiang},
  \citenamefont {Chou}, \citenamefont {Subasi},\ and\ \citenamefont
  {Hu}}]{hsiang20181}%
  \BibitemOpen
  \bibfield  {author} {\bibinfo {author} {\bibfnamefont {J.}~\bibnamefont
  {Hsiang}}, \bibinfo {author} {\bibfnamefont {C.}~\bibnamefont {Chou}},
  \bibinfo {author} {\bibfnamefont {Y.}~\bibnamefont {Subasi}}, \ and\ \bibinfo
  {author} {\bibfnamefont {B.}~\bibnamefont {Hu}},\ }\href {\doibase
  10.1103/PhysRevE.97.012135} {\bibfield  {journal} {\bibinfo  {journal} {Phys.
  Rev. E}\ }\textbf {\bibinfo {volume} {97}},\ \bibinfo {pages} {012135}
  (\bibinfo {year} {2018})}\BibitemShut {NoStop}%
\bibitem [{\citenamefont {Charalambous}\ \emph {et~al.}()\citenamefont
  {Charalambous}, \citenamefont {Garc\'ia-March}, \citenamefont {Lampo},
  \citenamefont {Mehboudi},\ and\ \citenamefont
  {Lewenstein}}]{charalambous20181}%
  \BibitemOpen
  \bibfield  {author} {\bibinfo {author} {\bibfnamefont {C.}~\bibnamefont
  {Charalambous}}, \bibinfo {author} {\bibfnamefont {M.}~\bibnamefont
  {Garc\'ia-March}}, \bibinfo {author} {\bibfnamefont {A.}~\bibnamefont
  {Lampo}}, \bibinfo {author} {\bibfnamefont {M.}~\bibnamefont {Mehboudi}}, \
  and\ \bibinfo {author} {\bibfnamefont {M.}~\bibnamefont {Lewenstein}},\
  }\href@noop {} {\bibinfo  {journal} {arXiv:1805.00709}\ }\BibitemShut
  {NoStop}%
\bibitem [{\citenamefont {Venkataraman}\ \emph {et~al.}(2014)\citenamefont
  {Venkataraman}, \citenamefont {Plato}, \citenamefont {Tufarelli},\ and\
  \citenamefont {Kim}}]{venkataraman20141}%
  \BibitemOpen
\bibfield  {journal} {  }\bibfield  {author} {\bibinfo {author} {\bibfnamefont
  {V.}~\bibnamefont {Venkataraman}}, \bibinfo {author} {\bibfnamefont
  {A.~D.~K.}\ \bibnamefont {Plato}}, \bibinfo {author} {\bibfnamefont
  {T.}~\bibnamefont {Tufarelli}}, \ and\ \bibinfo {author} {\bibfnamefont
  {M.~S.}\ \bibnamefont {Kim}},\ }\href
  {http://stacks.iop.org/0953-4075/47/i=1/a=015501} {\bibfield  {journal}
  {\bibinfo  {journal} {J. Phys. B: At., Mol. and Opt. Phys.}\ }\textbf
  {\bibinfo {volume} {47}},\ \bibinfo {pages} {015501} (\bibinfo {year}
  {2014})}\BibitemShut {NoStop}%
\bibitem [{\citenamefont {Ruggenthaler}\ \emph {et~al.}(2018)\citenamefont
  {Ruggenthaler}, \citenamefont {Tancogne-Dejean}, \citenamefont {Flick},
  \citenamefont {Appel},\ and\ \citenamefont {Rubio}}]{ruggenthaler20181}%
  \BibitemOpen
  \bibfield  {author} {\bibinfo {author} {\bibfnamefont {M.}~\bibnamefont
  {Ruggenthaler}}, \bibinfo {author} {\bibfnamefont {N.}~\bibnamefont
  {Tancogne-Dejean}}, \bibinfo {author} {\bibfnamefont {J.}~\bibnamefont
  {Flick}}, \bibinfo {author} {\bibfnamefont {H.}~\bibnamefont {Appel}}, \ and\
  \bibinfo {author} {\bibfnamefont {A.}~\bibnamefont {Rubio}},\ }\href
  {\doibase 10.1038/s41570-018-0118} {\bibfield  {journal} {\bibinfo  {journal}
  {Nat. Rev. Chem.}\ }\textbf {\bibinfo {volume} {2}},\ \bibinfo {pages} {0118}
  (\bibinfo {year} {2018})}\BibitemShut {NoStop}%
\bibitem [{\citenamefont {Rokaj}\ \emph {et~al.}(2018)\citenamefont {Rokaj},
  \citenamefont {Welakuh}, \citenamefont {Ruggenthaler},\ and\ \citenamefont
  {Rubio}}]{rokaj20181}%
  \BibitemOpen
  \bibfield  {author} {\bibinfo {author} {\bibfnamefont {V.}~\bibnamefont
  {Rokaj}}, \bibinfo {author} {\bibfnamefont {D.~M.}\ \bibnamefont {Welakuh}},
  \bibinfo {author} {\bibfnamefont {M.}~\bibnamefont {Ruggenthaler}}, \ and\
  \bibinfo {author} {\bibfnamefont {A.}~\bibnamefont {Rubio}},\ }\href
  {\doibase 10.1088/1361-6455/aa9c99} {\bibfield  {journal} {\bibinfo
  {journal} {J. Phys. B: At. Mol. Opt. Phys.}\ }\textbf {\bibinfo {volume}
  {51}},\ \bibinfo {pages} {034005} (\bibinfo {year} {2018})}\BibitemShut
  {NoStop}%
\bibitem [{\citenamefont {Kohler}\ and\ \citenamefont
  {Sols}(2013)}]{kohler20131}%
  \BibitemOpen
  \bibfield  {author} {\bibinfo {author} {\bibfnamefont {H.}~\bibnamefont
  {Kohler}}\ and\ \bibinfo {author} {\bibfnamefont {F.}~\bibnamefont {Sols}},\
  }\href {\doibase 10.1016/j.physa.2013.01.019} {\bibfield  {journal} {\bibinfo
   {journal} {Physica A}\ }\textbf {\bibinfo {volume} {392}},\ \bibinfo {pages}
  {1989} (\bibinfo {year} {2013})}\BibitemShut {NoStop}%
\bibitem [{\citenamefont {Valido}\ \emph
  {et~al.}(2013{\natexlab{b}})\citenamefont {Valido}, \citenamefont {Alonso},\
  and\ \citenamefont {Kohler}}]{valido20131}%
  \BibitemOpen
  \bibfield  {author} {\bibinfo {author} {\bibfnamefont {A.~A.}\ \bibnamefont
  {Valido}}, \bibinfo {author} {\bibfnamefont {D.}~\bibnamefont {Alonso}}, \
  and\ \bibinfo {author} {\bibfnamefont {S.}~\bibnamefont {Kohler}},\ }\href
  {\doibase 10.1103/PhysRevA.88.042303} {\bibfield  {journal} {\bibinfo
  {journal} {Phys. Rev. A}\ }\textbf {\bibinfo {volume} {88}} (\bibinfo {year}
  {2013}{\natexlab{b}}),\ 10.1103/PhysRevA.88.042303}\BibitemShut {NoStop}%
\bibitem [{\citenamefont {Rzazewski}\ and\ \citenamefont
  {Zakowicz}(1976)}]{rzazewski19761}%
  \BibitemOpen
  \bibfield  {author} {\bibinfo {author} {\bibfnamefont {K.}~\bibnamefont
  {Rzazewski}}\ and\ \bibinfo {author} {\bibfnamefont {W.}~\bibnamefont
  {Zakowicz}},\ }\href@noop {} {\bibfield  {journal} {\bibinfo  {journal}
  {J. Phys. A: Math. Gen,}\ }\textbf {\bibinfo {volume}
  {9}},\ \bibinfo {pages} {1159} (\bibinfo {year} {1976})}\BibitemShut
  {NoStop}%
\bibitem [{\citenamefont {Efimov}\ and\ \citenamefont
  {Vonwaldenfels}(1994)}]{efimov19941}%
  \BibitemOpen
  \bibfield  {author} {\bibinfo {author} {\bibfnamefont {G.}~\bibnamefont
  {Efimov}}\ and\ \bibinfo {author} {\bibfnamefont {W.}~\bibnamefont
  {Vonwaldenfels}},\ }\href {\doibase 10.1006/aphy.1994.1065} {\bibfield
  {journal} {\bibinfo  {journal} {Annals of Physics}\ }\textbf {\bibinfo
  {volume} {233}},\ \bibinfo {pages} {182} (\bibinfo {year}
  {1994})}\BibitemShut {NoStop}%
\bibitem [{\citenamefont {Reuther}\ \emph {et~al.}(2010)\citenamefont
  {Reuther}, \citenamefont {Zueco}, \citenamefont {Deppe}, \citenamefont
  {Hoffmann}, \citenamefont {Menzel}, \citenamefont {Wei\ss{}l}, \citenamefont
  {Mariantoni}, \citenamefont {Kohler}, \citenamefont {Marx}, \citenamefont
  {Solano}, \citenamefont {Gross},\ and\ \citenamefont
  {H\"anggi}}]{reuther20101}%
  \BibitemOpen
  \bibfield  {author} {\bibinfo {author} {\bibfnamefont {G.~M.}\ \bibnamefont
  {Reuther}}, \bibinfo {author} {\bibfnamefont {D.}~\bibnamefont {Zueco}},
  \bibinfo {author} {\bibfnamefont {F.}~\bibnamefont {Deppe}}, \bibinfo
  {author} {\bibfnamefont {E.}~\bibnamefont {Hoffmann}}, \bibinfo {author}
  {\bibfnamefont {E.~P.}\ \bibnamefont {Menzel}}, \bibinfo {author}
  {\bibfnamefont {T.}~\bibnamefont {Wei\ss{}l}}, \bibinfo {author}
  {\bibfnamefont {M.}~\bibnamefont {Mariantoni}}, \bibinfo {author}
  {\bibfnamefont {S.}~\bibnamefont {Kohler}}, \bibinfo {author} {\bibfnamefont
  {A.}~\bibnamefont {Marx}}, \bibinfo {author} {\bibfnamefont {E.}~\bibnamefont
  {Solano}}, \bibinfo {author} {\bibfnamefont {R.}~\bibnamefont {Gross}}, \
  and\ \bibinfo {author} {\bibfnamefont {P.}~\bibnamefont {H\"anggi}},\ }\href
  {\doibase 10.1103/PhysRevB.81.144510} {\bibfield  {journal} {\bibinfo
  {journal} {Phys. Rev. B}\ }\textbf {\bibinfo {volume} {81}},\ \bibinfo
  {pages} {144510} (\bibinfo {year} {2010})}\BibitemShut {NoStop}%
\bibitem [{\citenamefont {Cohen-Tannoudji}\ \emph {et~al.}(1997)\citenamefont
  {Cohen-Tannoudji}, \citenamefont {Dupont-Roc},\ and\ \citenamefont
  {Grynberg}}]{cohen19971}%
  \BibitemOpen
  \bibfield  {author} {\bibinfo {author} {\bibfnamefont {C.}~\bibnamefont
  {Cohen-Tannoudji}}, \bibinfo {author} {\bibfnamefont {J.}~\bibnamefont
  {Dupont-Roc}}, \ and\ \bibinfo {author} {\bibfnamefont {G.}~\bibnamefont
  {Grynberg}},\ }\href@noop {} {\emph {\bibinfo {title} {Photons and
  Atoms-Introduction to Quantum Electrodynamics}}}\ (\bibinfo  {publisher}
  {Wiley-VCH},\ \bibinfo {year} {1997})\BibitemShut {NoStop}%
\bibitem [{\citenamefont {Landau}\ and\ \citenamefont
  {Lifshitz}(1971)}]{landau19711}%
  \BibitemOpen
  \bibfield  {author} {\bibinfo {author} {\bibfnamefont {L.~D.}\ \bibnamefont
  {Landau}}\ and\ \bibinfo {author} {\bibfnamefont {E.~M.}\ \bibnamefont
  {Lifshitz}},\ }\href@noop {} {\emph {\bibinfo {title} {The classical theory
  of fields}}}\ (\bibinfo  {publisher} {Pergamon},\ \bibinfo {year}
  {1971})\BibitemShut {NoStop}%
\bibitem [{\citenamefont {Jackiw}(1990)}]{jackiw19902}%
  \BibitemOpen
  \bibfield  {author} {\bibinfo {author} {\bibfnamefont {R.}~\bibnamefont
  {Jackiw}},\ }\href {\doibase 10.1016/0920-5632(90)90648-E} {\bibfield
  {journal} {\bibinfo  {journal} {Nucl. Phys. B}\ }\textbf {\bibinfo {volume}
  {18A}},\ \bibinfo {pages} {107} (\bibinfo {year} {1990})}\BibitemShut
  {NoStop}%
\bibitem [{\citenamefont {Dunne}(1999)}]{dunne19991}%
  \BibitemOpen
  \bibfield  {author} {\bibinfo {author} {\bibfnamefont {G.~V.}\ \bibnamefont
  {Dunne}},\ }in\ \href@noop {} {\emph {\bibinfo {booktitle} {Aspects
  topologiques de la physique en basse dimension. Topological aspects of low
  dimensional systems}}}\ (\bibinfo  {publisher} {Springer},\ \bibinfo {year}
  {1999})\ pp.\ \bibinfo {pages} {177--263}\BibitemShut {NoStop}%
\bibitem [{\citenamefont {Deser}\ \emph {et~al.}(1982)\citenamefont {Deser},
  \citenamefont {Jackiw},\ and\ \citenamefont {Templeton}}]{deser19821}%
  \BibitemOpen
  \bibfield  {author} {\bibinfo {author} {\bibfnamefont {S.}~\bibnamefont
  {Deser}}, \bibinfo {author} {\bibfnamefont {R.}~\bibnamefont {Jackiw}}, \
  and\ \bibinfo {author} {\bibfnamefont {S.}~\bibnamefont {Templeton}},\ }\href
  {\doibase 10.1016/0003-4916(82)90164-6} {\bibfield  {journal} {\bibinfo
  {journal} {Ann. Phys.}\ }\textbf {\bibinfo {volume} {140}},\ \bibinfo {pages}
  {372} (\bibinfo {year} {1982})}\BibitemShut {NoStop}%
\bibitem [{\citenamefont {Matsuyama}(1990)}]{matsuyama19901}%
  \BibitemOpen
  \bibfield  {author} {\bibinfo {author} {\bibfnamefont {T.}~\bibnamefont
  {Matsuyama}},\ }\href {\doibase 10.1103/PhysRevD.42.3469} {\bibfield
  {journal} {\bibinfo  {journal} {Phys. Rev. D}\ }\textbf {\bibinfo {volume}
  {42}},\ \bibinfo {pages} {3469} (\bibinfo {year} {1990})}\BibitemShut
  {NoStop}%
\bibitem [{\citenamefont {Dunne}\ \emph {et~al.}(1990)\citenamefont {Dunne},
  \citenamefont {{R. Jackiw}},\ and\ \citenamefont
  {Trugenberger}}]{dunne19901}%
  \BibitemOpen
  \bibfield  {author} {\bibinfo {author} {\bibfnamefont {G.~V.}\ \bibnamefont
  {Dunne}}, \bibinfo {author} {\bibnamefont {{R. Jackiw}}}, \ and\ \bibinfo
  {author} {\bibfnamefont {C.~A.}\ \bibnamefont {Trugenberger}},\ }\href
  {\doibase 10.1103/PhysRevD.41.661} {\bibfield  {journal} {\bibinfo  {journal}
  {Phys. Rev. D}\ }\textbf {\bibinfo {volume} {41}},\ \bibinfo {pages} {661}
  (\bibinfo {year} {1990})}\BibitemShut {NoStop}%
\bibitem [{\citenamefont {Horvathy}\ and\ \citenamefont
  {Zhang}(2009)}]{horvathy20091}%
  \BibitemOpen
  \bibfield  {author} {\bibinfo {author} {\bibfnamefont {P.~A.}\ \bibnamefont
  {Horvathy}}\ and\ \bibinfo {author} {\bibfnamefont {P.}~\bibnamefont
  {Zhang}},\ }\href {\doibase 10.1016/j.physrep.2009.07.003} {\bibfield
  {journal} {\bibinfo  {journal} {Physics Reports}\ }\textbf {\bibinfo {volume}
  {481}},\ \bibinfo {pages} {83} (\bibinfo {year} {2009})}\BibitemShut
  {NoStop}%
\bibitem [{\citenamefont {Pachos}(2012)}]{pachos20121}%
  \BibitemOpen
  \bibfield  {author} {\bibinfo {author} {\bibfnamefont {J.~K.}\ \bibnamefont
  {Pachos}},\ }\href@noop {} {\emph {\bibinfo {title} {Introduction to
  topological quantum computation}}}\ (\bibinfo  {publisher} {Cambridge
  University Press},\ \bibinfo {year} {2012})\BibitemShut {NoStop}%
\bibitem [{\citenamefont {Campisi}\ \emph {et~al.}(2012)\citenamefont
  {Campisi}, \citenamefont {Denisov},\ and\ \citenamefont
  {H\"anggi}}]{campisi20121}%
  \BibitemOpen
  \bibfield  {author} {\bibinfo {author} {\bibfnamefont {M.}~\bibnamefont
  {Campisi}}, \bibinfo {author} {\bibfnamefont {S.}~\bibnamefont {Denisov}}, \
  and\ \bibinfo {author} {\bibfnamefont {P.}~\bibnamefont {H\"anggi}},\ }\href
  {\doibase 10.1103/PhysRevA.86.032114} {\bibfield  {journal} {\bibinfo
  {journal} {Phys. Rev. A}\ }\textbf {\bibinfo {volume} {86}},\ \bibinfo
  {pages} {032114} (\bibinfo {year} {2012})}\BibitemShut {NoStop}%
\bibitem [{\citenamefont {Guo}\ and\ \citenamefont {Poletti}(2016)}]{guo20161}%
  \BibitemOpen
  \bibfield  {author} {\bibinfo {author} {\bibfnamefont {C.}~\bibnamefont
  {Guo}}\ and\ \bibinfo {author} {\bibfnamefont {D.}~\bibnamefont {Poletti}},\
  }\href {\doibase 10.1103/PhysRevA.94.033610} {\bibfield  {journal} {\bibinfo
  {journal} {Phys. Rev. A}\ }\textbf {\bibinfo {volume} {94}},\ \bibinfo
  {pages} {033610} (\bibinfo {year} {2016})}\BibitemShut {NoStop}%
\bibitem [{\citenamefont {Yao}\ \emph {et~al.}(2017)\citenamefont {Yao},
  \citenamefont {Tang},\ and\ \citenamefont {Ao}}]{yao20171}%
  \BibitemOpen
  \bibfield  {author} {\bibinfo {author} {\bibfnamefont {Y.}~\bibnamefont
  {Yao}}, \bibinfo {author} {\bibfnamefont {Y.}~\bibnamefont {Tang}}, \ and\
  \bibinfo {author} {\bibfnamefont {P.}~\bibnamefont {Ao}},\ }\href {\doibase
  10.1103/PhysRevB.96.134414} {\bibfield  {journal} {\bibinfo  {journal}
  {Physical Review B}\ }\textbf {\bibinfo {volume} {96}} (\bibinfo {year}
  {2017}),\ 10.1103/PhysRevB.96.134414}\BibitemShut {NoStop}%
\bibitem [{\citenamefont {Callan}\ and\ \citenamefont
  {Freed}(1992)}]{callan19921}%
  \BibitemOpen
  \bibfield  {author} {\bibinfo {author} {\bibfnamefont {C.~G.}\ \bibnamefont
  {Callan}}\ and\ \bibinfo {author} {\bibfnamefont {D.}~\bibnamefont {Freed}},\
  }\href {\doibase 10.1016/0550-3213(92)90400-6} {\bibfield  {journal}
  {\bibinfo  {journal} {Nucl. Phys. B}\ }\textbf {\bibinfo {volume} {374}},\
  \bibinfo {pages} {543} (\bibinfo {year} {1992})}\BibitemShut {NoStop}%
\bibitem [{\citenamefont {Novais}\ \emph {et~al.}(2005)\citenamefont {Novais},
  \citenamefont {Guinea},\ and\ \citenamefont {{Castro Neto}}}]{novais20051}%
  \BibitemOpen
  \bibfield  {author} {\bibinfo {author} {\bibfnamefont {E.}~\bibnamefont
  {Novais}}, \bibinfo {author} {\bibfnamefont {F.}~\bibnamefont {Guinea}}, \
  and\ \bibinfo {author} {\bibfnamefont {A.~H.}\ \bibnamefont {{Castro
  Neto}}},\ }\href {\doibase 10.1103/PhysRevLett.94.170401} {\bibfield
  {journal} {\bibinfo  {journal} {Phys. Rev. Lett.}\ }\textbf {\bibinfo
  {volume} {94}},\ \bibinfo {pages} {170401} (\bibinfo {year}
  {2005})}\BibitemShut {NoStop}%
\bibitem [{\citenamefont {Anton}\ \emph {et~al.}(2013)\citenamefont {Anton},
  \citenamefont {Birenbaum}, \citenamefont {Bolkhovsky}, \citenamefont {Braje},
  \citenamefont {Fitch}, \citenamefont {Neeley}, \citenamefont {Hilton},
  \citenamefont {Irwin}, \citenamefont {Wellstood}, \citenamefont {Oliver},
  \citenamefont {Shnirman},\ and\ \citenamefont {Clarke}}]{anton20131}%
  \BibitemOpen
  \bibfield  {author} {\bibinfo {author} {\bibfnamefont {S.~M.}\ \bibnamefont
  {Anton}}, \bibinfo {author} {\bibfnamefont {J.~S.}\ \bibnamefont
  {Birenbaum}}, \bibinfo {author} {\bibfnamefont {V.}~\bibnamefont
  {Bolkhovsky}}, \bibinfo {author} {\bibfnamefont {D.~A.}\ \bibnamefont
  {Braje}}, \bibinfo {author} {\bibfnamefont {G.}~\bibnamefont {Fitch}},
  \bibinfo {author} {\bibfnamefont {M.}~\bibnamefont {Neeley}}, \bibinfo
  {author} {\bibfnamefont {G.~C.}\ \bibnamefont {Hilton}}, \bibinfo {author}
  {\bibfnamefont {K.~D.}\ \bibnamefont {Irwin}}, \bibinfo {author}
  {\bibfnamefont {F.~C.}\ \bibnamefont {Wellstood}}, \bibinfo {author}
  {\bibfnamefont {W.~D.}\ \bibnamefont {Oliver}}, \bibinfo {author}
  {\bibfnamefont {A.}~\bibnamefont {Shnirman}}, \ and\ \bibinfo {author}
  {\bibfnamefont {J.}~\bibnamefont {Clarke}},\ }\href {\doibase
  10.1103/PhysRevLett.110.147002} {\bibfield  {journal} {\bibinfo  {journal}
  {Phys. Rev. Lett.}\ }\textbf {\bibinfo {volume} {110}},\ \bibinfo {pages}
  {147002} (\bibinfo {year} {2013})}\BibitemShut {NoStop}%
\bibitem [{\citenamefont {Lee}\ and\ \citenamefont {Romalis}(2008)}]{lee20081}%
  \BibitemOpen
  \bibfield  {author} {\bibinfo {author} {\bibfnamefont {S.-K.}\ \bibnamefont
  {Lee}}\ and\ \bibinfo {author} {\bibfnamefont {M.~V.}\ \bibnamefont
  {Romalis}},\ }\href {\doibase 10.1063/1.2885711} {\bibfield  {journal}
  {\bibinfo  {journal} {J. Appl. Phys.}\ }\textbf {\bibinfo {volume} {103}},\
  \bibinfo {pages} {084904} (\bibinfo {year} {2008})}\BibitemShut {NoStop}%
\bibitem [{\citenamefont {Gupta}\ and\ \citenamefont
  {Bandyopadhyay}(2011)}]{gupta20111}%
  \BibitemOpen
  \bibfield  {author} {\bibinfo {author} {\bibfnamefont {S.}~\bibnamefont
  {Gupta}}\ and\ \bibinfo {author} {\bibfnamefont {M.}~\bibnamefont
  {Bandyopadhyay}},\ }\href {\doibase 10.1103/PhysRevE.84.041133} {\bibfield
  {journal} {\bibinfo  {journal} {Phys. Rev. E}\ }\textbf {\bibinfo {volume}
  {84}},\ \bibinfo {pages} {041133} (\bibinfo {year} {2011})}\BibitemShut
  {NoStop}%
\bibitem [{\citenamefont {Czopnik}\ and\ \citenamefont
  {Garbaczewski}(2001)}]{czopnik20011}%
  \BibitemOpen
  \bibfield  {author} {\bibinfo {author} {\bibfnamefont {R.}~\bibnamefont
  {Czopnik}}\ and\ \bibinfo {author} {\bibfnamefont {P.}~\bibnamefont
  {Garbaczewski}},\ }\href {\doibase 10.1103/PhysRevE.63.021105} {\bibfield
  {journal} {\bibinfo  {journal} {Phys. Rev. E}\ }\textbf {\bibinfo {volume}
  {63}},\ \bibinfo {pages} {021105} (\bibinfo {year} {2001})}\BibitemShut
  {NoStop}%
\bibitem [{\citenamefont {Li}\ \emph {et~al.}(1990)\citenamefont {Li},
  \citenamefont {Ford},\ and\ \citenamefont {O'Connell}}]{li19901}%
  \BibitemOpen
  \bibfield  {author} {\bibinfo {author} {\bibfnamefont {X.~L.}\ \bibnamefont
  {Li}}, \bibinfo {author} {\bibfnamefont {G.~W.}\ \bibnamefont {Ford}}, \ and\
  \bibinfo {author} {\bibfnamefont {R.~F.}\ \bibnamefont {O'Connell}},\ }\href
  {\doibase 10.1103/PhysRevA.41.5287} {\bibfield  {journal} {\bibinfo
  {journal} {Phys. Rev. A}\ }\textbf {\bibinfo {volume} {41}},\ \bibinfo
  {pages} {5287} (\bibinfo {year} {1990})}\BibitemShut {NoStop}%
\bibitem [{\citenamefont {Chun}\ \emph {et~al.}(2018)\citenamefont {Chun},
  \citenamefont {Durang},\ and\ \citenamefont {Noh}}]{chun20181}%
  \BibitemOpen
  \bibfield  {author} {\bibinfo {author} {\bibfnamefont {H.~M.}\ \bibnamefont
  {Chun}}, \bibinfo {author} {\bibfnamefont {X.}~\bibnamefont {Durang}}, \ and\
  \bibinfo {author} {\bibfnamefont {J.~D.}\ \bibnamefont {Noh}},\ }\href
  {\doibase 10.1103/PhysRevE.97.032117} {\bibfield  {journal} {\bibinfo
  {journal} {Phys. Rev. E}\ }\textbf {\bibinfo {volume} {97}},\ \bibinfo
  {pages} {032117} (\bibinfo {year} {2018})},\ \Eprint
  {http://arxiv.org/abs/1710.04676} {1710.04676} \BibitemShut {NoStop}%
\bibitem [{\citenamefont {Cobanera}\ \emph {et~al.}(2016)\citenamefont
  {Cobanera}, \citenamefont {Kristel},\ and\ \citenamefont
  {Smith}}]{cobanera20161}%
  \BibitemOpen
  \bibfield  {author} {\bibinfo {author} {\bibfnamefont {E.}~\bibnamefont
  {Cobanera}}, \bibinfo {author} {\bibfnamefont {P.}~\bibnamefont {Kristel}}, \
  and\ \bibinfo {author} {\bibfnamefont {C.~M.}\ \bibnamefont {Smith}},\ }\href
  {\doibase 10.1103/PhysRevB.93.245422} {\bibfield  {journal} {\bibinfo
  {journal} {Phys. Rev. B}\ }\textbf {\bibinfo {volume} {93}},\ \bibinfo
  {pages} {245422} (\bibinfo {year} {2016})}\BibitemShut {NoStop}%
\bibitem [{\citenamefont {Diehl}\ \emph {et~al.}(2011)\citenamefont {Diehl},
  \citenamefont {Rico}, \citenamefont {Baranov},\ and\ \citenamefont
  {Zoller}}]{diehl20111}%
  \BibitemOpen
  \bibfield  {author} {\bibinfo {author} {\bibfnamefont {S.}~\bibnamefont
  {Diehl}}, \bibinfo {author} {\bibfnamefont {E.}~\bibnamefont {Rico}},
  \bibinfo {author} {\bibfnamefont {M.~A.}\ \bibnamefont {Baranov}}, \ and\
  \bibinfo {author} {\bibfnamefont {P.}~\bibnamefont {Zoller}},\ }\href@noop {}
  {\bibfield  {journal} {\bibinfo  {journal} {Nature Physics}\ }\textbf
  {\bibinfo {volume} {7}},\ \bibinfo {pages} {971} (\bibinfo {year}
  {2011})}\BibitemShut {NoStop}%
\bibitem [{\citenamefont {Viyuela}\ \emph {et~al.}(2012)\citenamefont
  {Viyuela}, \citenamefont {Rivas},\ and\ \citenamefont
  {Martin-Delgado}}]{viyuela20121}%
  \BibitemOpen
  \bibfield  {author} {\bibinfo {author} {\bibfnamefont {O.}~\bibnamefont
  {Viyuela}}, \bibinfo {author} {\bibfnamefont {A.}~\bibnamefont {Rivas}}, \
  and\ \bibinfo {author} {\bibfnamefont {M.~A.}\ \bibnamefont
  {Martin-Delgado}},\ }\href {\doibase 10.1103/PhysRevB.86.155140} {\bibfield
  {journal} {\bibinfo  {journal} {Phys. Rev. B}\ }\textbf {\bibinfo {volume}
  {86}},\ \bibinfo {pages} {155140} (\bibinfo {year} {2012})}\BibitemShut
  {NoStop}%
\bibitem [{\citenamefont {Jackiw}\ and\ \citenamefont
  {Pi}(1990)}]{jackiw19901}%
  \BibitemOpen
  \bibfield  {author} {\bibinfo {author} {\bibfnamefont {R.}~\bibnamefont
  {Jackiw}}\ and\ \bibinfo {author} {\bibfnamefont {S.-Y.}\ \bibnamefont
  {Pi}},\ }\href {\doibase 10.1103/PhysRevD.42.3500} {\bibfield  {journal}
  {\bibinfo  {journal} {Phys. Rev. D}\ }\textbf {\bibinfo {volume} {42}},\
  \bibinfo {pages} {3500} (\bibinfo {year} {1990})}\BibitemShut {NoStop}%
\bibitem [{\citenamefont {Devecchi}\ \emph {et~al.}(1995)\citenamefont
  {Devecchi}, \citenamefont {Fleck}, \citenamefont {Girotti}, \citenamefont
  {Gomes},\ and\ \citenamefont {{Da Silva}}}]{devecchi19951}%
  \BibitemOpen
  \bibfield  {author} {\bibinfo {author} {\bibfnamefont {F.}~\bibnamefont
  {Devecchi}}, \bibinfo {author} {\bibfnamefont {M.}~\bibnamefont {Fleck}},
  \bibinfo {author} {\bibfnamefont {H.}~\bibnamefont {Girotti}}, \bibinfo
  {author} {\bibfnamefont {M.}~\bibnamefont {Gomes}}, \ and\ \bibinfo {author}
  {\bibfnamefont {A.}~\bibnamefont {{Da Silva}}},\ }\href {\doibase
  10.1006/aphy.1995.1081} {\bibfield  {journal} {\bibinfo  {journal} {Ann.
  Phys.}\ }\textbf {\bibinfo {volume} {242}},\ \bibinfo {pages} {275} (\bibinfo
  {year} {1995})}\BibitemShut {NoStop}%
\bibitem [{\citenamefont {Iengo}\ and\ \citenamefont
  {Lechner}(1992)}]{iengo19921}%
  \BibitemOpen
  \bibfield  {author} {\bibinfo {author} {\bibfnamefont {R.}~\bibnamefont
  {Iengo}}\ and\ \bibinfo {author} {\bibfnamefont {K.}~\bibnamefont
  {Lechner}},\ }\href {\doibase 10.1016/0370-1573(92)90039-3} {\bibfield
  {journal} {\bibinfo  {journal} {Phys. Rep.}\ }\textbf {\bibinfo {volume}
  {213}},\ \bibinfo {pages} {179} (\bibinfo {year} {1992})}\BibitemShut
  {NoStop}%
\bibitem [{\citenamefont {Bazeia}\ and\ \citenamefont
  {Nascimento}(1997)}]{bazeia19971}%
  \BibitemOpen
  \bibfield  {author} {\bibinfo {author} {\bibfnamefont {D.}~\bibnamefont
  {Bazeia}}\ and\ \bibinfo {author} {\bibfnamefont {J.~R.~S.}\ \bibnamefont
  {Nascimento}},\ }\href {\doibase 10.1103/PhysRevD.55.1105} {\bibfield
  {journal} {\bibinfo  {journal} {Phys. Rev. D}\ }\textbf {\bibinfo {volume}
  {55}},\ \bibinfo {pages} {1105} (\bibinfo {year} {1997})}\BibitemShut
  {NoStop}%
\bibitem [{\citenamefont {Moura-Melo}\ and\ \citenamefont
  {Helay{\"{e}}l-Neto}(2001)}]{moura20011}%
  \BibitemOpen
  \bibfield  {author} {\bibinfo {author} {\bibfnamefont {W.~A.}\ \bibnamefont
  {Moura-Melo}}\ and\ \bibinfo {author} {\bibfnamefont {J.~A.}\ \bibnamefont
  {Helay{\"{e}}l-Neto}},\ }\href {\doibase 10.1103/PhysRevD.63.065013}
  {\bibfield  {journal} {\bibinfo  {journal} {Phys. Rev. D}\ }\textbf {\bibinfo
  {volume} {63}},\ \bibinfo {pages} {065013} (\bibinfo {year} {2001})},\
  \Eprint {http://arxiv.org/abs/0004143} {0004143 [hep-th]} \BibitemShut
  {NoStop}%
\bibitem [{\citenamefont {Ford}\ \emph {et~al.}(1985)\citenamefont {Ford},
  \citenamefont {Lewis},\ and\ \citenamefont {O'Connell}}]{ford19851}%
  \BibitemOpen
  \bibfield  {author} {\bibinfo {author} {\bibfnamefont {G.}~\bibnamefont
  {Ford}}, \bibinfo {author} {\bibfnamefont {J.}~\bibnamefont {Lewis}}, \ and\
  \bibinfo {author} {\bibfnamefont {R.}~\bibnamefont {O'Connell}},\ }\href
  {\doibase 10.1103/PhysRevLett.55.2273} {\bibfield  {journal} {\bibinfo
  {journal} {Phys. Rev. Lett.}\ }\textbf {\bibinfo {volume} {55}},\ \bibinfo
  {pages} {2273} (\bibinfo {year} {1985})}\BibitemShut {NoStop}%
\bibitem [{\citenamefont {Li}\ and\ \citenamefont {Tuchin}(2018)}]{li20181}%
  \BibitemOpen
  \bibfield  {author} {\bibinfo {author} {\bibfnamefont {Y.}~\bibnamefont
  {Li}}\ and\ \bibinfo {author} {\bibfnamefont {K.}~\bibnamefont {Tuchin}},\
  }\href {\doibase 10.1016/j.physletb.2017.11.063} {\bibfield  {journal}
  {\bibinfo  {journal} {Phys. Lett. B}\ }\textbf {\bibinfo {volume} {776}},\
  \bibinfo {pages} {270} (\bibinfo {year} {2018})}\BibitemShut {NoStop}%
\bibitem [{\citenamefont {Coleman}\ and\ \citenamefont
  {Norton}(1962)}]{coleman19621}%
  \BibitemOpen
  \bibfield  {author} {\bibinfo {author} {\bibfnamefont {S.}~\bibnamefont
  {Coleman}}\ and\ \bibinfo {author} {\bibfnamefont {R.~E.}\ \bibnamefont
  {Norton}},\ }\href {\doibase 10.1103/PhysRev.125.1422} {\bibfield  {journal}
  {\bibinfo  {journal} {Phys. Rev.}\ }\textbf {\bibinfo {volume} {125}},\
  \bibinfo {pages} {1422} (\bibinfo {year} {1962})}\BibitemShut {NoStop}%
\bibitem [{\citenamefont {Gradshteyn}\ and\ \citenamefont
  {Ryzhik}(2014)}]{gradshteyn20141}%
  \BibitemOpen
  \bibfield  {author} {\bibinfo {author} {\bibfnamefont {I.~S.}\ \bibnamefont
  {Gradshteyn}}\ and\ \bibinfo {author} {\bibfnamefont {I.~M.}\ \bibnamefont
  {Ryzhik}},\ }\href@noop {} {\emph {\bibinfo {title} {Table of integrals,
  series, and products}}}\ (\bibinfo  {publisher} {Academic press},\ \bibinfo
  {year} {2014})\BibitemShut {NoStop}%
\bibitem [{\citenamefont {Caruso}\ \emph {et~al.}(2013)\citenamefont {Caruso},
  \citenamefont {Helay{\"{e}}l-Neto}, \citenamefont {Martins},\ and\
  \citenamefont {Oguri}}]{caruso20131}%
  \BibitemOpen
  \bibfield  {author} {\bibinfo {author} {\bibfnamefont {F.}~\bibnamefont
  {Caruso}}, \bibinfo {author} {\bibfnamefont {J.~A.}\ \bibnamefont
  {Helay{\"{e}}l-Neto}}, \bibinfo {author} {\bibfnamefont {J.}~\bibnamefont
  {Martins}}, \ and\ \bibinfo {author} {\bibfnamefont {V.}~\bibnamefont
  {Oguri}},\ }\href {\doibase 10.1140/epjb/e2013-40282-1} {\bibfield  {journal}
  {\bibinfo  {journal} {Eur. Phys. J. B}\ }\textbf {\bibinfo {volume} {86}},\
  \bibinfo {pages} {324} (\bibinfo {year} {2013})}\BibitemShut {NoStop}%
\bibitem [{\citenamefont {Pagel}\ \emph {et~al.}(2013)\citenamefont {Pagel},
  \citenamefont {Alvermann},\ and\ \citenamefont {Fehske}}]{pagel20131}%
  \BibitemOpen
  \bibfield  {author} {\bibinfo {author} {\bibfnamefont {D.}~\bibnamefont
  {Pagel}}, \bibinfo {author} {\bibfnamefont {A.}~\bibnamefont {Alvermann}}, \
  and\ \bibinfo {author} {\bibfnamefont {H.}~\bibnamefont {Fehske}},\ }\href
  {\doibase 10.1103/PhysRevE.87.012127} {\bibfield  {journal} {\bibinfo
  {journal} {Phys. Rev. E}\ }\textbf {\bibinfo {volume} {87}},\ \bibinfo
  {pages} {012127} (\bibinfo {year} {2013})}\BibitemShut {NoStop}%
\bibitem [{\citenamefont {Chandrasekharan}(2012)}]{chandrasekharan20121}%
  \BibitemOpen
  \bibfield  {author} {\bibinfo {author} {\bibfnamefont {K.}~\bibnamefont
  {Chandrasekharan}},\ }\href@noop {} {\emph {\bibinfo {title} {Classical
  Fourier Transforms}}}\ (\bibinfo  {publisher} {Springer Science \& Business
  Media},\ \bibinfo {year} {2012})\BibitemShut {NoStop}%
\bibitem [{\citenamefont {Dhar}\ and\ \citenamefont {Wagh}(2007)}]{dhar20071}%
  \BibitemOpen
  \bibfield  {author} {\bibinfo {author} {\bibfnamefont {A.}~\bibnamefont
  {Dhar}}\ and\ \bibinfo {author} {\bibfnamefont {K.}~\bibnamefont {Wagh}},\
  }\href {\doibase 10.1209/0295-5075/79/60003} {\bibfield  {journal} {\bibinfo
  {journal} {Europhys. Lett. (EPL)}\ }\textbf {\bibinfo {volume} {79}},\
  \bibinfo {pages} {60003} (\bibinfo {year} {2007})}\BibitemShut {NoStop}%
\bibitem [{\citenamefont {Joichi}\ \emph {et~al.}(1997)\citenamefont {Joichi},
  \citenamefont {Matsumoto},\ and\ \citenamefont {Yoshimura}}]{joichi19971}%
  \BibitemOpen
  \bibfield  {author} {\bibinfo {author} {\bibfnamefont {I.}~\bibnamefont
  {Joichi}}, \bibinfo {author} {\bibfnamefont {S.}~\bibnamefont {Matsumoto}}, \
  and\ \bibinfo {author} {\bibfnamefont {M.}~\bibnamefont {Yoshimura}},\ }\href
  {\doibase 10.1143/PTP.98.9} {\bibfield  {journal} {\bibinfo  {journal}
  {Progress of Theoretical Physics}\ }\textbf {\bibinfo {volume} {98}},\
  \bibinfo {pages} {9} (\bibinfo {year} {1997})}\BibitemShut {NoStop}%
\bibitem [{\citenamefont {Petersen}\ and\ \citenamefont
  {Pedersen}(2012)}]{petersen20121}%
  \BibitemOpen
  \bibfield  {author} {\bibinfo {author} {\bibfnamefont {K.}~\bibnamefont
  {Petersen}}\ and\ \bibinfo {author} {\bibfnamefont {M.}~\bibnamefont
  {Pedersen}},\ }\href@noop {} {\emph {\bibinfo {title} {The Matrix
  Cookbook}}}\ (\bibinfo  {publisher} {Technical University of Denmark},\
  \bibinfo {year} {2012})\BibitemShut {NoStop}%
\bibitem [{\citenamefont {Xu}\ \emph {et~al.}(2014)\citenamefont {Xu},
  \citenamefont {Xiang},\ and\ \citenamefont {He}}]{xu20141}%
  \BibitemOpen
  \bibfield  {author} {\bibinfo {author} {\bibfnamefont {Z.}~\bibnamefont
  {Xu}}, \bibinfo {author} {\bibfnamefont {S.}~\bibnamefont {Xiang}}, \ and\
  \bibinfo {author} {\bibfnamefont {G.}~\bibnamefont {He}},\ }\href {\doibase
  10.1016/j.cam.2013.08.031} {\bibfield  {journal} {\bibinfo  {journal}
  {Journal of Computational and Applied Mathematics}\ }\textbf {\bibinfo
  {volume} {258}},\ \bibinfo {pages} {57} (\bibinfo {year} {2014})}\BibitemShut
  {NoStop}%
\bibitem [{\citenamefont {Spiegel}(1993)}]{spiegel19931}%
  \BibitemOpen
  \bibfield  {author} {\bibinfo {author} {\bibfnamefont {M.~R.}\ \bibnamefont
  {Spiegel}},\ }\href@noop {} {\emph {\bibinfo {title} {Theory and Problems of
  Complex Variables (Schaum's Outline)}}}\ (\bibinfo  {publisher}
  {McGraw-Hill},\ \bibinfo {year} {1993})\BibitemShut {NoStop}%
\bibitem []{ao19991}%
  \BibitemOpen
  \bibfield  {author} {\bibinfo {author} {\bibfnamefont {P.}~\bibnamefont
  {Ao}}, \ and\
  \bibinfo {author} {\bibfnamefont {X.-M.}~\bibnamefont {Zhu}},\ }\href
  {\doibase 10.1103/PhysRevB.60.6850} {\bibfield  {journal} {\bibinfo
  {journal} {Phys. Rev. B}\ }\textbf {\bibinfo {volume} {60}},\ \bibinfo
  {pages} {6850} (\bibinfo {year} {1999})}\BibitemShut {NoStop}%
\bibitem [{\citenamefont {Marino}(1993)}]{marino19931}%
  \BibitemOpen
  \bibfield  {author} {\bibinfo {author} {\bibfnamefont {E.~C.}\ \bibnamefont
  {Marino}},\ }\href {\doibase 10.1016/0550-3213(93)90379-4} {\bibfield
  {journal} {\bibinfo  {journal} {Nucl. Phys. B}\ }\textbf {\bibinfo {volume}
  {408}},\ \bibinfo {pages} {551} (\bibinfo {year} {1993})}\BibitemShut
  {NoStop}%
\bibitem [{\citenamefont {Schwartz}(1982)}]{schwartz19821}%
  \BibitemOpen
  \bibfield  {author} {\bibinfo {author} {\bibfnamefont {C.}~\bibnamefont
  {Schwartz}},\ }\href {\doibase 10.1063/1.525317} {\bibfield  {journal}
  {\bibinfo  {journal} {J. Math. Phys.}\ }\textbf {\bibinfo {volume} {23}},\
  \bibinfo {pages} {2266} (\bibinfo {year} {1982})}\BibitemShut {NoStop}%
\bibitem [{\citenamefont {Abramowitz}\ and\ \citenamefont
  {Stegun}(1964)}]{abramowitz19641}%
  \BibitemOpen
  \bibfield  {author} {\bibinfo {author} {\bibfnamefont {M.}~\bibnamefont
  {Abramowitz}}\ and\ \bibinfo {author} {\bibfnamefont {I.~A.}\ \bibnamefont
  {Stegun}},\ }\href@noop {} {\emph {\bibinfo {title} {Handbook of mathematical
  functions: with formulas, graphs, and mathematical tables}}}\ (\bibinfo
  {publisher} {Courier Corporation},\ \bibinfo {year} {1964})\BibitemShut
  {NoStop}%
\end{thebibliography}%

%

\end{document}